\documentclass[a4paper,11pt]{article}
\pdfoutput=1
\usepackage{jheppub}

\usepackage[utf8]{inputenc}

\allowdisplaybreaks
\usepackage{amsmath}
\usepackage{amssymb}
\usepackage{amsmath}
\usepackage{breqn}
\usepackage{braket}
\usepackage{xcolor}
\usepackage{graphicx}
\usepackage{slashed}
\usepackage{slashbox}

\allowdisplaybreaks

\DeclareMathOperator{\Tr}{Tr}

\newcommand{\ud}{\mathrm{d}}

\def\as{a_s}

\preprint{{MSUHEP-23-002, ZU-TH 02/23} }

\title{Renormalization of twist-two operators in covariant gauge to three loops in QCD}
\author[1]{Thomas Gehrmann,}
\emailAdd{thomas.gehrmann@uzh.ch}
\author[2]{Andreas von Manteuffel,}
\emailAdd{vmante@msu.edu}
\author[1,2]{and Tong-Zhi Yang}
\emailAdd{toyang@physik.uzh.ch}

\affiliation[1]{Physik-Institut, Universit\"at Z\"urich, Winterthurerstrasse 190, CH-8057 Z\"urich, Switzerland}
\affiliation[2]{Department of Physics and Astronomy, Michigan State University, East Lansing, MI 48824, USA}

\abstract{The leading short-distance contributions to hadronic hard-scattering cross sections in the operator product expansion are 
described by twist-two quark and gluon operators. The anomalous dimensions of these operators determine the 
splitting functions that govern the scale evolution of parton distribution functions. 
In massless QCD, these anomalous dimensions can be determined through the calculation of off-shell operator matrix elements, typically performed in a covariant gauge, where the physical operators mix with gauge-variant operators of the same 
quantum numbers. We derive a new method to systematically extract the counterterm Feynman rules resulting from these gauge-variant operators. As a first application of the new method, we rederive the unpolarized three-loop singlet anomalous dimensions, independently confirming previous results obtained with other methods.
Employing a general covariant gauge, we observe the explicit cancellation of the gauge parameter dependence in these results.
}
\keywords{Operator Product Expansion, QCD, Renormalization, Splitting Functions}

\begin{document}

\maketitle

\clearpage

\section{Introduction}
\label{sec:introduction}

The operator product expansion (OPE)~\cite{Wilson:1969zs,Frishman:1973pp} provides an elegant method to separate short-distance from long-distance contributions in quantum field theory. Its early application to deeply inelastic lepton-nucleon scattering processes~\cite{Gross:1974cs} in 
quantum chromodynamics (QCD) successfully predicted 
the violation of Bjorken scaling~\cite{Bjorken:1968dy,Bjorken:1969ja}, thereby 
enabling the development of the QCD-improved parton 
model~\cite{Altarelli:1977zs}. The anomalous dimensions 
of quark and gluon operators in the OPE are 
directly related to the Altarelli-Parisi splitting 
functions~\cite{Altarelli:1977zs,Dokshitzer:1977sg,Gribov:1972ri} of the QCD-improved parton model by an inverse Mellin transformation. 

The splitting functions determine the scale evolution of the parton distributions,
which are an essential ingredient to all quantitative predictions for high-energy hadron collider processes.
The precise determination of parton distributions requires the iterated comparison of highly accurate experimental data for a multitude of processes
with theoretical predictions at a comparable level of precision. These predictions require higher order perturbative corrections~\cite{Heinrich:2020ybq} to the underlying hard scattering processes as well as to the splitting functions. 

Splitting functions are currently known to three-loop order in QCD~\cite{Moch:2004pa,Vogt:2004mw}, which enables a consistent description of hadron collider processes at next-to-next-to-leading order (NNLO) in perturbative QCD.  
Despite its early success and its computational simplicity, the OPE method has not played a significant role in this progress towards precision QCD for collider observables. 
Its applicability at higher loop orders is limited by the currently incomplete understanding of the renormalization of singlet quark and gluon operators, which involves the mixing with so-called gauge-variant (GV) operators, that are unphysical operators resulting from the gauge fixing in QCD. Although the existence of such operators has been known~\cite{Dixon:1974ss,Kluberg-Stern:1974iel} since the initial applications of the OPE in QCD, it has not been possible to determine the number and the form of these GV operators---or even only the renormalization counterterms that result from them---beyond what is required for two-loop calculations~\cite{Hamberg:1991qt}. 

In this paper, we revisit the long-standing question of the renormalization of the leading-twist quark and gluon operators, whose anomalous dimensions determine the scale evolution of parton distribution functions. We devise a new method to extract the Feynman rules for renormalization counterterms that result from GV operators, through the computation of multi-leg operator matrix elements. We apply the newly developed method to determine all counterterms required in the OPE up to three loops in QCD. 
Section~\ref{sec:notation}
establishes our notation and briefly summarizes the OPE of QCD. We describe the calculation of 
operator matrix elements (OMEs) up to three loops in Section~\ref{sec:ComputationalMethods}. Our method for the determination of Feynman rules for operators and counterterms is developed in Section~\ref{sec:DeriveGVOperators} and, in 
Section~\ref{sec:GVFeynmanRules}, applied to compute all 
counterterms that are required for the three-loop renormalization. These counterterms are used in Section~\ref{sec:threel} to rederive the three-loop anomalous dimensions, thereby rigorously establishing their independence of the QCD gauge parameter. We conclude with an extensive summary of our method and key results in Section~\ref{sec:conc}.

 \section{Notation and formalism}
 \label{sec:notation}

 \subsection{Lagrangian density in massless QCD}

The dynamics of quarks and gluons are controlled by the QCD Lagrangian. The classical gauge invariant Lagrangian of QCD is given by
 \begin{equation}
 \mathcal{L}_{\text{c}} = \bar{\psi}_i  i \gamma^\mu (D_\mu)_{ij}  \psi_{j}  -\frac{1}{4} G^a_{\mu \nu} G_a^{\mu \nu} \,,
\label{eq:Lclass}
\end{equation}  
where $\psi_i$ is the quark field in the fundamental representation of SU(3) gauge group
and $i$ is the color index. Further, $(D_\mu)_{ij}$ is the gauge-covariant derivative in the fundamental representation,
\begin{equation}
\label{eq:derivFund}
(D_\mu)_{ij} = \partial_\mu \delta_{ij} - i g_s (T^a)_{ij} A^a_\mu \,,
\end{equation} 
and $G^{a}_{\mu \nu}$  is the gluon field strength tensor, 
\begin{equation}
G^{a}_{\mu \nu} = \partial_\mu A^a_{\nu} - \partial_\mu A^a_{\mu} + g_s f^{abc} A^b_{\mu} A^c_{\nu}\,,
\end{equation}
with gluon fields $A^a_{\mu}$ in the adjoint representation of the SU(3) gauge group. 

The classical QCD Lagrangian is invariant under the following infinitesimal gauge transformations,
\begin{equation}
A^a_\mu \to A^a_\mu + D^{ab}_\mu \theta^b \,, 
\end{equation}
where the covariant derivative $D^{ab}_\mu$ in the adjoint representation is given by 
\begin{equation}
\label{eq:derivAdj}
D^{ab}_\mu = \partial_\mu \delta^{ab} - g_s f^{abc} A^c_\mu \,.
\end{equation} 

The gauge freedom causes difficulties when trying to quantize gauge field theories. One way to deal with it is the introduction of a gauge fixing term in the Lagrangian to eliminate the problematic gauge freedom. This procedure leads to a ghost term, which cancels the unphysical longitudinal degrees of the gauge field.
The commonly used gauge-fixing and ghost terms in covariant gauge are
\begin{equation}
\label{eq:GFghost}
\mathcal{L}_{\text{GF} + \text{ghost}} = -\frac{1}{2 \xi} (\partial^\mu A^a_\mu)^2 - \bar{c}^a \partial^\mu D^{ab}_\mu c^b \,, 
\end{equation} 
where $c^a$ is the ghost field, $\bar{c}^a$ is the anti-ghost field and $\xi$ is the gauge parameter with $\xi=1$ being the 't Hooft-Feynman gauge. The full QCD Lagrangian is thus
\begin{equation}
\mathcal{L}_{\text{QCD}} = \mathcal{L}_c +\mathcal{L}_{\text{GF} + \text{ghost}}\,.
\label{eq:Lfix}
\end{equation}
in covariant gauge.

\subsection{Twist-two operators}
\label{subsec:twist2O}

To study the collinear behavior of QCD, we consider the operator product expansion.
In the following, we focus entirely on twist-two operators, which encode the collinear physics at leading power. According to the flavour group, the twist-two operators are divided into non-singlet and singlet parts. The non-singlet operators of spin $n$ and twist two are given by
\begin{equation}
\label{eq:nonsingletOP}
O^{\mu_1 \cdots \mu_n}_{q,k} = \frac{i^{n-1}}{2}  \mathcal{S} \bigg[ \bar{\psi}_{i_1} \gamma^{\mu_1} D^{\mu_2}_{i_1 i_2} D^{\mu_3}_{i_2 i_3}\cdots D^{\mu_n}_{i_{n-1} i_n} \,\frac{\lambda_k}{2} \psi_{i_n} - \text{trace terms} \bigg],\, k = 3, \cdots n_f^2-1 \,, 
\end{equation}  
where $\lambda_k/2$ is a diagonal generator of the flavour group $\text{SU}(n_f)$ and $\mathcal{S}$ denotes the symmetrization of Lorentz indices $\mu_1 \cdots \mu_n$.
Since only the traceless part of the operator is of relevance, trace terms are subtracted to render the operator definition traceless.

There are two operators in the flavor singlet case. The twist-two singlet quark and gluon operators that are obtained from the OPE are
\begin{align}
\label{eq:singletOP}
O^{\mu_1 \cdots \mu_n}_q &= \frac{i^{n-1}}{2}  \mathcal{S} \bigg[ \bar{\psi}_{i_1} \gamma^{\mu_1} D^{\mu_2}_{i_1 i_2} D^{\mu_3}_{i_2 i_3}\cdots D^{\mu_n}_{i_{n-1} i_n} \psi_{i_n} - \text{trace terms} \bigg] \,, \nonumber  \\
O^{\mu_1 \cdots \mu_n}_g &=-\frac{i^{n-2}}{2} \mathcal{S} \bigg[ G^{\mu_1}_{a_1, \mu} D^{\mu_2}_{a_1 a_2} \cdots D^{\mu_{n-1}}_{a_{n-2} a_{n-1}} G^{\mu_n \mu}_{a_{n-1} a_n}  - \text{trace terms} \bigg] \,,  
\end{align}
where the covariant derivatives in the quark and gluon operators are  defined in \eqref{eq:derivFund} and \eqref{eq:derivAdj} respectively. No momentum transfer is associated 
with the operators. 

Since the operators are traceless, we can extract the information of interest by contracting the operators with an external source
\begin{equation}
J_{\mu_1 \cdots \mu_n} = \Delta_{\mu_1} \Delta_{\mu_2} \cdots \Delta_{\mu_n}
\end{equation}
where $\Delta$ is a light-like vector $\Delta^2 = 0$.
We thus define the following spin-$n$ twist-two operators 
\begin{align}
O_{q,k} &= O^{\mu_1 \cdots \mu_n}_{q,k}  J_{\mu_1 \cdots \mu_n} \,, \nonumber \\ 
O_{q} &= O^{\mu_1 \cdots \mu_n}_{q}  J_{\mu_1 \cdots \mu_n} \,, \nonumber \\ 
O_{g} &= O^{\mu_1 \cdots \mu_n}_{g}  J_{\mu_1 \cdots \mu_n} \,,  
\end{align}
Here and in the following, the dependence of $O_{q,k},\,O_{q}$ and $O_{g}$ on $n$ is understood.

To compute matrix elements of these operators for a given set of external parton states, one needs the corresponding operator Feynman rules, which can be 
obtained from (\ref{eq:singletOP}) by a functional variation. In addition to the primary $q\bar q$ and $gg$ states for $O_q$ and $O_g$ respectively, the covariant derivatives in (\ref{eq:singletOP}) lead to Feynman rules for an arbitrary number of additional gluons.
Since each additional gluon contributes a factor $g_s$, only a finite 
number of these operator Feynman rules need to be considered at a given order in perturbation theory. 
These Feynman rules can be cast into an all-$n$ form~\cite{Ablinger:2012qm}, as outlined in Section~\ref{sec:ComputationalMethods} below.

\subsection{Renormalization of the twist-two operators}

The non-singlet operator $O_{q,k}$ is distinguished from singlet operators by quark flavour, and can therefore be renormalized separately.
The renormalized non-singlet operator is defined by
\begin{equation}
O^{\text{R}}_{q,k}  = Z_{\text{ns}} O^{\text{B}}_{q,k} \,.
\end{equation} 
where $Z_{\text{ns}}$ is a multiplicative renormalization constant.
Here and in the following, the superscripts R and B denote the renormalized and bare operators, respectively.

The two singlet operators belong to the same irreducible representation and mix with each other under renormalization. Due to the absence of 
further physical operators in the same representation, one could naively expect the following operator renormalization to hold: 
\begin{align}
\label{eq:mixingOqOg}
\left( \begin{array}{c} 
O_q \\
O_g
\end{array} \right)^{\text{R,naive}} =  \left( \begin{array}{cc} 
Z_{qq} & Z_{qg} \\
Z_{gq} & Z_{gg} 
\end{array} \right)   \left( \begin{array}{c} 
O_q \\
O_g 
\end{array} \right)^{\text{B}}\,. 
\end{align}
As we will discuss further below, this picture needs to be extended to include further, non-physical operators.

First let us note that, for non-singlet and singlet operators, the anomalous dimension $\gamma$ can be extracted from the renormalization constant according to
\begin{equation}
\label{eq:renormalizaitonEq}
 \frac{d Z }{d \ln \mu} = -2  \gamma_{ } \cdot Z  \,.
\end{equation}
For the non-singlet case, we have $Z = Z_{\text{ns}}$ and $\gamma=\gamma_{\text{ns}}$.
For the singlet case, both $Z$ and $\gamma$ are two-by-two matrices due to operator mixing:
\begin{align}
  Z =  \left( \begin{array}{cc} 
Z_{qq} & Z_{qg} \\
Z_{gq} & Z_{gg} 
\end{array} \right),
\quad
\gamma =  \left( \begin{array}{cc} 
\gamma_{qq} & \gamma_{qg} \\
\gamma_{gq} & \gamma_{gg} 
\end{array} \right). \label{eq:Z}  
\end{align} 
We expand the renormalization constants perturbatively according to
\begin{align}
Z = \sum_{i=0}^{\infty} \as^i \, Z^{(i)} \,,
\label{eq:zexp}
\end{align}
where we have defined $\as = {\alpha_s}/{(4\pi)}$.
For the anomalous dimensions, we use
\begin{align}
\gamma =  \sum_{i=0}^\infty  \as^{i+1} \, \gamma^{(i)} \,.
\end{align} 
By using the definition of the $d$-dimensional QCD $\beta$ function 
\begin{align}
 \beta(\as,\,\epsilon) = \frac{d \as }{ d \ln \mu } = -2 \epsilon \, \as - 2 \as \sum_{i=0}^{\infty} \as^{i+1} \beta_i \,, 
\end{align}
with $\epsilon$ being the dimensional regulator $\epsilon = {(4-d)}/{2}$, we can express the renormalization constant $Z$ in terms of the anomalous dimension.
In non-singlet case, the renormalization constant is 
\begin{align}
\label{eq:ZfactorIntermsofGammaNS}
Z_{\text{ns}} = &1 + a_s \frac{\gamma^{(0)}_{\text{ns}}}{\epsilon}+a_s^2 \Bigg( \frac{\gamma_{\text{ns}}^{(1)}}{2 \epsilon} + \frac{1}{2 \epsilon^2} \bigg[ -\beta_0 \gamma_{\text{ns}}^{(0)} +  \big( \gamma_{\text{ns}}^{(0)}\big)^2  \bigg] \Bigg) \nonumber 
\\
& + a_s^3  \Bigg( \frac{1}{3 \epsilon} \gamma_{\text{ns}}^{(2)} + \frac{1}{6 \epsilon^2} \bigg[ -2 \beta_1 \gamma_{\text{ns}}^{(0)} - 2 \beta_0 \gamma_{\text{ns}}^{(1)}+ 3  \gamma^{(0)}_{\text{ns}} \gamma^{(1)}_{\text{ns}}    \bigg] \nonumber 
\\
& \quad + \frac{1}{6\epsilon^3} \bigg[ 2 \beta_0^2 \gamma_{\text{ns}}^{(0)}  - 3 \beta_0  \big( \gamma_{\text{ns}}^{(0)} \big)^2  + \big( \gamma_{\text{ns}}^{(0)} \big)^3   \bigg] \Bigg)+ \mathcal{O}(a_s^4)\,. 
\end{align}
In the singlet case, one has
\begin{align}
\label{eq:ZfactorIntermsofGamma}
Z_{ij} = &\delta_{ij} + \as \frac{\gamma^{(0)}_{ij}}{\epsilon}+\as^2 \Bigg( \frac{\gamma_{ij}^{(1)}}{2 \epsilon} + \frac{1}{2 \epsilon^2} \bigg[ -\beta_0 \gamma_{ij}^{(0)} + \sum_{k=q,\,g} \gamma^{(0)}_{ik} \gamma^{(0)}_{kj}  \bigg] \Bigg) \nonumber 
\\
& + \as^3  \Bigg( \frac{1}{3 \epsilon} \gamma_{ij}^{(2)} + \frac{1}{6 \epsilon^2} \bigg[ -2 \beta_1 \gamma_{ij}^{(0)} - 2 \beta_0 \gamma_{ij}^{(1)} +2 \sum_{k=q,\,g} \gamma^{(1)}_{ik} \gamma^{(0)}_{kj}  +\sum_{k} \gamma^{(0)}_{ik} \gamma^{(1)}_{kj}    \bigg] \nonumber 
\\
& \quad + \frac{1}{6\epsilon^3} \bigg[ 2 \beta_0^2 \gamma_{ij}^{(0)}  - 3 \beta_0 \sum_{k=q,\,g} \gamma_{ik}^{(0)} \gamma_{kj}^{(0)} + \sum_{k=q,\,g} \sum_{l=q,\,g} \gamma^{(0)}_{ik} \gamma^{(0)}_{kl} \gamma^{(0)}_{lj}   \bigg] \Bigg)+ \mathcal{O}(a_s^4)\,,
\end{align}
where $i,j=q,g$.
By extracting the anomalous dimensions, one can subsequently determine the splitting functions, see the end of section~\ref{sec:threel}.

It was already pointed out by Gross and Wilczek in the first calculation of the one-loop singlet anomalous dimensions~\cite{Gross:1974cs} that the 
quark and gluon operators may mix with further GV operators under renormalization. These GV operators originate from the interplay of gauge-fixing 
at the QCD Lagrangian level and the OPE, with the twist-two operator basis (\ref{eq:singletOP}) being obtained from the classical gauge-invariant QCD Lagrangian (\ref{eq:Lclass}) prior to gauge fixing. 

The derivation of a consistent OPE 
on the basis of the gauge-fixed QCD Lagrangian (\ref{eq:Lfix}) has been investigated extensively in the literature. In the seminal work of Dixon and Taylor~\cite{Dixon:1974ss}, the set of GV operators relevant at order $g_s$ was constructed explicitly. These results demonstrated that the 
 naive renormalization of the singlet operators is consistent at the one-loop level, and, subsequently, enabled the construction of the correct singlet 
 renormalization at two loops~\cite{Hamberg:1991qt}, thereby resolving earlier inconsistencies~\cite{Floratos:1978ny,Gonzalez-Arroyo:1979qht}. All-order renormalization
 conditions  for the gauge-invariant operators $O_q$ and 
 $O_g$ were derived by Joglekar and Lee~\cite{Joglekar:1975nu}, confirming the results of~\cite{Dixon:1974ss} but lacking a procedure for the 
 construction of GV operators at higher orders in $g_s$.  An independent approach to the renormalization of the singlet operators 
 was based on the renormalization of the QCD energy-momentum tensor~\cite{Kluberg-Stern:1974nmx,Kluberg-Stern:1975ebk}, 
 equally yielding~\cite{Collins:1994ee} the GV operators at order $g_s$. This order is sufficient for the extraction  of the 
 anomalous dimension matrix of the singlet operators up to two loops. Previous calculations of the three-loop anomalous dimensions 
 did not employ the OPE, but used the forward scattering amplitude in deep inelastic scattering~\cite{Vogt:2004mw}, inclusive hadron collider cross sections~\cite{Mistlberger:2018etf,Duhr:2020seh} or beam functions~\cite{Luo:2019szz,Ebert:2020yqt,Ebert:2020unb,Luo:2020epw,Baranowski:2022vcn} to determine the three-loop mass factorization counterterms of parton distributions, which contain the required anomalous dimensions. 
 
Most recently, Falcioni and Herzog~\cite{Falcioni:2022fdm} translated the conditions formulated by Joglekar and Lee~\cite{Joglekar:1975nu} into a set of constraint equations, that allow to infer the GV  operators and their associated renormalization constants 
 at fixed $n$, order-by-order in the number of loops and powers of the strong coupling constant. They demonstrated their method in the derivation 
 of the three-loop anomalous dimensions for $n\leq 6$ and the four-loop anomalous dimensions for $n \leq 4$.
 
 In the following, we propose to overcome the current lack of understanding of the full structure of the GV operators by devising a procedure for the direct 
 extraction of the  all-$n$ counterterm Feynman rules resulting from these operators. 
We start from the generic form of the renormalization of the singlet operators, including  their mixing with the GV operators. 

The most general form of the renormalization for $O_q$ and $O_g$ can be written as follows,
\begin{eqnarray}
O^{\text{R}}_q &=& Z_{qq} O^{\text{B}}_q + Z_{qg} O^{\text{B}}_g + \sum_{i=1}^\infty \sum_{j=1}^{N_i}
\left( Z_{qA_{i,j}} O^{\text{B}}_{A_{i,j}} + Z_{qB_{i,j}} O^{\text{B}}_{B_{i,j}} + Z_{qC_{i,j}} O^{\text{B}}_{C_{i,j}}\right)\,,  \nonumber
 \\
O^{\text{R}}_g &=& Z_{gq} O^{\text{B}}_q + Z_{gg} O^{\text{B}}_g + \sum_{i=1}^\infty \sum_{j=1}^{N_i}
\left(  Z_{gA_{i,j}} O^{\text{B}}_{A_{i,j}} + Z_{gB_{i,j}} O^{\text{B}}_{B_{i,j}} + Z_{gC_{i,j}} O^{\text{B}}_{C_{i,j}}\right)\,.
\label{eq:RenorGV}
\end{eqnarray} 
In the above equations, the GV operators carry subscripts $A\,,B$ and $C$ to distinguish three kinds of operators: $O_{A_{i,j}}$ involve gluon fields only, $O_{B_{i,j}}$ two quark fields plus gluon fields, and $O_{C_{i,j}}$ two ghost fields plus gluon fields. The index $i$ assigned to the operators is used to indicate the order in 
the strong coupling constant at which each operator 
starts to contribute to the renormalization of the 
physical quark and gluon operators, 
such that the renormalization constants $Z_{qA_{i,j}}, Z_{qB_{i,j}}, Z_{qC_{i,j}}$ start at $\mathcal{O}{(a_s^{i+1})}$, and $Z_{gA_{i,j}}, Z_{gB_{i,j}}, Z_{gC_{i,j}}$ start at $\mathcal{O}{(a_s^{i})}$. In principle, the index $i$ extends to infinity.  However, we only need operators up to finite $i$ for practical computations at a given fixed loop order. For example, we need $i\leq 2$  to extract three-loop splitting functions and $i\leq 3$ to extract four-loop splitting functions. We also assign another index $j$ to enumerate $N_i$ different operator structures, which contribute to the same index $i$ but require their own ($n$ dependent) renormalization constant.
We note that the number $N_i$ is in general not known and could a priori be infinite.

Our key idea is to directly extract the counterterm Feynman rules resulting from a linear combination of operators instead of determining the operators themselves or determining the Feynman rules resulting from each operator separately. From explicit computations described in Section~\ref{sec:GVFeynmanRules} below, we find that $N_1=1$ and $N_2 > 1$; it even seems possible that $N_{\geq 2} = \infty$.
In other words, we can disentangle the operator from its corresponding renormalization constant for $i=1$, but not for $i\geq 2$. Furthermore, we find that the type A, B, and C renormalization constants for $i=1$ are identical at the lowest order in $a_s$. We expect this to hold to all orders, such that the following relations are fulfilled:
\begin{align}
\label{eq:conjectureZRelation}
    & Z_{qA_{1,1}} = Z_{qB_{1,1}} = Z_{qC_{1,1}} \equiv  Z_{qA}\,, \nonumber \\
    & Z_{gA_{1,1}} = Z_{gB_{1,1}} = Z_{gC_{1,1}} \equiv  Z_{gA} \,.
\end{align}
It is then sufficient to consider the following combination of $i=1$ type operators,
\begin{align}
\label{eq:combination}
    O_{ABC} = O_{A} + O_{B}+ O_C \,,  
\end{align}
where we have abbreviated $O_{A_{1,1}} \equiv O_A$, $O_{B_{1,1}} \equiv O_B$, and $O_{C_{1,1}} \equiv O_C$.  
It is possible to prove the relations~\eqref{eq:conjectureZRelation} in the context of BRST symmetry, noting that only the combination shown in \eqref{eq:combination} is compatible with the requirement of transversity of the physical gluon fields. Given the simplicity  of the $i=1$ type operators, as compared to all $i\geq 2$ type operators, it is natural to treat them separately. By defining the following counterterm operators as the linear combination of all $i\geq 2$ type GV operators,  
\begin{eqnarray}
  \left[Z O\right]_{q}^{\textrm{GV}} =  \sum_{i=2}^\infty \sum_{j=1}^{N_i}
\left( Z_{qA_{i,j}} O^{\text{B}}_{A_{i,j}} + Z_{qB_{i,j}} O^{\text{B}}_{B_{i,j}} + Z_{qC_{i,j}} O^{\text{B}}_{C_{i,j}}\right) \,, \nonumber \\ 
 \left[Z O\right]_{g}^{\textrm{GV}} =  \sum_{i=2}^\infty \sum_{j=1}^{N_i}
\left( Z_{gA_{i,j}} O^{\text{B}}_{A_{i,j}} + Z_{gB_{i,j}} O^{\text{B}}_{B_{i,j}} + Z_{gC_{i,j}} O^{\text{B}}_{C_{i,j}}\right)\,,
\end{eqnarray}
equation \eqref{eq:RenorGV} can be simplified as follows:
\begin{eqnarray}
O^{\text{R}}_q &=& Z_{qq} O^{\text{B}}_q + Z_{qg} O^{\text{B}}_g + Z_{qA} \left( O^{\text{B}}_{A}+O^{\text{B}}_{B}+O^{\text{B}}_{C} \right)+  \left[Z O\right]_{q}^{\textrm{GV}}\,, 
 \label{eq:RenorGVSimQ} \\
O^{\text{R}}_g &=& Z_{gq} O^{\text{B}}_q + Z_{gg} O^{\text{B}}_g + Z_{gA} \left( O^{\text{B}}_{A} +O^{\text{B}}_{B}+O^{\text{B}}_{C} \right)+ \left[Z O\right]_{g}^{\textrm{GV}}\,.
\label{eq:RenorGVSimG}
\end{eqnarray}
We used symbols $\left[Z O\right]_{q}^{\textrm{GV}}$ and $\left[Z O\right]_{g}^{\textrm{GV}}$ for the reason that we can not disentangle the renormalization constant from the corresponding operator, otherwise, we should write them as $Z_{qV}\, O_V$ and $Z_{gV} \, O_V$, where $V$ stands for any $i\geq 2$ type GV operator. The counterterm operators can be further decomposed as follows according to the number of loops,
\begin{align}
    \label{eq:CODec}
    \left[Z O\right]_{q}^{\textrm{GV}}  = \sum_{l=3}^\infty a_s^{l} \left[Z O\right]_{q}^{\textrm{GV},\,\left(l\right)} \,, \quad
   \left[Z O\right]_{g}^{\textrm{GV}}  = \sum_{l=2}^\infty a_s^l \left[Z O\right]_{g}^{\textrm{GV},\,\left(l\right)}\,,
\end{align}
where the expansions apply to the renormalization constants, for example 
\begin{align}
\left[Z O\right]_{g}^{\textrm{GV},\,(2)} = \sum_{j=1}^{N_2} \left( Z_{gA_{2,j}}^{(2)} O_{A_{2,j}} +  Z_{gB_{2,j}}^{(2)} O_{B_{2,j}}  +  Z_{gC_{2,j}}^{(2)} O_{C_{2,j}} \right)\,. 
\end{align}

As we stated above, the renormalization of physical operators  mixes with GV operators. However, the renormalization of GV operators can not mix with physical operators, as shown by Joglekar and Lee~\cite{Joglekar:1975nu}. It means that the eigenvalues of the corresponding mixing matrix factorize into physical and non-physical parts. In our context, it implies that the inclusion of GV operators affects only the determination of physical renormalization constants $Z_{kj}$ with $k,j=q \text{ or } g$, but not the extraction of physical anomalous dimensions from the renormalization constants, such that \eqref{eq:ZfactorIntermsofGamma} remains valid even with the presence of GV operators. By also considering the renormalization of the $i=1$ type GV operators, the equations \eqref{eq:RenorGVSimQ} and \eqref{eq:RenorGVSimG} are written as the following matrix form, 
\begin{align}
\label{eq:mixingOqOgOA}
\left( \begin{array}{c} 
O_q \\
O_g \\
O_{ABC}\\
\end{array} \right)^{\text{R}} =  \left( \begin{array}{ccc} 
Z_{qq} & Z_{qg} & Z_{qA} \\
Z_{gq} & Z_{gg} & Z_{gA}  \\ 
0 & 0 & Z_{AA}  \\
\end{array} \right)   \left( \begin{array}{c} 
O_q \\
O_g \\
O_{ABC}\\
\end{array} \right)^{\text{B}} + 
\left( 
\begin{array}{c}
\left[Z O\right]_{q}^{\textrm{GV}}\\
\left[Z O\right]_{g}^{\textrm{GV}}\\
\left[Z O\right]_{A}^{\textrm{GV}}
\end{array}
\right)^{\text{B}} 
,
\end{align}
where we have introduced another counterterm operator $\left[ZO\right]_{A}^{\textrm{GV}}$ renormalizing the operator $O_{ABC}$ defined in \eqref{eq:combination}.

We emphasize again 
that the distinction between operators of type $i=1$ and of types 
$i\geq 2$ in (\ref{eq:mixingOqOgOA}) is a choice made by us for the sake of 
computational simplicity, and will be justified in detail in Section~\ref{sec:GVFeynmanRules} below. Alternatively, one could choose not to single out the $i=1$ operator $O_{ABC}$ and include its contributions in the counterterm operators used for $i \geq 2$.

\subsection{Operator matrix elements}

To extract the renormalization constants as well as to derive the
counterterm Feynman rules of the GV operators, we need to introduce the concept of OMEs which are defined as correlation functions or matrix elements with an operator insertion
\begin{align}
\label{eq:OMEsDe}
A^{}_{ij} = \braket{j(p)|O_i|j(p)}^{},
\end{align}   
where $O_i$ is a twist-two operator, $j$ denotes a quark, gluon or ghost state, and $p$ is the momentum of $j$. Up to three-loop order, the results for the above OMEs can be conveniently expressed in terms of Casimir invariants $C_A$, $C_F$, and $d_{abc} d^{abc}$, and the number of massless quark flavours $N_f$.
For an $SU(N_c)$ gauge group, the Casimir invariants or color factors read
\begin{align}
     C_A &= N_c\,,
    \nonumber\\
     C_F &= \frac{N_c^2-1}{2 N_c}\,,
    \nonumber\\
     d_{abc} d^{abc} &= 4 \Tr[( T_a T_b +T_b T_a ) T_c ] \,\Tr[( T^a T^b +T^b T^a ) T^c ] = \frac{(N_c^2-1)(N_c^2-4)}{N_c}\,,
\end{align}
with $N_c=3$ for QCD. We use the normalization
\begin{align}
    \Tr(T^a T^b) = T_F \delta^{ab} = \frac{1}{2} \delta^{ab}\,.
\end{align}

The OMEs with two-quark external states can be decomposed into their physical part and an equation-of-motion (EOM) part,
which are described by form factors $A^{\text{Phy}}_{iq}$ and $A^{\text{EOM}}_{iq}$, respectively,
\begin{align}
A_{iq}  = \braket{q(p)|O_i|q(p)} =  A^{\text{Phy}}_{iq}  (\Delta 
 \cdot p )^{n-1}  \slashed{\Delta} + A^{\text{EOM}}_{iq} (\Delta 
 \cdot p )^n\slashed{p}\,.  
\label{eq:Aiq}
\end{align}
The OMEs with two-ghost external states involve only a single form factor, 
\begin{align}
A_{ic}  = \braket{c(p)|O_i|c(p)} =   A^{\text{}}_{ic,\,1}\,  (\Delta 
 \cdot p )^n \,, 
\label{eq:Aic}
\end{align}
where we use $c$ to represent a ghost state and the index 1 in $A_{ic,\,1}$ is used to distinguish $A_{ic,\,1}$ from $A_{ic}$. Similarly, the decomposition of OMEs with two-gluon external states is given in terms of four form factors,  
\begin{align}
A^{\mu \nu}_{ig}  = \braket{g(p)|O_i|g(p)}^{\mu \nu} =  \sum_{k=1}^4  A_{ig,\,k} T_k^{\mu \nu} \,,      
\label{eq:Aig}
\end{align}
with the four tensor structures
\begin{align}
\label{eq:tensorStructure}
T^{\mu \nu}_1 &= \frac{1+(-1)^n}{2} \bigg[ (\Delta \cdot p)^2 g^{\mu \nu} - \Delta \cdot p \left( p^\mu \Delta^\nu + \Delta^\mu p^\nu \right) + \Delta^\mu \Delta^\nu p^2 \bigg] (\Delta \cdot p )^{n-2}  \,, \nonumber \\
T^{\mu \nu}_2 &= \frac{1+(-1)^n}{2} \bigg[ \frac{p^\mu p^\nu}{p^2} (\Delta \cdot p )^2 - \Delta \cdot p \left( p^\mu \Delta^\nu + \Delta^\mu p^\nu \right) + \Delta^\mu \Delta^\nu p^2 \bigg] (\Delta \cdot p)^{n-2} \,, \nonumber \\
T^{\mu \nu}_3 &= \frac{1+(-1)^n}{2} \bigg[-\Delta \cdot p \left(p^\mu \Delta^\nu + \Delta^\mu p^\nu  \right) + 2 \Delta^\mu \Delta^\nu p^2 \bigg] (\Delta \cdot p)^{n-2} \,, \nonumber \\
T^{\mu \nu}_4 &= \frac{1+(-1)^n}{2} \bigg[ \Delta \cdot p \left(p^\mu \Delta^\nu + \Delta^\mu p^\nu  \right) + 2 \Delta^\mu \Delta^\nu p^2 \bigg] (\Delta \cdot p)^{n-2} \,, 
\end{align}
where $n\geq 2$. The above tensor structures satisfy the relations
\begin{align}
p_\mu T^{\mu \nu}_i &=0 \quad ( i=1,2), & p_\mu T^{\mu \nu}_i &\neq 0 \quad( i=3,4) \,, \\ 
p_\mu p_\nu T^{\mu \nu}_i &=0 \quad(i=1,2,3), & p_\mu p_\nu T^{\mu \nu}_i &\neq 0 \quad( i=4) \,.  
\end{align}  

Once we determined all GV counterterm Feynman rules for the renormalization of the physical twist-two operators, the physical, EOM, or non-physical form factors in \eqref{eq:Aiq} and \eqref{eq:Aig} should be renormalized independently. For the purpose of extracting the splitting functions, it is sufficient to consider the renormalization of the following form factors, 
\begin{align}
\label{eq:sumoverSpinColorF}
   &\mathcal{F}_{iq} = \frac{1}{ 2 N_c } \cdot \frac{\text{Tr}\left( \slashed{p} A_{iq} \right)}{  (\Delta \cdot p )^n}  \,, 
   \nonumber\\
   & \mathcal{F}_{ig} = \frac{1
   }{(d-2) (N_c^2-1)}  \cdot \frac{(-g_{\mu \nu} A_{ig}^{\mu \nu})}{ (\Delta \cdot p )^n}   \,, 
   \nonumber\\
   &  \mathcal{F}_{ic} = \frac{1}{(d-2) (N_c^2-1) } \cdot \frac{  A_{ic}}{ (\Delta \cdot p )^n} \,, 
\end{align}
where we averaged and summed over spins and colors of incoming and outgoing states, respectively.
We employ the perturbative expansion
\begin{equation}
\mathcal{F} = \sum_{l=0}^\infty \mathcal{F}^{(l)} a_s^l
\end{equation}
for the form factors.
With the above definitions, both $\mathcal{F}_{qq}$ and $\mathcal{F}_{gg}$ are normalized to unity at lowest order,
\begin{align}
    \mathcal{F}_{qq}^{(0)} =  \mathcal{F}_{gg}^{(0)} = 1\,.
\end{align}

If the external state $j$ in \eqref{eq:OMEsDe} is on-shell,
the OMEs of the GV operators do not mix under renormalization with those of the physical operators and 
the naive renormalization procedure \eqref{eq:mixingOqOg} remains valid. Likewise, the GV operators do not mix with 
the polarized analogues of the singlet operators
\eqref{eq:singletOP}, which involve the difference of 
the quark or gluon spin states instead of their sums. 
This can be understood from the structure of the Fock space in a covariant gauge, where the ghost degrees of freedom mix only with the unphysical gauge field polarization states. In unpolarized OMEs, all polarization states (including the unphysical ones) are summed over, while polarized OMEs are constructed as differences between the two physical polarization states, thereby decoupling the unphysical sector of the Fock space.  
Consequently, the renormalization of polarized 
operators
also fulfils \eqref{eq:mixingOqOg}. 

Another approach to avoid GV operators 
is to perform the quantization in an axial gauge, which 
requires to introduce a light-like reference
direction $n^\mu$ for the quantization. In an axial gauge,
the Fock spaces of physical and unphysical polarizations 
decouple completely, again allowing for the naive operator
renormalization \eqref{eq:mixingOqOg}.

The form factors appearing on the right-hand sides of 
\eqref{eq:Aiq}, \eqref{eq:Aic}, and \eqref{eq:Aig} are Lorentz scalars. By 
dimensional analysis, one finds that they vanish trivially
in massless QCD for external on-shell states if dimensional regularization is applied. To generate a mass scale, one either needs to insert an internal mass scale 
or to consider off-shell external states $p^2 < 0$. 
Working with an internal quark mass, three-loop on-shell OME 
have been used to extract heavy-flavour contributions 
to the three-loop anomalous 
dimensions~\cite{Ablinger:2014nga,Ablinger:2014vwa,Behring:2019tus,Ablinger:2022wbb,Bierenbaum:2022biv}, 
as well as for the computation of a subset of the genuine
massless anomalous dimensions at three loops~\cite{Ablinger:2017tan}. 

In terms of computational simplicity, the 
calculations of purely massless 
off-shell OMEs are preferable over on-shell OMEs with 
an internal mass, since the underlying 
all-massless Feynman integrals are considerably 
simpler than corresponding integrals involving an 
internal mass scale. 

Massless off-shell OMEs have been 
computed up to three-loop order
and used for the determination 
of the three-loop 
non-singlet anomalous dimensions~\cite{Blumlein:2021enk}
and for the polarized singlet anomalous 
dimensions at two loops~\cite{Mertig:1995ny,Vogelsang:1995vh} 
and three 
loops~\cite{Blumlein:2021ryt}. In all these cases, 
the GV operators do not contribute. 

The first successful extraction
of the two-loop singlet anomalous dimensions~\cite{Curci:1980uw,Furmanski:1980cm,Ellis:1996nn} was based on performing QCD quantization in an axial 
gauge (where the GV operators decouple~\cite{Bassetto:1998uv}) and computing all-massless off-shell OMEs. Owing to the presence of the gauge vector $n^\mu$, the resulting Feynman 
integrals are considerably more 
complicated than in a covariant gauge, such that the use 
of an axial gauge is not a viable option at the higher loop 
orders.

\section{Computation of bare OMEs up to three loops}
 \label{sec:ComputationalMethods}

Before working out the 
Feynman rules for the GV operators, we first explain how to compute the bare OMEs for the physical operators $O_q$ and $O_g$ to three-loop order. The Feynman rules 
resulting from these operators are summarized in 
Appendix~\ref{sec:FeynOqOg}. 

\begin{figure}
\begin{center}
\begin{minipage}{4.5cm}
\includegraphics[scale=0.9]{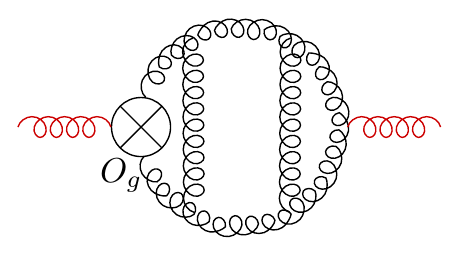} 
\end{minipage}
\begin{minipage}{4.5cm}
\includegraphics[scale=0.9]{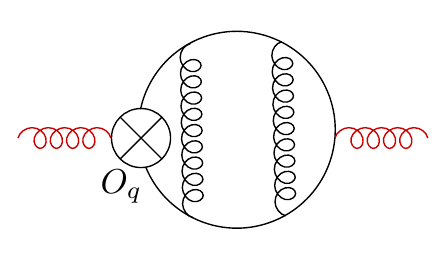} 
\end{minipage}
\begin{minipage}{4.5cm}
\includegraphics[scale=0.9]{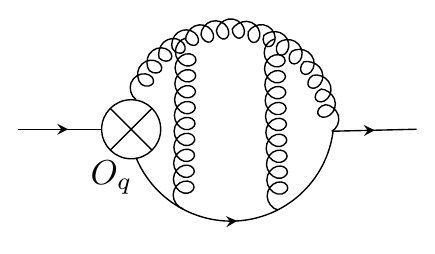} 
\end{minipage}
\begin{minipage}{4.5cm}
\includegraphics[scale=0.9]{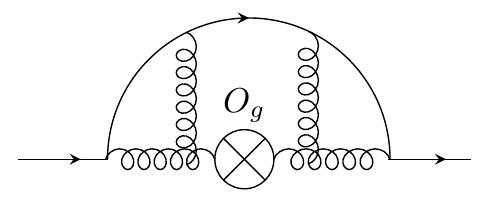} 
\end{minipage}
\begin{minipage}{4.5cm}
\includegraphics[scale=0.9]{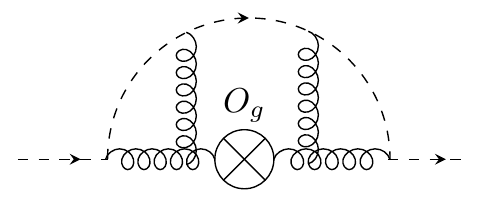} 
\end{minipage}
\begin{minipage}{4.5cm}
\includegraphics[scale=0.9]{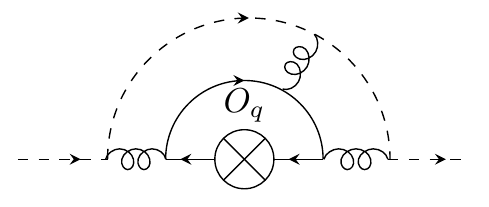} 
\end{minipage}
\caption{Representative 3-loop Feynman diagrams with physical operator insertions to extract 3-loop splitting functions.}
\label{fig:dia3loop2legs}
\end{center}
\end{figure}

As the first step, we generate all relevant Feynman diagrams to three-loop order using \texttt{QGRAF}~\cite{Nogueira:1991ex}. Some representative diagrams are shown in Fig.~\ref{fig:dia3loop2legs}. To translate the diagrams into expressions, one needs the standard QCD Feynman rules as well as the Feynman rules for the twist-two operators. As shown in the Appendix~\ref{sec:FeynOqOg}, the Feynman rules for the twist-two operators are non-standard, in the sense that they involve terms like $(\Delta \cdot p)^n$ with $n$ being an arbitrary non-negative integer. If $n$ is fixed to a specific integer, as adopted in~\cite{Moch:2021qrk,Falcioni:2022fdm}, the Feynman rules become standard and the relevant Feynman integrals are the two-point integrals. 

Here, we want to keep $n$ arbitrary, in order to 
directly obtain the all-$n$ splitting functions. However, it is not immediately obvious how to perform the integration-by-parts (IBP) reductions~\cite{Chetyrkin:1981qh} with terms like $(\Delta \cdot p)^n$ present. In the following, we adopt a method first proposed in~\cite{Ablinger:2012qm,Ablinger:2014nga}. The method sums terms proportional to $(\Delta \cdot p)^n$ to linear propagators with the help of a tracing parameter $x$, for example,  
 \begin{align}
 \label{eq:sumXmethod}
&(\Delta \cdot p)^n \to \sum_{n=0}^\infty x^n (\Delta \cdot p)^n = \frac{1}{1-x \Delta \cdot p } \,, \nonumber \\
 &  \sum^{n-3}_{j=0} (\Delta \cdot p_1)^{n-3-j} (\Delta \cdot p_2)^{j} \to \sum^{\infty}_{n=3} x^n \sum^{n-3}_{j=0} (\Delta \cdot p_1)^{n-3-j} (\Delta \cdot p_2)^{j} = \frac{x^3}{(1-x \Delta\cdot p_1)(1- x \Delta \cdot p_2)} \,.    
 \end{align}
The above method translates the non-standard terms depending on $n$ into standard  linear propagators depending on $x$, which can be easily handled by standard IBP algorithms. To revert back to $n$ space, we symbolically extract the coefficient of $x^n$ from the results in the $x$-parameter space. It should be noted that the limit $x\to 0$ trivializes the linear propagators and that the corresponding Feynman integrals
converge to massless two-point functions in this limit. 
The auxiliary parameter $x$ should not be confused with the Bjorken-$x$ variable, which relates to $n$ by an inverse Mellin transformation.

We work in parameter-$x$ space throughout, starting at the level of the Feynman rules, which are all transformed from 
$n$ space to parameter-$x$ space. Mathematica is used to substitute the standard QCD Feynman rules and the effective Feynman rules in parameter $x$-space into the Feynman diagrams. The Dirac and color algebra is performed with \texttt{FORM}~\cite{Vermaseren:2000nd}. Subsequently, the Feynman integrals are classified into different integral families with 
an in-house code invoking \texttt{Reduze 2}~\cite{vonManteuffel:2012np} and the latest version of \texttt{FeynCalc}~\cite{Shtabovenko:2016sxi,Shtabovenko:2021hjx}. During the topology classification, partial fractions with respect to the Feynman propagators are also needed, and they are performed using \texttt{Apart}~\cite{Feng:2012iq}.

For the singlet operator insertions, we find 1, 2, and 7 integral families at one, two, and three loop order, respectively.
Here, an integral family is defined as a complete set of quadratic and linear denominator polynomials, such that any scalar product of a loop momentum can be expressed in terms of them.
In general, one integral family may describe more than one ``top-level topology''.
As an example, we give the 2 integral families at two-loop order, 
\begin{align}
\left(
\begin{array}{ccccccc}
 1-x \Delta \cdot l_1\,, & 1-x \Delta \cdot l_2 \,,& l_1^2 \,,& \left(l_1-p\right){}^2 \,,&
   l_2^2 \,,& \left(l_2-p\right){}^2 \,,& \left(l_1-l_2\right){}^2\\
 1-x \Delta \cdot \left(l_2-l_1\right)\,, & 1-x \Delta \cdot l_2 \,,& l_1^2 \,,&
   \left(l_1-p\right){}^2 \,,& l_2^2 \,,& \left(l_2-p\right){}^2 \,,&
   \left(l_1-l_2\right){}^2 \\
\end{array}
\right)\,,
\end{align}
where $l_1,\,l_2$ are the loop momenta. Each integral family contains 7 propagators with the first two being the linear propagators obtained through \eqref{eq:sumXmethod}. We perform the IBP reduction using the combination of \texttt{LiteRed}~\cite{Lee:2012cn} and \texttt{FIRE6}~\cite{Smirnov:2019qkx}, \texttt{Reduze 2}~\cite{vonManteuffel:2012np} and \texttt{Kira}~\cite{Klappert:2020nbg}, which are 
based on implementations of the Laporta 
algorithm~\cite{Laporta:2000dsw}.

To directly compute the master integrals in parameter-$x$ space, we derive differential 
equations (DEs)~\cite{Gehrmann:1999as} with respect to $x$ for the master integrals. The DE system is turned into canonical form~\cite{Henn:2013pwa} using a combination of \texttt{CANONICA}~\cite{Meyer:2017joq, Meyer:2016slj} and \texttt{Libra}~\cite{Lee:2014ioa,Lee:2020zfb}. Due to \eqref{eq:sumXmethod}, the master integrals are regular in the limit $x \to 0$ and reduce to  two-point integrals. The two-point integrals are known to four-loop order~\cite{Baikov:2010hf,Lee:2011jt} and serve as the boundary conditions for the DEs. In this way, we manage to express all master integrals up to three loops in terms of harmonic polylogarithms (HPLs) \cite{Remiddi:1999ew}, which are defined by
\begin{align}
H(a_1,\,a_2,\,\cdots,\, a_m;x) &= \int_{0}^x \ud t\, f_{a_1}(t) H(a_2,\,\cdots,\,a_m;t)\,,\nonumber\\
H(\vec{0}_m,;x) &= \frac{\ln^n x}{n!}\,,\nonumber\\
H(x)&=1\,,
\end{align}
where $a_i$ is $0$, $1$, or $-1$, and the kernel $f_{a}(t)$ is defined as 
\begin{align}
f_1(t) = \frac{1}{1-t}\,,\quad f_0(t) = \frac{1}{t}\,, \quad f_{-1}(t) = \frac{1}{1+t}\,.
\end{align}
Inserting IBP relations and the solutions for the master integrals into the integrand, we obtain the final expressions for the bare OMEs. 

As the last step before renormalization, we turn the results  from the parameter-$x$ space into $n$ space. The HPLs can be expanded around $x=0$, 
\begin{align}
\label{eq:HPLtoHS}
H(a_1,\,a_2,\,\cdots; x) = \sum_{n=0}^{\infty} b_n x^n \,,
\end{align}  
with $b_n$ being composed of the harmonic sums~\cite{Vermaseren:1998uu,Blumlein:1998if} of argument $n$, which are  defined recursively by
\begin{align}
\label{eq:HarmonicDefinition}
S_{\pm m_1, \,m_2,\,\cdots m_d}(n) &= \sum_{j=1}^{n} (\pm 1)^{j} j^{-m_1} S_{m_2,\,\cdots m_d}(j) \quad(m_i \in \mathbb{N}),\nonumber\\
S_\emptyset(n)&=1\,.
\end{align}
We use the Mathematica package \texttt{HarmonicSums}~\cite{Ablinger:2009ovq,Ablinger:2012ufz,Ablinger:2014rba,Ablinger:2011te,Ablinger:2013cf,Ablinger:2014bra} to perform the expansion \eqref{eq:HPLtoHS} of the HPLs.
In order to extract the relevant coefficients of $x^n$, one needs to take into account also the rational prefactors multiplying the HPLs.
In our non-singlet computation, we find powers of $x$, $1/x$, and $1/(1-x)$ after partial fraction decomposition. 
For the singlet case, we also encounter the factor $1/(1+x)$, which can, however, always be mapped to $1/(1-x)$ due to the fact that only even Mellin moments contribute. At this stage, an OME $G(x)$ in parameter-$x$ space can be written as 
\begin{align}
\label{eq:AmplitudeXspaceTot}
G(x) =  \sum_{m=0}^{m_{\text{max}}} F_m(x)  \,,
\end{align} 
with $F_m(x)$ being defined as
\begin{align}
\label{eq:AmplitudeXspace}
F_m(x) = \frac{ B_m(x)}{(1-x)^m}  \,,
\end{align} 
where $B_m(x)$ is a linear combination of HPLs with coefficients involving powers of $x$ only, and $m$ is a non-negative integer.
Prior to simplifications, we find $m_{\text{max}}=3$ for individual contributions.
We expand both $F_m(x)$ and $B_m(x)$ of \eqref{eq:AmplitudeXspace}, 
\begin{align}
\label{eq:AmplitudeXspaceExpansion}
F_m(x) = \sum f_{n} x^n =   \frac{1}{(1-x)^m}\bigg[ {\sum^{\alpha-1}_{j_1=0} h_{j_1} x^{j_1} + \sum^{\infty}_{j_1=\alpha} c_{j_1} x^{j_1} } \bigg]  \,, 
\end{align} 
where $f_n$ is the
the desired result in Mellin space and $\alpha$ is a small integer that is used to separate $B(x)$ into two parts. The coefficients $h_{j_1}$ in the first part 
are stated explicitly for each value of $j_1$ (since for them typically 
no closed form for symbolic $j_1$ can be obtained),  
while $c_{j_1}$ can be written in terms of rational functions or harmonic sums depending on the symbol $j_1$. For example, if $B_0(x)$ equals $ (1+x)\ln (1-x)$, we can express it as follows,
\begin{align}
(1+x) \ln(1-x) = \sum^{\alpha-1}_{j_1=0} h_{j_1} x^{j_1} + \sum^{\infty}_{j_1=\alpha} c_{j_1} x^{j_1}\,,
\end{align}
with $$\alpha=2,\, h_0 = 0,\, h_1=-1,\, c_{j_1} = \frac{-1}{j_1}+\frac{-1}{j_1-1}\,,$$
where $c_0$ and $c_1$ are not well defined. 
Taking into account the expansion of the $1/(1-x)^m$ factor, the resulting $f_n$ for generic $n$ in \eqref{eq:AmplitudeXspaceExpansion} can be written down recursively,  
\begin{align}
\label{eq:solutionN}
&f_{n}\big|_{m=1} = f_\alpha + \sum_{j_1=\alpha+1}^n c_{j_1} \,, \nonumber \\
&f_{n}\big|_{m=2} = f_\alpha +\sum_{j_1=\alpha+1}^n \left[ - f_\alpha + f_{\alpha +1 } + \sum_{j_2=\alpha+2}^{j_1} c_{j_2}  \right]\,, \nonumber 
\\
&f_{n}\big|_{m=3} = \bigg(f_{n}\big|_{m=2}\bigg)\bigg|_{c_{j_2} \to  f_\alpha - 2 f_{\alpha+1} + f_{\alpha+2} + \sum_{j_3=\alpha+3}^{j_2} c_{j_3} } \,, \nonumber \\
&\vdots  \nonumber \\
&f_{n}\big|_{m} =  \bigg(f_{n}\big|_{m-1}\bigg)\bigg|_{c_{j_{m-1}} \to \sum_{k=0}^{m-1} \binom{k}{m-1} (-1)^k f_{\alpha-k+m-1}    +  \sum_{j_m=\alpha+ m}^{j_{m-1}} c_{j_m}  }\,.
\end{align}
The multiple sums appearing on the right side of the above formula can be again transformed to harmonic sums by \texttt{HarmonicSums}.
We note that, after simplification, $B_m$ in \eqref{eq:AmplitudeXspace} is non-zero only for $m=0,1$ for all partonic channels at three loops.
Following the steps outlined above, 
we are able to get all bare three-loop OMEs 
of the singlet quark and gluon operators 
in terms of harmonic sums.

\section{GV counterterm Feynman rules from renormalization conditions}
\label{sec:DeriveGVOperators}

We are now ready to derive the counterterm Feynman rules resulting from the GV operators. Several ingredients are needed: the renormalization equations in \eqref{eq:RenorGVSimQ},~\eqref{eq:RenorGVSimG}, the OMEs in \eqref{eq:OMEsDe} generalized to any number of external states, and the renormalization conditions.

First, we consider equations \eqref{eq:RenorGVSimQ},~\eqref{eq:RenorGVSimG}. For simplicity, we restrict our discussion to \eqref{eq:RenorGVSimG}, which can be used to determine the Feynman rules for operators $O_{ABC}$ as well as for the counterterm operator $\left[Z O\right]_{g}^{\textrm{GV}}$. Feynman rules for the other counterterm operator $\left[Z O\right]_{q}^{\textrm{GV}}$ can be extracted from \eqref{eq:RenorGVSimQ} in a similar way. The main point is to consider the off-shell one-particle-irreducible (1PI) OMEs for both sides of equation \eqref{eq:RenorGVSimG} and impose the renormalization conditions. 

To determine the vertex Feynman rules for the GV operators, it is sufficient to consider the following off-shell 1PI OMEs with external states consisting of two particles of type $j$ and $m$ gluons,
\begin{align}
\label{eq:2jMgluonState} 
&\braket{j|O_g|j+ m\,g}^{\mu_1\cdots\mu_m,\,\text{R}}_{\text{1PI}} =  Z_j (\sqrt{Z_g})^m \bigg[  \braket{j| (Z_{gq} O_q+ Z_{gg} O_g  ) |j+m \,g} ^{\mu_1\cdots\mu_m,\,\text{B}}_{ \text{1PI}} \bigg]  \nonumber  \\
 & \qquad + Z_j (\sqrt{Z_g})^m \bigg[  Z_{gA}  \braket{j|O_{ABC}|j+m\,g} ^{\mu_1\cdots\mu_m,\,\text{B}}_{ \text{1PI}} + \braket{j|\left[Z O\right]_{g}^{\textrm{GV}}|j+ m\,g }^{\mu_1\cdots\mu_m,\,\text{B}}_{ \text{1PI}}  \bigg]\,.
\end{align} 
Here, $j$ could denote quarks($q$), gluons($g$), or ghosts($c$), $\sqrt{Z_j}$ is the corresponding field renormalization constant, and the renormalization of the strong coupling constant is implicitly understood on the right-hand side. 
To make the extraction of counterterm Feynman rules from the above equation transparent, we expand the OMEs according to the number of loops and legs,
\begin{align}
 \braket{j|O|j+ m\,g}^{\mu_1\cdots\mu_m\text{}} = \sum_{l=1}^{\infty}\left[ \braket{j|O|j+ m \,g}^{\mu_1\cdots\mu_m, \,(l),\,(m)\text{}}_{\text{}} \right] a_s^l  g_s^m \,,
\end{align} 
where $O$ denotes a generic operator.
For fixed $m$, we can compute the off-shell OMEs in the first line of the right-hand side of \eqref{eq:2jMgluonState} order-by-order in a loop expansion. Since the left-hand side of \eqref{eq:2jMgluonState} is an ultraviolet renormalized and infrared finite quantity, the sum of terms on the right-hand side should also be finite. If the GV operators were not to contribute, the divergences should cancel within the OMEs of the physical operators. Otherwise, the remaining divergences should be absorbed into the contribution from GV operators. Requiring the finiteness of the right-hand side of \eqref{eq:2jMgluonState}, we can extract the counterterm Feynman rules from the GV operators (or even the Feynman rules for the GV operators themselves) order-by-order in a loop expansion. 

\subsection{\texorpdfstring
{General formulae to determine the Feynman rules for the $O_{ABC}$ operators}
{General formulae to determine the Feynman rules for the O_ABC operators}}

Due to the hierarchy of GV operators, 
the counterterm operators $\left[ZO\right]_{j}^{\textrm{GV}}$ contribute to the 
renormalization of the OMEs only from two loops onwards. Consequently, the 
 Feynman rules for the $O_{ABC}$ operators can be extracted from the evaluation of one-loop off-shell 1PI OMEs. 

To extract the Feynman rules for $O_C$, we consider the one-loop OMEs  in \eqref{eq:2jMgluonState} with two ghosts, $j=c$, and $m$ gluons in the external states. In this case, by employing the renormalization conditions, equation \eqref{eq:2jMgluonState} simplifies to the form  
\begin{align}
\label{eq:formulaOC}
Z_{gA}^{(1)} \,\textcolor{black}{{\braket{c|O_C|c+ m\,g}^{\mu_1\cdots\mu_m, \,(0),\,(m)}_{\text{1PI}}=- \left[\braket{c|O_g|c+ m\,g}^{\mu_1\cdots\mu_m, \,(1),\,(m),\,\text{B}}_{\text{1PI}} \right]_{\text{div}} \,,}}
\end{align}
where the subscript '$\text{div}$' denotes the divergent contribution, that is, the poles in $\epsilon$. We notice that the left-hand side of the above equation is the multiplication of a renormalization constant and the vertex Feynman rules for the $O_C$ operator. By evaluating the right-hand side for $m=0$ and factorizing the dependence on kinematics, $Z_{gA}^{(1)}$ can be determined up to an overall $m$-independent constant 
\begin{align}
\label{eq:zgA1}
Z_{gA}^{(1)}= -\frac{1}{\epsilon}\frac{C_A}{ n (n-1)}\,. 
\end{align}
For $m=1$, a similar method was used to fix the unknown coefficients in an ansatz constrained by a generalized BRST symmetry in~\cite{Hamberg:1991qt}.

To determine the Feynman rule for the $O_B$ operator, we consider the OMEs with two quarks and $m$ gluon external states. Setting $j$ to a quark state and expanding to one-loop order, \eqref{eq:2jMgluonState} becomes 
\begin{align}
\label{eq:formulaOB}
{Z_{gA}^{(1)}}\, \braket{q|O_B|q+ m\,g}&^{\mu_1\cdots\mu_m, \,(0),\,(m)}_{\text{1PI}} = -\bigg\{ Z^{(1)}_{gq}  \left[\braket{q|O_q|q+ m\,g}^{\mu_1\cdots\mu_m, \,(0),\,(m)\text{}}_{\text{1PI}} \right] & \nonumber \\
 &+  \left[\braket{q|O_g|q+ m\,g}^{\mu_1\cdots\mu_m, \,(1),\,(m),\,\text{B}}_{\text{1PI}} \right]_{\text{div}} \bigg\}
  \,,
\end{align}
with $Z^{(1)}_{gq}$ being the first order renormalization constant for the  $q \to g$ transition
\begin{align}
Z^{(1)}_{gq} = \frac{1}{\epsilon} \left[\frac{4}{n}-\frac{2}{n+1}-\frac{4}{n-1}\right] C_F\,. 
\end{align}
The results of the OMEs $\left[\braket{q|O_g|q+ m\,g}^{\mu_1\cdots\mu_m, \,(1),\,(m),\,\text{B}}_{\text{1PI}} \right]_{\text{div}}$ contain both $C_F$ and $C_A$ color factors. Since $Z_{gA}^{(1)}$ is proportional to $C_A$, the $C_F$ from the one-loop OMEs must cancel against the one from $Z_{gq}^{(1)}$ in the above equation.  

Finally, to determine the Feynman rule for $O_A$ operator, we consider the OMEs with $m+2$ gluon states in a similar way. Setting $j$ to a gluon state and expanding to one-loop order, \eqref{eq:2jMgluonState} becomes 
\begin{align}
\label{eq:formulaOA}
{Z_{gA}^{(1)}} \, &\braket{g|O_A|g+ m\,g}^{\mu\nu\mu_1\cdots\mu_m,\,(0),\,(m)\text{}}_{\text{1PI}}  = - \bigg\{  \left[ \braket{g|O_g|g+ m\,g}^{\mu\nu\mu_1\cdots\mu_m,\,(1),\,(m),\,\text{B}}_{\text{1PI}} \right]_{\text{div}}  \nonumber \\
& +  \left[ Z^{(1)}_{gg} - \frac{m}{2 \epsilon} \beta_0 +  \frac{m+2}{2} Z^{(1)}_g \right] \braket{g|O_g|g+ m\,g}^{\mu\nu\mu_1\cdots\mu_m,\,(0),\,(m)\text{}}_{\text{1PI}}   \bigg\}\,,  
\end{align}
where $\beta_0$ is the one-loop QCD beta function
\begin{align}
\beta_0 = \frac{11 C_A}{3} - \frac{2 N_f}{3} \,, 
\end{align}
$Z_{g}^{(1)}$ is the first-order gluon field renormalization constant 
\begin{align}
Z^{(1)}_g =  \frac{1}{\epsilon} \left[\frac{13 C_A}{6}-\frac{ C_A \xi }{2}-\frac{2 N_f}{3} \right] \,, 
\end{align}
and $Z_{gg}^{(1)}$ is the first order renormalization constant for the $g \to g$ transition.
\begin{align}
\label{eq:Z1ggResults}
Z^{(1)}_{gg} =\frac{1}{\epsilon}\left[  C_A \left(4
  S_1(n)-\frac{4}{n-1}+\frac{4}{n}-\frac{4}{n+1}+\frac{4}{n+2}-\frac{11}{3}\right) +  \frac{2 N_f}{3} \right] \,.
\end{align}
For a one-loop, purely gluonic scattering process involving the operator $O_g$, one expects that no $N_f$ factor appears. Indeed, through a simple calculation, we found that the $N_f$ factor cancels out in the combination $Z^{(1)}_{gg} - \frac{m}{2 \epsilon} \beta_0 +  \frac{m+2}{2} Z^{(1)}_g $ of equation~\eqref{eq:formulaOA}. 
 
 Through equations~\eqref{eq:formulaOC}, \eqref{eq:formulaOB} and \eqref{eq:formulaOA}, deriving the Feynman rules for the unknown operators $O_{ABC}$ is equivalent to the computation of the divergent contributions to one-loop off-shell 1PI OMEs with an operator insertion of $O_g$. Since the Feynman rules resulting from $O_g$ are known to all multiplicities, as shown in \eqref{eq:singletOP}, the Feynman rules for $O_{ABC}$ could in principle also be calculated to all multiplicities. 

\subsection{\texorpdfstring{Determining the counterterm Feynman rules for $\left[Z O\right]_{g}^{\textrm{GV},\,(2)}$ and beyond}
{Determining the counterterm Feynman rules for [Z O]_g^{GV,(2)} and beyond}}
\label{subsec:FormulaZOgVV}

In the last subsection, the Feynman rules for the $O_{ABC}$ operators were extracted from one-loop OMEs in \eqref{eq:2jMgluonState}. Similarly, the $l$-loop corrections of \eqref{eq:2jMgluonState} can be used to extract the 
counterterm Feynman rules for $\left[Z O\right]_{g}^{\textrm{GV},\,(l)}$ with $l \geq 2$. Firstly, we work out the explicit formulae for the Feynman rules of $\left[Z O\right]_{g}^{\textrm{GV},\,(2)}$. Considering two ghost plus $m$ gluon external states, to two-loop order, \eqref{eq:2jMgluonState} reads
\begin{align}
\label{eq:counterterm2cmg}
    &\braket{c|\left[Z O\right]_{g}^{\textrm{GV},\,(2)}|c+ m \,g}^{\mu_1\cdots\mu_m,\,(0),\,(m)\text{}}_{\text{1PI}} =-\bigg\{\bigg[ \braket{c|O_g|c+ m\,g}^{\mu_1\cdots\mu_m, \,(2),\,(m),\,\text{B}}_{\text{1PI}}  \nonumber 
    \\
    & \qquad + \Big(Z_c^{(1)}+\frac{m Z_g^{(1)}}{2}+Z_{gg}^{(1)}-\frac{\beta _0 (m+2)}{2 \epsilon }\Big) \braket{c|O_g|c+ m\,g}^{\mu_1\cdots\mu_m, \,(1),\,(m),\,\text{B}}_{\text{1PI}} \nonumber 
    \\
    & \qquad + \Big(Z_c^{(1)}
   Z_{gA}^{(1)}+\frac{1}{2} m
   Z_g^{(1)}
   Z_{gA}^{(1)}-\frac{\beta _0 m
   Z_{gA}^{(1)}}{2 \epsilon
   }+Z_{gA}^{(2)}\Big)  \braket{c|O_{C}|c+ m\,g}^{\mu_1\cdots\mu_m, \,(0),\,(m),\,\text{B}}_{\text{1PI}}
    \nonumber 
    \\
    & \qquad+ Z_{gA}^{(1)} \braket{c|O_{AC}|c+ m\,g}^{\mu_1\cdots\mu_m, \,(1),\,(m),\,\text{B}}_{\text{1PI}}  \nonumber
    \\
    & \qquad + Z_g^{(1)} \sum_{t=1}^s  \xi^t \,t \, \braket{c|O_g|c+ m\,g}^{\mu_1\cdots\mu_m, \,(1),\,(m),\, (t),\,\text{B}}_{\text{1PI}}  \bigg]_{\text{div}} \bigg\}\bigg|_{\xi^{\text{B} } \to \xi }   \,,
\end{align}
where the last term originates from the gauge parameter renormalization
\begin{align}
    \xi^{\text{B}} = Z_\xi \xi, 
\end{align}
with $Z_\xi = Z_g$ in covariant gauge. We further decompose the OMEs according to the power of the gauge parameter, 
\begin{align}
    \braket{c|O_g|c+ m\,g}^{\mu_1\cdots\mu_m, \,(1),\,(m),\,\text{B}}_{\text{1PI}} = \sum_{t=0}^{s} (\xi^{\text{B}} )^t \braket{c|O_g|c+ m\,g}^{\mu_1\cdots\mu_m, \,(1),\,(m),\, (t),\,\text{B}}_{\text{1PI}}\,,
\end{align}
where $s$ is a small positive integer.

For the two quark plus $m$ gluon external states, to two-loop order, \eqref{eq:2jMgluonState} becomes
\begin{align}
\label{eq:counterterm2qmg}
    &\braket{q|\left[Z O\right]_{g}^{\textrm{GV},\,(2)}|q+ m \,g}^{\mu_1\cdots\mu_m,\,(0),\,(m)\text{}}_{\text{1PI}} =-\bigg\{\bigg[ \braket{q|O_g|q+ m\,g}^{\mu_1\cdots\mu_m, \,(2),\,(m),\,\text{B}}_{\text{1PI}}  \nonumber 
    \\
    & \qquad + \Big(\frac{m
   Z_g^{(1)}}{2}+Z_{gg}^{(1)}+Z_q^
   {(1)}-\frac{\beta _0 (m+2)}{2 \epsilon
   }\Big)   \braket{q|O_g|q+ m\,g}^{\mu_1\cdots\mu_m, \,(1),\,(m),\,\text{B}}_{\text{1PI}} \nonumber 
    \\
    & \qquad + \Big(\frac{1}{2} m Z_g^{(1)}
   Z_{gA}^{(1)}-\frac{\beta _0 m
   Z_{gA}^{(1)}}{2 \epsilon
   }+Z_{gA}^{(1)}
   Z_q^{(1)}+Z_{gA}^{(2)}\Big)  \braket{q|O_{B}|q+ m\,g}^{\mu_1\cdots\mu_m, \,(0),\,(m),\,\text{B}}_{\text{1PI}}
    \nonumber 
    \\
    & \qquad + \Big(\frac{1}{2} m Z_g^{(1)}
   Z_{gq}^{(1)}-\frac{\beta _0 m
   Z_{gq}^{(1)}}{2 \epsilon
   }+Z_{gq}^{(1)}
   Z_q^{(1)}+Z_{gq}^{(2) }\Big) \braket{q|O_{q}|q+ m\,g}^{\mu_1\cdots\mu_m, \,(0),\,(m),\,\text{B}}_{\text{1PI}}
    \nonumber
    \\
    & \qquad + Z_{gA}^{(1)} \braket{q|O_{AB}|q+ m\,g}^{\mu_1\cdots\mu_m, \,(1),\,(m),\,\text{B}}_{\text{1PI}} + Z_{gq}^{(1)} \braket{q|O_{q}|q+ m\,g}^{\mu_1\cdots\mu_m, \,(1),\,(m),\,\text{B}}_{\text{1PI}}  \nonumber
    \\
    & \qquad + Z_g^{(1)} \sum_{t=1}^s  \xi^t \,t \, \braket{q|O_g|q+ m\,g}^{\mu_1\cdots\mu_m, \,(1),\,(m),\, (t),\,\text{B}}_{\text{1PI}}  \bigg]_{\text{div}} \bigg\}\bigg|_{\xi^{\text{B} } \to \xi }    \,.
\end{align}
Similarly, for $m+2$ gluon external states, to two-loop order, \eqref{eq:2jMgluonState} reads
\begin{align}
\label{eq:counterterm2gmg}
    &\braket{g|\left[Z O\right]_{g}^{\textrm{GV},\,(2)}|g+ m \,g}^{\mu\nu\mu_1\cdots\mu_m,\,(0),\,(m)\text{}}_{\text{1PI}} =-\bigg\{ \bigg[ \braket{g|O_g|g+ m\,g}^{\mu\nu\mu_1\cdots\mu_m, \,(2),\,(m),\,\text{B}}_{\text{1PI}}  \nonumber 
    \\
    & \qquad + \Big(\frac{1}{2} (m+2)
   Z_g^{(1)}+Z_{gg}^{(1)}-\frac{\beta _0 (m+2)}{2 \epsilon }\Big)   \braket{g|O_g|g+ m\,g}^{\mu\nu\mu_1\cdots\mu_m, \,(1),\,(m),\,\text{B}}_{\text{1PI}} \nonumber 
    \\
    & \qquad + \Big(\frac{1}{2} (m+2)
   Z_g^{(1)}
   Z_{gA}^{(1)}-\frac{\beta _0 m
   Z_{gA}^{(1)}}{2 \epsilon
   }+Z_{gA}^{(2)}\Big)   \braket{g|O_{A}|g+ m\,g}^{\mu\nu\mu_1\cdots\mu_m, \,(0),\,(m),\,\text{B}}_{\text{1PI}} \nonumber
    \\
    & \qquad + Z_{gA}^{(1)} \braket{g|O_{ABC}|g+ m\,g}^{\mu\nu\mu_1\cdots\mu_m, \,(1),\,(m),\,\text{B}}_{\text{1PI}} + Z_{gq}^{(1)} \braket{g|O_{q}|g+ m\,g}^{\mu\nu\mu_1\cdots\mu_m, \,(1),\,(m),\,\text{B}}_{\text{1PI}}
    \nonumber 
    \\
    & \qquad + \Big(\frac{-\frac{1}{4} \beta _0 m (m+2)
   Z_g^{(1)}-\frac{1}{2} \beta_0 m
   Z_{gg}^{(1)}-\frac{\beta _1
   m}{4}}{\epsilon } +\frac{1}{2} (m+2) \big(Z_g^{(1)}
   Z_{gg}^{(1)}+Z_g^{(2)} \big)+Z_{gg}^{(2)}  \nonumber 
   \\
   & \qquad +\frac{1}{8} m (m+2)
   (Z_g^{(1)})^2  +\frac{\beta_0^2 m (m+2)}{8 \epsilon ^2}\Big)  \braket{g|O_{g}|g+ m\,g}^{\mu\nu\mu_1\cdots\mu_m, \,(0),\,(m),\,\text{B}}_{\text{1PI}}  \nonumber
    \\
    & \qquad + Z_g^{(1)} \sum_{t=1}^s  \xi^t \,t \, \braket{g|O_g|g+ m\,g}^{\mu\nu\mu_1\cdots\mu_m, \,(1),\,(m),\, (t),\,\text{B}}_{\text{1PI}}  \bigg]_{\text{div}} \bigg\}\bigg|_{\xi^{\text{B} } \to \xi }     \,.
\end{align}
Here, we introduced some short-hand notations, $O_{AB} = O_A+ O_B,\, O_{AC} = O_A+ O_C$. The above equations express the Feynman rules for the counterterm operator $\left[Z O\right]_{g}^{\textrm{GV},\,(2)}$ through the divergent terms of two-loop off-shell OMEs with $O_g$ insertion plus contributions from lower-loop OMEs with other operator insertions. We noticed that the one-loop OMEs need to be evaluated to order $\epsilon^0$.

In addition to the renormalization constants appearing in \eqref{eq:formulaOC},~\eqref{eq:formulaOB},~\eqref{eq:formulaOA}, we need several further renormalization constants in the above equations, including the first order quark and ghost field renormalization constants $Z_q^{(1)}$, $Z_c^{(1)}$, the second order gluon field renormalization constant $Z_g^{(2)}$, the 2-loop QCD beta function $\beta_1$ as well as the second order renormalization constants $Z_{gA}^{(2)},\,Z_{gq}^{(2)},\,Z_{gg}^{(2)}$. The field renormalization constants and beta function are known from the computations of diagrams without a twist-two operator insertion, they are summarized in 
Appendix~\ref{sec:QCDren}. However, $Z_{gA}^{(2)},\,Z_{gq}^{(2)},\,Z_{gg}^{(2)}$ are supposed to be extracted from two-point two-loop OMEs in the above equations for $m=0$, and thus should be regarded unknown. Therefore, it seems that the equations \eqref{eq:counterterm2cmg},~\eqref{eq:counterterm2qmg} and~\eqref{eq:counterterm2gmg} can not be used to determine the Feynman rules for $\left[Z O\right]_{g}^{\textrm{GV},\,(2)}$. 

However, as shown in Appendix~\ref{sec:FeynmanRulesGV2les}, the 
two-point vertex Feynman rules ($m=0$) for all GV operators except 
$O_A$ and $O_C$ are zero. 
Consequently, 
\begin{eqnarray}
&\braket{c|\left[Z O\right]_{g}^{\textrm{GV},\,(2)}|c}^{(0),\,(0)\text{}}_{\text{1PI}} = 0 \,,  \nonumber
\\
&   \braket{q|\left[Z O\right]_{g}^{\textrm{GV},\,(2)}|q}^{(0),\,(0)\text{}}_{\text{1PI}} = 0 \,,  \nonumber
\\
&   \braket{g|\left[Z O\right]_{g}^{\textrm{GV},\,(2)}|g}^{\mu\nu,\,(0),\,(0)\text{}}_{\text{1PI}} = 0 \,.
\label{eq:2legsFeyZOgVV}
\end{eqnarray}
 These equations allow us to easily determine $Z_{gA}^{(2)},\,Z_{gq}^{(2)},\,Z_{gg}^{(2)}$ by separately evaluating \eqref{eq:counterterm2cmg}, \eqref{eq:counterterm2qmg} and \eqref{eq:counterterm2gmg} for $m=0$. 
 With all renormalization constants determined, the counterterm Feynman rules for $\left[Z O\right]_{g}^{\textrm{GV},\,(2)}$ can, in principle, be  determined to an arbitrary number of legs from \eqref{eq:counterterm2cmg},~\eqref{eq:counterterm2qmg} and~\eqref{eq:counterterm2gmg}, similarly to the case of $O_{ABC}$. Since the vertex Feynman rules with two legs for $\left[Z O\right]_{q}^{\textrm{GV},\,(3)}$ and $\left[Z O\right]_{g}^{\textrm{GV},\,(3)}$ are zero, the counterterm operators $\left[Z O\right]_{q}^{\textrm{GV},\,(3)}$ and $\left[Z O\right]_{g}^{\textrm{GV},\,(3)}$ contribute to the splitting functions only from four loops onwards. 
 We leave them for future study.

\section{Computation of Feynman rules for the GV operators}   
\label{sec:GVFeynmanRules}

We demonstrated in the above section that, in order to derive the Feynman rules of GV operators, one needs to work with general kinematics and keep all relevant Lorentz structures when computing the multi-loop, multi-particle OMEs. This task is in general non-trivial. For a $k$ particle scattering process with $O_g$ insertion, the external kinematics introduces a reference vector $\Delta$ and $k-1$ independent momenta $p_1\,,\,\cdots\,,\, p_{k-1}$ of the scattering particles. Therefore, the total number of mass scales is
\begin{equation} 
\frac{1}{2} (k-1) (k+2)\,.
\end{equation}
For example, for a three-particle scattering, one has the 5 scales 
\begin{align}
p_1 \cdot p_2,\quad p_1^2,\quad  p_2^2,\quad \Delta \cdot p_1, \quad \Delta \cdot p_2\,.
\end{align}
For four-particle and five-particle scatterings, we need 9 and 14 scales respectively.
The presence of many mass scales renders the computations complicated, and the number of independent Lorentz structures increases quickly especially for pure gluon scatterings. Before considering the symmetries, the 5 Lorentz structures for 2-gluon scattering can be listed as
\begin{align}
\Delta^\mu \Delta^\nu , \quad \Delta^\mu p^\nu,\quad p^\mu \Delta^\mu, \quad g^{\mu \nu},\quad p^\mu p^\nu \,. 
\end{align} 
For a 3-gluon scattering, we divide the Lorentz structures into different categories according to the number of $\Delta$, 
\begin{align}
\label{eq:3gluonLorentz}
& 3\, \Delta: \,  \Delta^{\mu_1} \Delta^{\mu_2} \Delta^{\mu_3} \nonumber 
\\
&2 \, \Delta : \, \Delta^{\mu_1} \Delta^{\mu_2} p_1^{\mu_3}, \, \Delta^{\mu_1} \Delta^{\mu_2} p_2^{\mu_3} ,\, \ldots \nonumber 
\\
& 1 \, \Delta: \, \Delta^{\mu_1} g^{\mu_2 \mu_3}, \, \Delta^{\mu_1} p_1^{\mu_2} p_1^{\mu_3} , \, \Delta^{\mu_1} p_1^{\mu_2} p_2^{\mu_3},\, \Delta^{\mu_1} p_2^{\mu_2} p_2^{\mu_3}, \, \ldots \nonumber 
\\ 
& 0 \,\Delta: \, g^{\mu_1 \mu_2} p_1^{\mu_3} , \, g^{\mu_1 \mu_2} p_2^{\mu_3},\, p_1^{\mu_1} p_1^{\mu_2} p_1^{\mu_3}, \, p_1^{\mu_1} p_1^{\mu_2} p_2^{\mu_3}, \, p_1^{\mu_1} p_2^{\mu_2} p_2^{\mu_3}, \, p_2^{\mu_1} p_2^{\mu_2}p_2^{\mu_3},\, \ldots
\end{align} 
where we have eliminated $p_3$ using momentum conservation $p_3=-p_1-p_2$, and the dots represent Lorentz structures that can be obtained from the permutations of the Lorentz indices $\mu_1,\,\mu_2,\,\mu_3$. In total, we have 36 Lorentz structures for 3-gluon scattering, 353 Lorentz structures for 4-gluon scattering, and 4400 for 5-gluon scattering. A naive projection method will make the computations extremely complicated. However, we observe that many Lorentz structures do not appear in the Feynman rules of the physical operators $O_q$ and $O_g$, as shown in Appendix~\ref{sec:FeynOqOg}. Since the GV operators are also twist-two operators, the observation may be used to simplify the computations. To make things explicit, we work out all allowed Lorentz structures for a general twist-two operator in the following.

\subsection{Lorentz structures for the Feynman rules of a general twist-two operator}
\label{subsec:Lorentzstructure}

For the sake of simplicity and without loss of generality, we focus on the Lorentz structures of Feynman rules for GV operators $O_{ABC}$. The same conclusion also applies to a general twist-two operator. Our guiding principles are the properties of  twist-two operators and the fact that no inverse mass scale can be generated in the Feynman rule of a vertex. 

We first analyze the Lorentz structures of $m$-gluon Feynman rules for $O_A$. For the twist-two $O_A$ operator, the mass dimension is $n+2$ and the spin is $n$. By factoring out $m$ gluon fields, the mass dimension becomes $n+2-m$. If we denote the number of $\Delta^{\mu}$, $p_i^{\mu}$ and $g^{\mu \nu}$ appearing in a Lorentz structure by $n_1,\,n_2,\,n_3$ respectively, then they should sum up to the total number of Lorentz indices $m$, 
\begin{equation}
\label{eq:relationLIndex}
m = n_1 +n_2 + 2 n_3\,. 
\end{equation}
Next, we count the mass dimension for the coefficient of a Lorentz structure. The coefficient is composed of Lorentz scalar products formed by the reference vector $\Delta$ and the momentum $p_i$ of scattering particles.
The dependence of the coefficients on the scalar products is of polynomial form, in particular, propagator-type terms (for example ${1}/({p_1 \cdot p_2})$) can not appear in the Feynman rule of a vertex.
Each monomial is further divided into two factors, the first one involves the contractions of $\Delta$ and the momenta $p_i$ (for example $\Delta \cdot p_1$), and the second one includes the contractions within the momenta $p_i$ (for example $p_1 \cdot p_2$). The mass dimension from the first factor must be $n-n_1$ since the operator should always include a total number $n$ of $\Delta$ vectors. We set the mass dimension of the second factor to a non-negative integer $2 n_4$.

By adding up the mass dimensions from the different ingredients, we should recover the total mass dimensions,
\begin{equation}
\label{eq:TotalMassDimension}
(n-n_1) + n_2 + 2 n_4 = n+2-m\,.
\end{equation}
Substituting \eqref{eq:relationLIndex} into \eqref{eq:TotalMassDimension} gives us 
\begin{equation}
\label{eq:conditionOfOA}
n_2+n_3+n_4=1\,.
\end{equation}
Since all $n_2,\,n_3,\,n_4$ should be non-negative, the above equation tells us that each of them should not be larger than 1 and that their sum must be 1. The equation drastically constrains the allowed Lorentz structures. For example, the structure $\Delta^{\mu_1} p_1^{\mu_2} p_1^{\mu_3}$ in \eqref{eq:3gluonLorentz} can not appear since $n_2=2>1$. As another example, $g^{\mu_1 \mu_2} p_1^{\mu_3}$ can not appear due to $n_2=1,\,n_3=1$. 

With the constraint in \eqref{eq:conditionOfOA}, it is easy to write down the general ansatz for the 3-gluon Feynman rule of $O_A$,
\begin{align}
\label{eq:ansatz3Gluon}
 &\left[\braket{g|O_A|g g}^{\mu_1\mu_2\mu_3,\,(0),\,(1)}_{\text{1PI}} \right] = a_1  \Delta^{\mu_1} \Delta^{\mu_2} \Delta^{\mu_3} + a_2 \Delta^{\mu_1} \Delta^{\mu_2} p_1^{\mu_3} + a_3 \Delta^{\mu_1} \Delta^{\mu_3} p_1^{\mu_2} \nonumber \\
 & \quad +  a_4 \Delta^{\mu_2} \Delta^{\mu_3} p_1^{\mu_1}+a_5 \Delta^{\mu_1} \Delta^{\mu_2} p_2^{\mu_3} + a_6 \Delta^{\mu_1} \Delta^{\mu_3} p_2^{\mu_2}  +  a_7 \Delta^{\mu_2} \Delta^{\mu_3} p_2^{\mu_1} \nonumber \\
 & \quad + a_8 \Delta^{\mu_1} g^{\mu_2 \mu_3}+ a_9 \Delta^{\mu_2} g^{\mu_1 \mu_3}+ a_{10} \Delta^{\mu_3} g^{\mu_1 \mu_2}\,. 
\end{align}
The number of Lorentz structures is reduced from 36 to 10 for the 3-gluon Feynman rule in the above equation. Similarly, the general ansatz for the 4-gluon Feynman rule of $O_A$ is obtained as
\begin{align}
\label{eq:4gAnsatz}
 &\left[\braket{g|O_A|g g g}^{\mu_1\mu_2\mu_3\mu_4,\,(0),\,(2)}_{\text{1PI}} \right] = b_1  \Delta^{\mu_1} \Delta^{\mu_2} \Delta^{\mu_3} \Delta^{\mu_4} + b_2 \Delta^{\mu_1} \Delta^{\mu_2} \Delta^{\mu_3} p_1^{\mu_4} \nonumber \\
 & \quad + b_3 \Delta^{\mu_1} \Delta^{\mu_2} \Delta^{\mu_4} p_1^{\mu_3} +b_4 \Delta^{\mu_1} \Delta^{\mu_3} \Delta^{\mu_4} p_1^{\mu_2} +b_5 \Delta^{\mu_2} \Delta^{\mu_3} \Delta^{\mu_4} p_1^{\mu_1} \nonumber \\
 & \quad+ b_6 \Delta^{\mu_1} \Delta^{\mu_2} \Delta^{\mu_3} p_2^{\mu_4} + b_7 \Delta^{\mu_1} \Delta^{\mu_2} \Delta^{\mu_4} p_2^{\mu_3} +b_8 \Delta^{\mu_1} \Delta^{\mu_3} \Delta^{\mu_4} p_2^{\mu_2} +b_9 \Delta^{\mu_2} \Delta^{\mu_3} \Delta^{\mu_4} p_2^{\mu_1} \nonumber \\
 & \quad + b_{10} \Delta^{\mu_1} \Delta^{\mu_2} \Delta^{\mu_3} p_3^{\mu_4} + b_{11} \Delta^{\mu_1} \Delta^{\mu_2} \Delta^{\mu_4} p_3^{\mu_3} +b_{12} \Delta^{\mu_1} \Delta^{\mu_3} \Delta^{\mu_4} p_3^{\mu_2} +b_{13} \Delta^{\mu_2} \Delta^{\mu_3} \Delta^{\mu_4} p_3^{\mu_1} \nonumber \\ 
 & \quad +   b_{14} \Delta^{\mu_1} \Delta^{\mu_2} g^{\mu_3 \mu_4}+b_{15} \Delta^{\mu_1} \Delta^{\mu_3} g^{\mu_2 \mu_4}+b_{16} \Delta^{\mu_1} \Delta^{\mu_4} g^{\mu_2 \mu_3}+b_{17} \Delta^{\mu_2} \Delta^{\mu_3} g^{\mu_1 \mu_4}\nonumber \\
 & \quad +b_{18} \Delta^{\mu_2} \Delta^{\mu_4} g^{\mu_1 \mu_3}+b_{19} \Delta^{\mu_3} \Delta^{\mu_4} g^{\mu_1 \mu_2}\,,
\end{align}
where we need only 19 out of the initial list of 353 Lorentz structures. 

We also analyze the dependence of the scalar coefficients of Lorentz tensors in \eqref{eq:ansatz3Gluon} and \eqref{eq:4gAnsatz} on the kinematic invariants. For term $a_1 \Delta^{\mu_1} \Delta^{\mu_2} \Delta^{\mu_3}$, $n_2=n_3=0$, therefore, $n_4$ must be 1. It indicates that the coefficient $a_1$ must be linear in $p_1^2$, $p_2^2$, or $p_1\cdot p_2$. Except for $\Delta^{\mu_1} \Delta^{\mu_2} \Delta^{\mu_3}$ and $\Delta^{\mu_1} \Delta^{\mu_2} \Delta^{\mu_3} \Delta^{\mu_4}$, other tensor structures have either $n_2=1$ or $n_3=1$. Therefore, their corresponding coefficients $a_i,\,b_i$ with $i>1$ can be only constructed from scalar products involving $\Delta$.
Exploiting the above properties allows for a particularly efficient
determination of the coefficients with modern finite-field and function reconstruction techniques~\cite{vonManteuffel:2014ixa,Peraro:2016wsq}.

Since Lorentz indices can not be carried by the quark fields and ghost fields, the Lorentz structures appearing in the Feynman rules for the operators $O_B$ and $O_C$ are much more constrained. Regarding a vertex involving two quarks plus $m$ gluons, or two ghosts plus $m$ gluons, \eqref{eq:relationLIndex} is still valid by using the same convention as for the pure $m$-gluon vertex. The difference is the extra mass dimension from quark fields or ghost fields. Regarding the operator $O_C$,  \eqref{eq:TotalMassDimension} should be modified to 
\begin{align}
\label{eq:ghostMassDimension}
(n-n_1) + n_2 + 2 n_4 = n+2-[\bar{c} c]-m = n+2-2-m\,,
\end{align} 
where $[\bar{c} c]=2$ stands for the mass dimension of a pair of ghost and anti-ghost fields. For the operator $O_B$, one always needs a factor $\slashed{\Delta}$ to compensate the extra mass dimension from $\bar{\psi} \psi$ compared with $\bar{c}c$. Correspondingly,  \eqref{eq:TotalMassDimension} is modified to
\begin{align}
\label{eq:quarkMassDimension}
(n-n_1-1) + n_2 + 2 n_4 = n+2-[\bar{\psi} \psi]-m = n+2-3-m\,, 
\end{align} 
where $[\bar{\psi} \psi ]=3$ stands for the mass dimension of a pair of quark and anti-quark fields. The above equation is equivalent to the equation \eqref{eq:ghostMassDimension}. Solving \eqref{eq:relationLIndex} and \eqref{eq:ghostMassDimension}, the result reads,
\begin{align}
\label{eq:ghostQuarkCondition}
n_2+n_3+n_4=0\,.
\end{align}
The above equation indicates, for both Feynman rules of $O_B$ and $O_C$, the Lorentz tensors can be composed of $\Delta^\mu$ only,
\begin{align}
\label{eq:ansatzForOBOC}
&\left[\braket{c|O_C|c+ m\,g}^{\mu_1\cdots\mu_m, \,(0),\,(m) }_{\text{1PI}} \right]  = c_m \Delta^{\mu_1} \Delta^{\mu_2} \cdots \Delta^{\mu_m} \,, \nonumber \\
& \left[\braket{q|O_B|q+ m\,g}^{\mu_1\cdots\mu_m, \,(0),\,(m)}_{\text{1PI}} \right]  = d_m \Delta^{\mu_1} \Delta^{\mu_2} \cdots \Delta^{\mu_m} \,,   
\end{align}
where the coefficients $c_m$ and $d_m$ have the same property as the coefficients $a_i, b_i$ with $i>1$ in \eqref{eq:ansatz3Gluon} and \eqref{eq:4gAnsatz}. 

Since the counterterm operators $\left[Z O\right]_{q}^{\textrm{GV}}$ and $\left[Z O\right]_{g}^{\textrm{GV}}$ are also twist-two operators,
the above considerations remain valid if we replace $O_{A},\,O_{B},\,O_C$ in \eqref{eq:ansatz3Gluon}, \eqref{eq:4gAnsatz} and \eqref{eq:ansatzForOBOC} by $\left[Z O\right]_{q}^{\textrm{GV}}$ or $\left[Z O\right]_{g}^{\textrm{GV}}$. To compute the multi-loop, multi-particle OMEs, we project out the coefficients of the corresponding Lorentz structures in \eqref{eq:ansatz3Gluon}, \eqref{eq:4gAnsatz} and \eqref{eq:ansatzForOBOC}. In general, 
the Lorentz structures that are allowed in the Feynman rules do not 
form a complete basis of the vector space, such that 
we need extra Lorentz structures in the construction of the 
projectors. However, the number of independent projectors is the same as the number of required Lorentz structures in corresponding equations. For example, a single projector $p_{1,\,\mu_1} p_{1,\,\mu_2}\cdots p_{1,\,\mu_m}$ is enough to project out $c_m$ and $d_m$ in \eqref{eq:ansatzForOBOC}.

\subsection{\texorpdfstring
{Feynman rules for $O_{ABC}$ up to four legs}
{Feynman rules for O_ABC up to four legs}}
From \eqref{eq:formulaOC},~\eqref{eq:formulaOB} and~\eqref{eq:formulaOA}, we observe that the Feynman rules for $O_{ABC}$ are either proportional to the divergent terms of the one-loop OMEs, or expressed as a linear combination of one-loop OMEs and the Feynman rules for physical operators $O_q$ and $O_g$. Therefore, the same ansatz as in \eqref{eq:ansatz3Gluon}, \eqref{eq:4gAnsatz} and \eqref{eq:ansatzForOBOC} also applies to the divergent terms of the one-loop OMEs. Having worked out the simplified ansatz and projectors, we are ready to compute the one-loop, multi-particle OMEs by adopting the method in Section~\ref{sec:ComputationalMethods}. We work directly in parameter-$x$ space. Since we need just the divergent terms, with generic kinematics, only two kinds of master integrals contribute. One is the bubble integral, the other one is the bubble integral with a linear propagator insertion, 
\begin{align}
I_1 &= (\mu^2)^\epsilon \int \frac{d^dl}{i \pi^{d/2}} \frac{1}{(l-q_1)^2 l^2} \,, \nonumber
\\
I_2 &= (\mu^2)^\epsilon \int \frac{d^d l}{i \pi^{d/2}} \frac{1}{(l-q_1)^2 l^2  \big(1- x \Delta \cdot (l+q_2) \big) } \,, 
\end{align}
where $q_1\,,\,q_2$ are a linear combination of momenta $p_1\,,\,\cdots\,,\, p_{k-1}$ and $\mu$ is the 't Hooft scale.
The two master integrals can be computed easily, and the only special function that appears in their divergent parts is the logarithm.
In particular, we encounter $x$-dependent logarithms from the master integral $I_2$,
\begin{align}
I_2 = \frac{1}{\epsilon} \left[ \frac{\ln (1-x \Delta \cdot q_1  - x \Delta \cdot q_2) - \ln (1-x \Delta \cdot q_2)  }{- x \Delta \cdot q_1}  \right] + \mathcal{O}(\epsilon^0)\,.
\end{align} 
The spurious poles appearing in the results for the OMEs in parameter-$x$ space can be eliminated using the 
{\tt MultivariateApart}~\cite{Heller:2021qkz} and {\tt Singular\_pfd}~\cite{Boehm:2020ijp} packages implementing multivariate partial fraction algorithms. Using the method in Section~\ref{sec:ComputationalMethods} and the following type of replacement,  
\begin{align}
&\frac{x^3}{\left(1 -x \Delta \cdot p_1 \right) \left( 1- x \Delta \cdot  p_2 \right)   } = \sum_{n=3}^{\infty}  x^n \sum_{j_1=0}^{n-3}  \left( \Delta \cdot p_1 \right)^{n-3-j_1} \left( \Delta \cdot p_2\right)^{j_1} \nonumber 
\\
& \quad \quad \qquad \qquad  \to  \sum_{j_1=0}^{n-3}   \left( \Delta \cdot p_1 \right)^{n-3-j_1} \left( \Delta \cdot p_2\right)^{j_1}\,,   
\end{align}
we manage to express the divergent terms of one-loop OMEs in terms of a single harmonic sum $S_1(n)$ and multiple summations of the form shown in the last line of the above equation.    

\begin{figure}
\begin{center}
\begin{minipage}{4.6cm}
\includegraphics[scale=0.9]{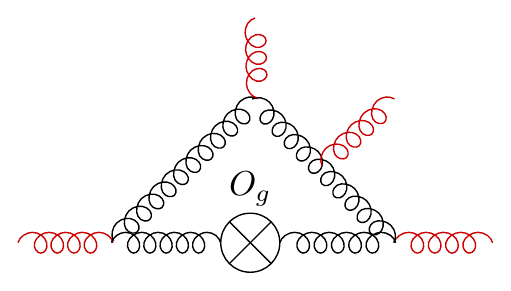} 
\end{minipage}
\begin{minipage}{4.6cm}
\includegraphics[scale=0.9]{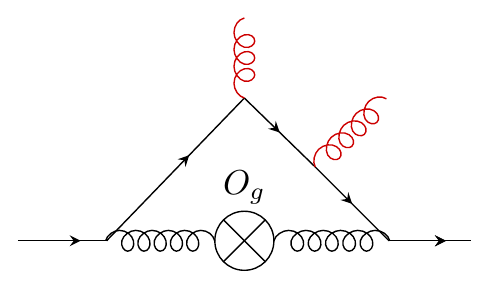} 
\end{minipage}
\begin{minipage}{4.6cm}
\includegraphics[scale=0.9]{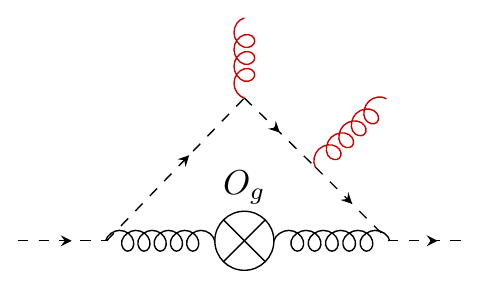} 
\end{minipage}
\caption{Sample diagrams to determine Feynman rules with $gggg$ vertex for $O_{A}$ (left), $q \bar{q} g g$ vertex for $O_{B}$ (middle) and $c \bar{c} g g$ vertex for $O_{C}$ (right). All Feynman diagrams contain an insertion of the physical operator $O_g$.}
\end{center}
\end{figure}

We  compute the divergent terms of the one-loop OMEs in \eqref{eq:formulaOC},~\eqref{eq:formulaOB} and~\eqref{eq:formulaOA} for $m=0\,,\,1\,,\,2$ with general $\xi$ dependence. The computation of the $m=3$ case is also straightforward, but becomes relevant only for the 4-loop splitting functions. In all cases, the right-hand sides are found to be proportional to $Z_{gA}^{(1)}$. The observation verifies our statement right above \eqref{eq:conjectureZRelation}. Explicitly for $m=0\,,\,1\,,\,2$, the OMEs involving ghosts have no dependence on $S_1(n)$ and are directly proportional to $Z_{gA}^{(1)}$. The OMEs involving quarks also have no dependence on $S_{1}(n)$ and their dependence on $C_F$ cancels against the $Z_{gq}^{(1)}$ terms. The dependence on $S_1(n)$ of OMEs involving only gluons cancels against the $Z_{gg}^{(1)}$ terms. By dividing out the common factor $Z_{gA}^{(1)}$ in the left-hand and right-hand side of \eqref{eq:formulaOC},~\eqref{eq:formulaOB} and~\eqref{eq:formulaOA}, we manage to obtain all-$n$ Feynman rules up to 4 legs for the operators $O_A,\, O_B$ and $O_C$. We emphasize that all Feynman rules are found to be $\xi$ independent. 

Up to three legs, the Feynman rules for the operators $O_A$ and $O_C$ were already available in~\cite{Hamberg:1991qt}, we find full agreement with them upon corrections of typographical errors according to~\cite{Blumlein:2022ndg}. The Feynman rule up to 3 legs for $O_B$ was given in~\cite{Matiounine:1998ky}, and we find full agreement in that case as well.
Our all-$n$ Feynman rules for operators $O_{ABC}$ with four legs are new. For completeness, we present all Feynman rules up to 4 legs in the following with the convention of all momenta flowing into the vertices. 

The all-$n$ Feynman rules for the $O_B$ operator with up to 4 legs are given by
\begin{align}
\label{eq:OB1Feynmanrules}
&\includegraphics[scale=1.0]{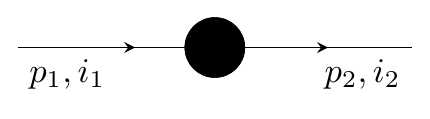}
\nonumber\\
&\quad \to 0  \,, \\[.5ex]
&\includegraphics[scale=1.0]{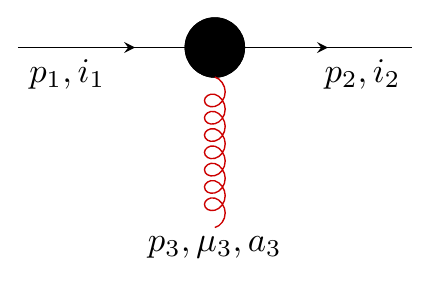}\nonumber\\
 \label{eq:FeynOB2q1g}
&\quad \to - \frac{1+(-1)^n}{2 } \,g_s  \, \Delta ^{\mu _3}  T^{a_3}_{i_2 i_1} \slashed{\Delta}_{} \big(\Delta
   \cdot \left(p_1+p_2\right)\big){}^{n-2} \,, \\[.5ex]
& \includegraphics[scale=1.0]{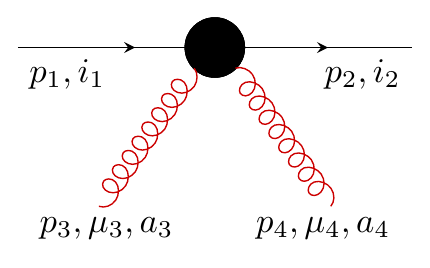}\nonumber\\
&\quad \to -\frac{1}{4} \frac{1+(-1)^n}{2 } g_s^2 \Delta ^{\mu _3} \Delta ^{\mu _4}
   \left( T^{a_3} T^{a_4} -   T^{a_4} T^{a_3}  \right)_{i_2 i_1} \slashed{\Delta}_{}
 \sum_{j_1=0}^{n-3}  \bigg(
     3 \left(\Delta \cdot \left(p_1+p_2\right)\right){}^{-j_1+n-3} \nonumber\\ 
     &\qquad
     \times \Big[  \left(-\Delta \cdot p_3\right){}^{j_1}  - \left(-\Delta \cdot p_4\right){}^{j_1} \Big] -\left(-\Delta \cdot p_4\right){}^{j_1}
   \left(\Delta \cdot p_3\right){}^{-j_1+n-3} \bigg) \,.  
   \label{eq:OA2q2g}
\end{align}

The all-$n$ Feynman rules for the $O_C$ operator with up to 4 legs are given by 
\begin{align}
&\includegraphics[scale=1.0]{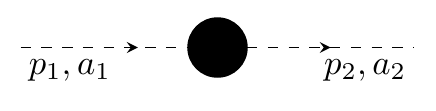} 
\nonumber \\
& \quad \to  \frac{1+(-1)^n }{2} \delta^{a_1 a_2}(\Delta \cdot p_1)^n \,, 
\label{eq:FeynOC2c}
\\
&\includegraphics[scale=1.0]{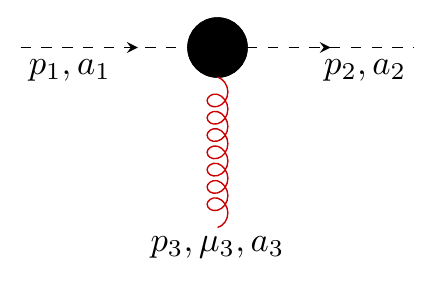} 
\nonumber \\
&\quad \to \frac{-i}{4}  \frac{1+(-1)^n}{2 } \Delta ^{\mu _3} g_s f^{a_1 a_2 a_3} \bigg(3 \Delta \cdot p_1 \Delta \cdot p_2
  \sum_{j_1 = 0}^{n-3}\left(\left(-\Delta \cdot p_2\right){}^{j_1} \left(\Delta \cdot p_1\right){}^{-j_1+n-3}\right) \nonumber \\
  & \qquad +\left(\Delta
  \cdot p_1-\Delta \cdot p_2\right) \left(\Delta \cdot \left(p_1+p_2\right)\right){}^{n-2}-\left(\Delta \cdot
  p_1\right){}^{n-1}+\left(\Delta \cdot p_2\right){}^{n-1}\bigg) \,, 
  \\
 & \includegraphics[scale=1.0]{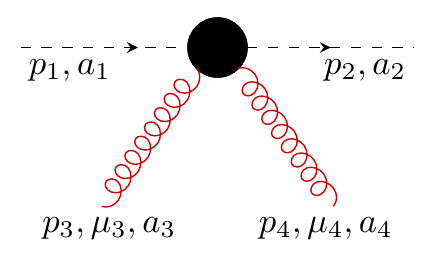}
\nonumber \\
& \quad \to \frac{1}{24}  \frac{1+(-1)^n}{2 }  g_s^2 \Delta ^{\mu _3} \Delta ^{\mu _4} \bigg\{ f^{a_1 a_3 a} f^{a_2 a_4 a} \bigg( 6
   \left(-\Delta \cdot p_4\right){}^{n-2}+6 \left(\Delta \cdot p_3\right){}^{n-2} \nonumber \\
   & \qquad +6 \left(\Delta \cdot
   \left(p_1+p_3\right)\right){}^{n-2}   +6 \left(\Delta \cdot \left(p_2+p_3\right)\right){}^{n-2}-
   \sum_{j_1=0}^{ n-2}   \bigg[ \nonumber 
   \\
   & \qquad +\big[ \left(-\Delta \cdot p_3\right){}^{j_1} + \left(-\Delta \cdot p_4\right){}^{j_1}\big] \big[ 3\left(\Delta \cdot p_1\right){}^{n-j_1-2} \nonumber
   \\
   &\qquad +3 \left(\Delta \cdot p_2\right){}^{n-j_1-2} +\left(\Delta \cdot
   \left(p_1+p_2\right)\right){}^{n-j_1-2} \big] \nonumber 
   \\
   & \qquad+9\big[ \left(\Delta \cdot
   p_1\right){}^{n-j_1-2}+  \left(-\Delta \cdot
   p_2\right){}^{n-j_1-2}  \big]\big[  \left(\Delta \cdot \left(-p_2-p_3\right)\right){}^{j_1}+ \left(\Delta
   \cdot \left(p_1+p_3\right)\right){}^{j_1} \big] \bigg]  \nonumber
   \\
   & \qquad 
   +13\sum_{j_1=0}^{n-2} \sum_{j_2=0}^{j_1}  \bigg[ \left(-\Delta
   \cdot p_2\right){}^{j_1-j_2} \left(\Delta \cdot p_1\right){}^{n-j_1-2}\big[ \left(\Delta \cdot
   \left(-p_2-p_3\right)\right){}^{j_2} +  \left(\Delta \cdot \left(p_1+p_3\right)\right){}^{j_2} \big]  \bigg]\bigg)  \nonumber 
   \\
   & \qquad 
   +f^{a_1 a_2 a} f^{a_3 a_4 a} \bigg(-6 \left(\Delta
   \cdot p_3\right){}^{n-2}-6 \left(\Delta \cdot \left(p_2+p_3\right)\right){}^{n-2}  \nonumber 
   \\
   & \qquad +\sum_{j_1=0}^{n-2}\bigg[ 3 \left(-\Delta \cdot
   p_4\right){}^{j_1} \left(\Delta \cdot p_1\right){}^{n-j_1-2}  \nonumber 
   \\
   & \qquad +3 \left(-\Delta \cdot
   p_3\right){}^{j_1} \left(\Delta \cdot p_2\right){}^{n-j_1-2}+\big[5 \left(-\Delta \cdot
   p_3\right){}^{j_1} -4\left(-\Delta
   \cdot p_4\right){}^{j_1}\big] \left(\Delta \cdot \left(p_1+p_2\right)\right){}^{n-j_1-2} \nonumber 
   \\
   & \qquad+9
  \big[ \left(\Delta \cdot p_1\right){}^{n-j_1-2}+ \left(-\Delta \cdot p_2\right){}^{n-j_1-2}  \big]\left(\Delta \cdot
   \left(-p_2-p_3\right)\right){}^{j_1}   \bigg]  -3 \Delta \cdot p_2 \sum_{j_1=0}^{n-3} \bigg[ \nonumber 
   \\
   & \qquad 3
   \big[ \left(-\Delta \cdot p_3\right){}^{j_1} -\left(-\Delta \cdot p_4\right){}^{j_1} \big] \left(\Delta \cdot
   \left(p_1+p_2\right)\right){}^{n-j_1-3}  -\left(-\Delta \cdot p_4\right){}^{j_1}
   \left(\Delta \cdot p_3\right){}^{n-j_1-3}\bigg]  \nonumber \\
   & \qquad +\sum_{j_1=0}^{n-2} \sum_{j_2=0}^{j_1} \bigg[ \left(-\Delta \cdot p_2\right){}^{j_1-j_2}
   \left(\Delta \cdot p_1\right){}^{n-j_1-2} \big[ \left(\Delta \cdot \left(p_1+p_3\right)\right){}^{j_2}   
-14 \left(\Delta \cdot \left(p_1+p_4\right)\right){}^{j_2} \big]  \bigg]   
   \bigg)   \nonumber
   \\
   & \qquad + \frac{6 d_A^{a_1 a_2 a_3 a_4}}{C_A}  \bigg(  - \sum_{j_1=0}^{n-2} \bigg[  \big[ \left(-\Delta \cdot
   p_3\right){}^{j_1} +  \left(-\Delta \cdot p_4\right){}^{j_1}   \big] \left(\Delta \cdot
   \left(p_1+p_2\right)\right){}^{-j_1+n-2} \bigg] \nonumber 
   \\
   & \qquad +  \sum_{j_1=0}^{n-2} \sum_{j_2=0}^{j_1}\bigg[ \left(-\Delta \cdot p_2\right){}^{j_1-j_2}
   \left(\Delta
   \cdot p_1\right){}^{-j_1+n-2} \big[ \left(\Delta \cdot
   \left(p_1+p_4\right)\right){}^{j_2} + \left(\Delta \cdot
   \left(p_1+p_3\right)\right){}^{j_2} \big] \bigg]    \bigg)  \bigg\} \,.
      \label{eq:OA2c2g}
\end{align}

Finally, the all-$n$ Feynman rules for the operator $O_{A}$ and up to 4 legs are given by
\begin{align}
&\includegraphics[scale=1]{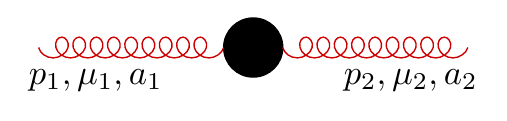} \nonumber
\\
&\quad \to   {\delta^{a_1 a_2}}  \frac{1+(-1)^n}{2} \bigg[-\Delta \cdot p_1 \left(p_1^{\mu_1} \Delta^{\mu_2} + \Delta^{\mu_1} p_1^{\mu_2}  \right) + 2 \Delta^{\mu_1} \Delta^{\mu_2} p_1 \cdot p_1 \bigg] (\Delta \cdot p_1)^{n-2}  \,,
\\[0.5ex] \label{eq:FeynOA2g}
&\includegraphics[scale=1.0]{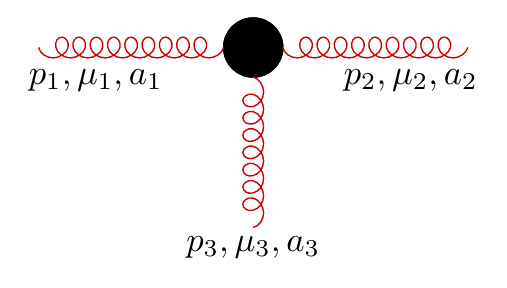} \nonumber \\
& \quad \to  \frac{-i}{4} \frac{1+(-1)^n}{2 } g_s f^{a_1 a_2 a_3} \bigg(-4 \Delta ^{\mu _3} g^{\mu _1 \mu _2} \Delta \cdot p_1 \left(\Delta \cdot
   \left(p_1+p_2\right)\right){}^{n-2}
\nonumber\\ &\qquad
   -3 \Delta ^{\mu _1} \Delta ^{\mu _3} p_2^{\mu _2} \sum_{j_1=0}^{n-2}\left(\left(-\Delta \cdot
   p_2\right){}^{j_1} \left(\Delta \cdot p_1\right){}^{-j_1+n-2}\right)
\nonumber \\& \qquad +2 \Delta ^{\mu _1} \Delta ^{\mu _2} \left(4
   p_2^{\mu _3}+p_3^{\mu _3}\right) \left(\Delta \cdot p_1\right){}^{n-2}
   -\Delta ^{\mu _1} \Delta ^{\mu _2} \Delta ^{\mu _3} \left(p_1\cdot p_1-p_1\cdot
   p_2+p_2\cdot p_2\right) \nonumber
   \\
   & \qquad \times \sum_{j_1=0}^{n-3}\left(\left(-\Delta \cdot p_2\right){}^{j_1} \left(\Delta \cdot
   p_1\right){}^{-j_1+n-3}\right) \bigg) + \text{{\it{permutations}}} \,,
\\[0.5ex]
&\includegraphics[scale=1.0]{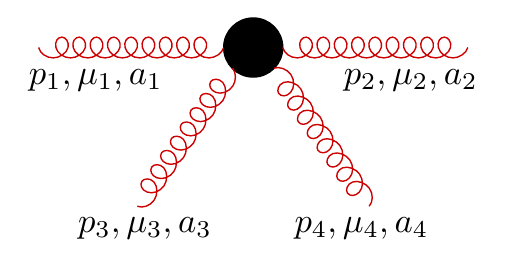}
\nonumber \\
& \quad \to \frac{1}{48}\frac{1+(-1)^n}{2 } g_s^2 \bigg\{3 \Delta ^{\mu _1} \Delta ^{\mu _2} g^{\mu _3 \mu _4} \bigg(8 \left(2 f^{a a_1 a_3}
   f^{a a_2 a_4} -f^{a a_1 a_2} f^{a a_3 a_4} \right) \left(\Delta \cdot
   p_1\right){}^{n-2}  \nonumber 
   \\
   &\qquad -f^{a a_1 a_2} f^{a a_3 a_4} \left(\Delta \cdot p_1+\Delta \cdot p_2+2 \Delta
   \cdot p_3\right)  \nonumber 
   \\
   &\qquad  \times \sum_{j_1=0}^{n-3} \bigg[ \big[6  \left(\Delta \cdot \left(-p_1-p_2\right)\right){}^{j_1}+\left(\Delta \cdot p_2\right){}^{j_1} \big] \left(-\Delta \cdot
   p_1\right){}^{-j_1+n-3}  \bigg] \bigg)\nonumber 
   \\
   &\qquad 
   + \Delta ^{\mu _1} \Delta ^{\mu _2} \Delta ^{\mu _3} \Delta ^{\mu _4}
   \bigg(\left(p_1\cdot p_1+p_2\cdot p_2+p_3\cdot p_3+p_4\cdot p_4\right)
   f^{a a_1 a_3} f^{a a_2 a_4} \nonumber 
   \\
   &\qquad+\left(13 p_1\cdot p_1-5 p_2\cdot p_2+13
   p_3\cdot p_3 -5 p_4\cdot p_4\right) f^{a a_1 a_2} f^{a a_3 a_4} \bigg)
   \nonumber 
   \\
   &\qquad \times \sum_{j_1=0}^{n-4} \sum_{j_2=0}^{j_1} \left(\left(-\Delta
   \cdot p_3\right){}^{j_2} \left(\Delta \cdot \left(p_1+p_2\right)\right){}^{j_1-j_2} \left(\Delta \cdot p_1\right){}^{-j_1+n-4}\right) \nonumber 
   \\
   &\qquad
   -\Delta ^{\mu _2} \Delta ^{\mu _4} \left( f^{a a_1 a_3}
   f^{a a_2 a_4}+13 f^{a a_1 a_2} f^{a a_3 a_4} \right) \left(\Delta ^{\mu _1} p_3^{\mu
   _3}-\Delta ^{\mu _3} p_1^{\mu _1}\right)\nonumber 
   \\
   &\qquad\times \sum_{j_1=0}^{n-3} \sum_{j_2=0}^{j_1}  \left(\left(\Delta \cdot \left(-p_1-p_2\right)\right){}^{j_1-j_2}
   \left(\Delta \cdot p_3\right){}^{j_2} \left(-\Delta \cdot p_1\right){}^{-j_1+n-3}\right)
   +3 \Delta ^{\mu _1} \Delta^{\mu _2} \Delta ^{\mu _4}\nonumber 
   \\
   &\qquad \times \left(4 p_1^{\mu _3}+p_3^{\mu _3}\right) f^{a a_1 a_3} f^{a a_2 a_4}
   \sum_{j_1=0}^{n-3} \bigg[ \big[  4 \left(\Delta \cdot \left(p_1+p_3\right)\right){}^{j_1} + \left(\Delta \cdot p_4\right){}^{j_1} \big]  \left(-\Delta \cdot
   p_2\right){}^{-j_1+n-3} \nonumber 
   \\
   &\qquad   +2  \left(\Delta \cdot p_4\right){}^{j_1} \left(\Delta \cdot
   \left(-p_1-p_3\right)\right){}^{-j_1+n-3} \bigg] +  \frac{12 d_A^{a_1 a_2 a_3 a_4}}{C_A}  \bigg( \nonumber 
   \\
   & \qquad  -\Delta^{\mu _2} \Delta^{\mu _3} \Delta^{\mu _4} p_1^{\mu _1}
   \sum_{j_1=0}^{n-3} \sum_{j_2=0}^{j_1} \bigg[ \left(-\Delta \cdot p_2\right){}^{j_1-j_2} \left(\Delta \cdot \left(p_1+p_4\right)\right){}^{j_2}
   \left(\Delta \cdot p_1\right){}^{-j_1+n-3} \bigg]   \nonumber 
   \\
   & \qquad  +  \Delta^{\mu _1} \Delta^{\mu _2} \Delta^{\mu _3} \Delta^{\mu _4} \left(p_1\cdot p_1+p_3\cdot p_3\right) \nonumber
   \\
   & \qquad \times \sum_{j_1=0}^{n-4} \sum_{j_2=0}^{j_1} \bigg[ \left(-\Delta \cdot p_2\right){}^{j_1-j_2} \left(\Delta \cdot \left(-p_2-p_3\right)\right){}^{j_2}
   \left(\Delta \cdot p_1\right){}^{-j_1+n-4} \bigg]   \bigg) \bigg\} + \text{{\it{permutations}}} \,,
\label{eq:OA4g}
\end{align}
where plus {\it{permutations}} indicates the summation over all the 
external gluon indices (simultaneous permutation of $\mu_i,a_i,p_i$).
In the above equations, the fully symmetric color structure $d_A^{a_1 a_2 a_3 a_4}$ in the adjoint representation is defined by
\begin{align}
    d_A^{a_1 a_2 a_3 a_4} = \frac{1}{4!} \bigg[ \text{Tr} \left( T_A^{a_1} T_A^{a_2} T_A^{a_3} T_A^{a_4} \right) + \text{symmetric permutations}  \bigg]\,, 
    \label{eq:d4ADef}
\end{align}
with $(T_A)^{a_1}_{a_2 a_3} = -i f^{a_1 a_2 a_3}$.

\subsection{\texorpdfstring
{Two-loop counterterm Feynman rules for $[Z O]_{g}^{\textrm{GV}}$}
{Two-loop counterterm Feynman rules for [ZO]_g^GV}}
As discussed in Subsection~\ref{subsec:FormulaZOgVV}, the second order renormalization constants $Z_{gA}^{(2)}$, $Z_{gq}^{(2)}$ and $Z_{gg}^{(2)}$ can be derived from \eqref{eq:counterterm2cmg},~\eqref{eq:counterterm2qmg} and~\eqref{eq:counterterm2gmg} for $m=0$. In practice, this can be done following the same computational method as described in Section~\ref{sec:ComputationalMethods}. As a reference, we write down the result for $Z_{gA}^{(2)}$ extracted from \eqref{eq:counterterm2cmg}, 
\begin{align}
\label{eq:zgA2Res}
    &Z_{gA}^{(2)} = \frac{1}{\epsilon^2\, n(n-1)}\bigg[C_A^2 \left(\frac{56 n^4+115 n^3-8 n^2-85 n+30}{12 (n-1) n (n+1) (n+2)}-\frac{11
   S_1(n)}{4}\right)-\frac{2 C_A N_f}{3}\bigg]   \nonumber
   \\
   & +  \frac{1}{\epsilon n(n-1)} \bigg[C_A^2 \Big(-\frac{5}{4} S_{1,1}(n)+\frac{\left(35 n^2-11 n-12\right) S_1(n)}{12 (n-1) n}+S_{-2}(n)+3 S_2(n)   \nonumber
   \\
   & -\frac{301 n^8+1195 n^7+539
   n^6-2102 n^5-1747 n^4+1159 n^3+1555 n^2-72 n-180}{36 (n-1)^2 n^2 (n+1)^2 (n+2)^2}\Big) \nonumber
   \\
   & +(1-\xi )
   C_A^2 \left(\frac{2 n^3+n^2-1}{8 (n-1) n (n+1)}-\frac{S_1(n)}{8}\right)+C_A N_f \left(\frac{16}{9}-\frac{2
   S_1(n)}{3}\right)\bigg]\,,
\end{align}
where two color structures, $C_A^2$ and $C_A N_f$, appear in the result. The three-loop correction to $Z_{gA}$ is documented in Appendix \ref{sec:zqAzgA}. Unlike the physical renormalization constants $Z_{ij}$ with $i,\,j = q \text{ or } g$, $Z_{gA}^{}$ depends on the gauge parameter $\xi$.

In the following, we focus on equations \eqref{eq:counterterm2cmg},~\eqref{eq:counterterm2qmg} and~\eqref{eq:counterterm2gmg} with $m=1$, especially on the computation of the off-shell, two-loop, three-leg OMEs. We follow closely the method described in Section~\ref{sec:ComputationalMethods} and work directly in parameter-$x$ space. However, the computations are much more involved compared to
the two-leg case, due to the larger number of scales:
\begin{align}
    p_1\cdot p_1,\, p_2\cdot p_2,\, p_1 \cdot p_2,\, \Delta \cdot p_1 =1,\, \Delta \cdot p_2 = z_1 \,, 
\end{align}
where we consider an off-shell, three-particle interaction with all momenta incoming, and eliminate $p_3$ using momentum conservation $p_3 = -p_1-p_2$. We also set $\Delta \cdot p_1$ to 1 and $\Delta \cdot p_2 $ to $z_1$. 

According to the analysis in Subsection~\ref{subsec:Lorentzstructure}, for a twist-two operator, the Feynman rules involving quarks or ghosts do not depend on the Mandelstam variables. Further, the Feynman rules containing only gluons are linear in the Mandelstam variables. Therefore, we can perform an IBP reduction by setting $ p_1\cdot p_1,\, p_2\cdot p_2,\, p_1 \cdot p_2$ to some random non-zero rational numbers. A single numeric sample is then sufficient to determine the Feynman rules involving quarks or ghosts. Another sample is used as a cross-check. For the case involving only gluons, we also need only one numeric sample to fix the single unknown parameter, due to the symmetry constraint resulting from gluons obeying Bose statistics.

Another simplification is due to the application of simplified Lorentz structures in the Feynman rules of a twist-two operator, as discussed in Subsection~\ref{subsec:Lorentzstructure}. In contrast to the last subsection, the application to the present case requires us to consider each of the right-hand sides of equations \eqref{eq:counterterm2cmg},~\eqref{eq:counterterm2qmg} and~\eqref{eq:counterterm2gmg} as a whole. To explain the reason, we analyze the Lorentz structures of each term in \eqref{eq:counterterm2cmg} for the case $m=1$. The same analysis also applies to \eqref{eq:counterterm2qmg} and \eqref{eq:counterterm2gmg}. The left-hand side of \eqref{eq:counterterm2cmg} is composed of one Lorentz structure only:
\begin{align}
\label{eq:m1zogvv}
    \braket{c|\left[Z O\right]_{g}^{\textrm{GV},\,(2)}|c \,g}^{\mu_1 ,\,(0),\,(1)\text{}}_{\text{1PI}} = d_1 \, \Delta^{\mu_1}\,.
\end{align}
The sum on the right-hand side of \eqref{eq:counterterm2cmg} also yields the same Lorentz structure, which is however not true for each term individually. For example, the following two-loop OMEs depend on three Lorentz structures
\begin{eqnarray}
\label{eq:2c1gstructure}
     \left[\braket{c|O_g|c \,g}^{\mu_1,\,(2),\,(1),\,\text{B}}_{\text{1PI}} \right]_{\text{div}}  = c_1 \,\Delta^{\mu_1} + c_2\, p_1^{\mu_1} +c_3 \,p_2^{\mu_1} \,,
\end{eqnarray}
where $c_2$ and $c_3$ are in general non-zero, but they  cancel with the corresponding structures from the $\mathcal{O}(\epsilon^0)$ part of the one-loop OMEs. Even if we do not care about the values of $c_2$ or $c_3$, three projectors are still needed to determine the value of $c_1$. To resolve this problem, we postpone the evaluation of $c_1$ and only evaluate the right-hand side of \eqref{eq:counterterm2cmg} as a whole. Explicitly, we contract $p_{1,\,\mu_1}$ with each term in the right-hand side of \eqref{eq:counterterm2cmg} and sum them up. The contributions from the last two Lorentz structures in \eqref{eq:2c1gstructure} cancel in the result. Therefore, $d_1$ in \eqref{eq:m1zogvv} is determined by dividing the result by $\Delta \cdot p_1$, i.e.,
\begin{align}
    d_1 = \frac{1} {\Delta \cdot p_1}   p_{1,\,\mu_1} \cdot \left[ \text{right-hand side of \eqref{eq:counterterm2cmg} for } m=1 \right]\,.
\end{align}

\begin{figure}
\begin{center}
\begin{minipage}{4.6cm}
\includegraphics[scale=0.9]{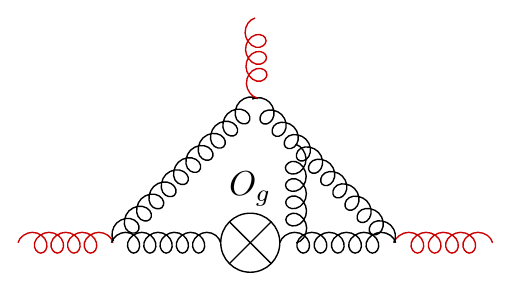} 
\end{minipage}
\begin{minipage}{4.6cm}
\includegraphics[scale=0.9]{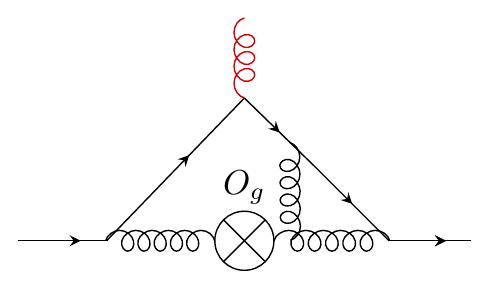} 
\end{minipage}
\begin{minipage}{4.6cm}
\includegraphics[scale=0.9]{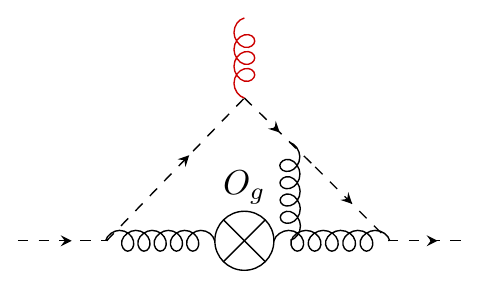} 
\end{minipage}
\caption{Sample 2-loop Feynman diagrams to determine the counterterm Feynman rules with 3 legs stemming from $\left[Z O\right]_{g}^{\textrm{GV},\,(2)}$.}
\end{center}
\end{figure}

With the above two simplifications, the computations become much more feasible. Since they otherwise follow a standard chain of steps described in Section~\ref{sec:ComputationalMethods}, we do not need to go into much further detail. We stress that the IBP reductions and the constructions of DEs are done by setting $p_1\cdot p_1,\,p_2\cdot p_2,\,p_1\cdot p_2$ to three different non-zero prime integers. About 500 master integrals appear in the result, for which we derive only the DEs with respect to $x$, not
with respect to $z_1$. For the solutions of the DEs, we fix the boundary conditions in the limit of $x \to 0$, where the integrals coincide with the all-off-shell, two-loop, 3-leg integrals without operator insertion~\cite{Birthwright:2004kk}. Only a few master integrals are needed to fix the boundary conditions, but their solutions are in terms of rather complicated Goncharov multiple polylogarithms (GPLs) or Bloch-Wigner functions. From Subsection~\ref{subsec:Lorentzstructure}, we know that Feynman rules should not contain GPLs with Mandelstam variables as their arguments. Moreover, as two-loop counterterm Feynman rules, the results are free from transcendental numbers for a fixed $n$, similarly to \eqref{eq:zgA2Res}. Thus we can simplify the boundary conditions by only keeping the transcendental-weight zero contributions, which involve the following three types of master integrals only,
\begin{align}
J_1 &=(\mu^2)^{2\epsilon}\int \frac{d^d l_1}{i \pi^{d/2}} \frac{d^d l_2}{i \pi^{d/2}} \,\frac{1}{l_1^2\, l_2^2\, (l_1+l_2+p_1)^2}= \frac{-p_1^2}{2 \epsilon (1-2 \epsilon) (2- 3 \epsilon) (1- 3 \epsilon)} + \mathcal{O}(\lambda)\,, \nonumber
\\
J_2 &= (\mu^2)^{2\epsilon}\int \frac{d^d l_1}{i \pi^{d/2}} \frac{d^d l_2}{i \pi^{d/2}}\, \frac{1}{l_1^2\, (l_1+p_1)^2 \, l_2^2 \,(l_2+p_2)^2} = \frac{1}{\epsilon^2(1-2 \epsilon)^2} + \mathcal{O}(\lambda)\,, \nonumber
\\
J_3 &= (\mu^2)^{2\epsilon}\int \frac{d^d l_1}{i \pi^{d/2}} \frac{d^d l_2}{i \pi^{d/2}}\, \frac{1}{l_1^2\, (l_1+p_1)^2 \,l_2^2 \,(l_2-l_1+p_2)^2} =  \frac{1}{2 \epsilon^2 (1-2 \epsilon)(1-3 \epsilon)} + \mathcal{O}(\lambda)\,,
\end{align}
where $\lambda^i$ indicates the contribution from polylogarithmic functions and numbers of transcendental weight $i$ and $\mu$ is the 't Hooft scale.
We do not try to solve the DEs in terms of special functions. Instead, we extract fixed Mellin moments of master integrals by solving the DEs in terms of a power series expansion in $x$.
For each master integrals $f_i$ we write
\begin{align}
     f_i(\epsilon,z_1)  = \sum_{j=0}^{\infty} a_{ij} (\epsilon, z_1) x^j \,,
\end{align}
where it is straightforward to expand to high powers in $x$.
We stress that we do not need to include $\ln (x)$ or $x^\epsilon$ terms due to our interpretation of $x$ as a tracing parameter, see \eqref{eq:sumXmethod}. In practice, we also expand in $\epsilon$ and set $z_1$ to some random prime numbers to speed up the expansion. The reconstruction of the $z_1$ dependence is then performed after combining different contributions to a Feynman rule for fixed $n$.

For the one-loop OMEs appearing in \eqref{eq:counterterm2cmg},~\eqref{eq:counterterm2qmg} and~\eqref{eq:counterterm2gmg}, we adopt the same method as described above, since they are required to be evaluated to order $\epsilon^0$. We obtain all counterterm Feynman rules with 3 legs for $\left[Z O\right]_{g}^{\textrm{GV},\,(2)}$ up to $n=96$.
In order to obtain all-$n$ Feynman rules, we match our results against a heuristic ansatz.
While we find that harmonic sums \eqref{eq:HarmonicDefinition} are not sufficient to express the Feynman rules, we were successful in expressing them using also generalized harmonic sums~\cite{Moch:2001zr} defined by
\begin{align}
&S_{\pm m} (y;n) = \sum_{j=1}^n  j^{\mp m}  y^{j} \,,\, S_0(y;n)=  \sum_{j=1}^n  y^{j} \,, \nonumber \\
    & S_{\pm m_1, m_2, \cdots m_d}(y_1,y_2,\cdots y_d;n)= \sum_{j=1}^n j^{\mp m_1} y_1^j S_{m_2, \cdots m_d}(y_2,\cdots y_d;j) \,. 
\end{align}
From the structure of the leading poles, we expect sums up to
weight two with integer $m_i$ and arguments
\begin{equation}
y_i \in \left\{ 1, \pm z_1, \pm (z_1 + 1), \pm \frac{1}{z_1}, \pm \frac{1}{z_1+1}, \pm \frac{z_1}{z_1+1} \right\}.
\end{equation}
We reconstruct the all-$n$ Feynman rules from fixed Mellin moments up to $n=76$, and then cross-check our symbolic result against the remaining numerical data for moments up to $n=96$. The package \texttt{FiniteFlow}~\cite{Peraro:2019svx} was used to speed up the reconstruction process.

As shown in \eqref{eq:2legsFeyZOgVV}, the two-leg counterterm Feynman rules for $\left[Z O\right]_{g}^{\textrm{GV},\,(2)}$ are zero,
\begin{flalign}
 &\,\includegraphics[scale=1.0]{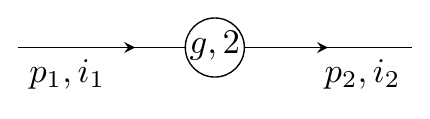} 
 \nonumber 
 \\
   & \qquad \to 0\,,  \label{eq:FeynOgVV2q0g} && 
   \\
 &\,\includegraphics[scale=1.0]{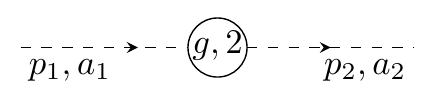}  
 \nonumber 
 \\
 &\qquad  \to 0 \,,  &&
 \\
  &\includegraphics[scale=01.0]{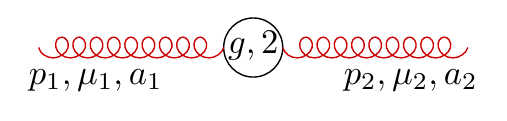} 
 \nonumber 
 \\
 & \qquad \to 0 \,. &&
 \end{flalign}
 The two-loop counterterm Feynman rules contributing to $\left[Z O\right]_{g}^{\textrm{GV}}$, i.e, the left-hand sides of \eqref{eq:counterterm2cmg},~\eqref{eq:counterterm2qmg},~\eqref{eq:counterterm2gmg}, are listed in the following, with the convention of all momenta flowing into the vertices,
 \begin{align}
 &\includegraphics[scale=1.0]{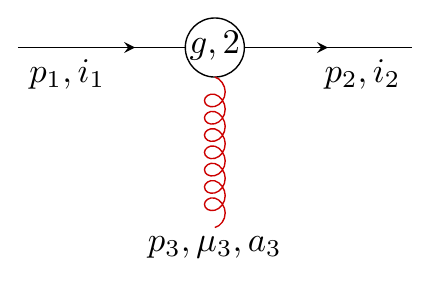} 
 \nonumber 
 \\
   & \quad \to 0\,,  \label{eq:FeynOgVV2q1g}
   \\
 &\includegraphics[scale=1.0]{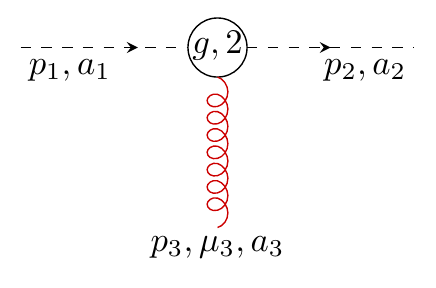} 
 \nonumber 
 \\&\quad  \to 2i g_s C_A^2 f^{a_1 a_2 a_3} \frac{1+(-1)^n }{256n(n-1)}  \left( \Delta \cdot p_1\right)^{n-1}\Delta^{\mu_3} \bigg\{ \frac{F_{-2,0} + \left( 1- \xi\right) F_{-2,1} }{\epsilon^2} \nonumber 
   \\
   & \qquad 
   + \frac{F_{-1,0} + \left( 1- \xi\right) F_{-1,1} }{\epsilon}  \bigg\}  \,,  \label{eq:FeynRulesccgZOgVV2}
 \\
 &\includegraphics[scale=1.0]{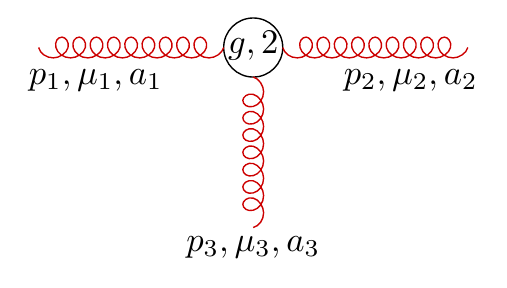}  
\nonumber
\\
   &\quad  \to 2i g_s C_A^2 f^{a_1 a_2 a_3}  \frac{1+(-1)^n }{256n(n-1)}  \frac{ \left( \Delta \cdot p_1\right)^{n-2}}{\Delta \cdot p_2}  
    \bigg( - \Delta^{\mu _1}
   \Delta^{\mu _2} \Delta^{\mu _3} \left(p_1\cdot p_1+p_2\cdot p_2+p_3\cdot p_3\right)    \nonumber 
   \\
   &\qquad +\Delta^{\mu _2} \Delta^{\mu _3} p_1^{\mu _1} \Delta \cdot p_1 +\Delta^{\mu _1} \Delta^{\mu _3} p_2^{\mu _2} \Delta \cdot p_2 +\Delta^{\mu _1} \Delta^{\mu _2} p_3^{\mu _3} \Delta \cdot p_3   \bigg) \nonumber 
   \\
   & \qquad \times
    \bigg\{\frac{F_{-2,0} + \left( 1- \xi\right) F_{-2,1} }{\epsilon^2} + \frac{F_{-1,0} + \left( 1- \xi\right) F_{-1,1} }{\epsilon} \bigg\} \,,
     \label{eq:FeynRules3gZOgVV2}
\end{align}
where the quark-quark-gluon counterterm Feynman rule for $\left[Z O\right]_{g}^{\textrm{GV},\,(2)}$ is zero. The 
counterterm Feynman rules for 3-gluon and ghost-ghost-gluon vertices are proportional to each other, indicating they are related by a generalized BRST symmetry. 
The scalar form factors in \eqref{eq:FeynRules3gZOgVV2} are as follows:
\begin{align}
& F_{-2,1}= -\frac{\left(z_1-1\right) (1+z_1)^n}{\left(z_1+1\right){}^2}+\frac{\left(z_1+2\right) z_1^n}{z_1 \left(z_1+1\right)}-\frac{\left(2 z_1+1\right)
   1^n}{z_1+1} \,, 
   \\
   & F_{-2,0}= z_1^n \bigg[-\frac{2 \left(z_1-2\right)
  }{z_1 \left(z_1+1\right)}  S_1\Big(-\frac{1}{z_1};n\Big)+\frac{2
   \left(3 z_1+2\right) }{z_1
   \left(z_1+1\right)}S_1\Big(\frac{z_1+1}{z_1};n\Big) \nonumber 
   \\
   &-\frac{2 \left(z_1+2\right) \left(2 n^2 z_1+2 n^2-2
   n z_1^2-9 n z_1-9 n+2 z_1^2+3 z_1+3\right)}{(n-1) n z_1
   \left(z_1+1\right){}^2}-\frac{6 \left(z_1+2\right) }{z_1
   \left(z_1+1\right)} S_1(n)\bigg]   \nonumber 
   \\
   &+(1+z_1)^n \bigg[\frac{2
   \left(z_1+3\right)
  }{\left(z_1+1\right){}^2}  S_1\Big(\frac{1}{z_1+1};n\Big)-\frac{2
   \left(3 z_1+1\right)
   }{\left(z_1+1\right){}^2} S_1\Big(\frac{z_1}{z_1+1};n\Big) \nonumber 
   \\
   &+\frac{2
   \left(z_1-1\right) \left(2 n^2 z_1+2 n z_1^2-5 n z_1+2 n-2
   z_1^2-z_1-2\right)}{(n-1) n z_1 \left(z_1+1\right){}^2}+\frac{6
   \left(z_1-1\right)
   }{\left(z_1+1\right){}^2} S_1(n)\bigg] \nonumber 
   \\
   &+1^n \bigg[-\frac{2
   \left(2 z_1-1\right) }{z_1+1} S_1\left(-z_1;n\right)-\frac{2 \left(2
   z_1+3\right) }{z_1+1} S_1\left(z_1+1;n\right)+\frac{6
   \left(2 z_1+1\right)}{z_1+1}  S_1(n) \nonumber 
   \\
   &+\frac{2 \left(2
   z_1+1\right) \left(2 n^2 z_1^2+2 n^2 z_1-9 n z_1^2-9 n z_1-2 n+3
   z_1^2+3 z_1+2\right)}{(n-1) n z_1 \left(z_1+1\right){}^2}\bigg]  \,, 
   \\
   & F_{-1,1} = z_1^n
   \bigg[-\frac{\left(z_1+2\right)
   \left(n^2 z_1+n^2-n z_1^2+2 n z_1+2 n+z_1^2-z_1-1\right)}{(n-1) n z_1
   \left(z_1+1\right){}^2}  \nonumber 
   \\
   &+\frac{S_1\Big(-\frac{1}{z_1};n\Big)}{z_1}+\frac{S_1\Big(\frac
   {z_1+1}{z_1};n\Big)}{z_1 \left(z_1+1\right)}\bigg] +(1+z_1)^n
   \bigg[\frac{S_1\Big(\frac{z_1}{z_1+1};n\Big)}{\left(z_1+1\right){}^2
   }-\frac{z_1
   S_1\Big(\frac{1}{z_1+1};n\Big)}{\left(z_1+1\right){}^2} \nonumber 
   \\
   &+\frac{\left(
   z_1-1\right) \left(n^2 z_1+n z_1^2+4 n z_1+n-z_1^2-3
   z_1-1\right)}{(n-1) n z_1 \left(z_1+1\right){}^2}\bigg] +1^n
   \bigg[-S_1\left(-z_1;n\right)\nonumber 
   \\
   &-\frac{z_1
   S_1\left(z_1+1;n\right)}{z_1+1}+\frac{\left(2 z_1+1\right) \left(n^2
   z_1^2+n^2 z_1+2 n z_1^2+2 n z_1-n-z_1^2-z_1+1\right)}{(n-1) n z_1
   \left(z_1+1\right){}^2}\bigg]  \,, 
   \\
   &F_{-1,0} = z_1^n \bigg[-\frac{4 n
  }{z_1 \left(z_1+1\right)}  S_2\Big(\frac{1}{z_1};n\Big)+\frac{8 n
    \left(z_1+2\right)}{z_1 \left(z_1+1\right)} S_{-2}(n) -\frac{4 (n+1)
    \left(z_1+2\right)}{z_1 \left(z_1+1\right)} S_2(n)\nonumber 
   \\
   &+\frac{6
   \left(z_1+2\right)}{z_1 \left(z_1+1\right)}  S_{1,1}(n) +\frac{4
    \left(n+z_1\right)}{z_1
   \left(z_1+1\right)} S_2\Big(\frac{z_1+1}{z_1};n\Big)-\frac{2 
   \left(3 z_1+2\right)}{z_1 \left(z_1+1\right)} S_{1,1}\Big(1,\frac{z_1+1}{z_1};n\Big)\nonumber 
   \\
   &+\frac{4
   \left(z_1 n+n-z_1\right)}{z_1
   \left(z_1+1\right)}  S_2\Big(-\frac{1}{z_1};n\Big) -\frac{4 n
   }{z_1} S_2\Big(-\frac{z_1+1}{z_1};n\Big) -\frac{4 n
   }{z_1} S_{1,1}\Big(-\frac{1}{z_1},1;n \Big) \nonumber 
   \\
   &-\frac{2 \left(z_1+2\right) \left(6 z_1 n^3+6 n^3-4
   z_1^2 n^2-15 z_1 n^2-15 n^2+6 z_1^2 n+16 z_1 n+16 n-2 z_1^2-3
   z_1-3\right)}{(n-1) n^2 z_1 \left(z_1+1\right){}^2}\nonumber 
   \\
   &-\frac{2 
   \left(z_1+2\right) \left(2 z_1^2 n^2+z_1 n^2+n^2-2 z_1^2 n-7 z_1 n-7
   n+2 z_1+2\right)}{(n-1) n z_1 \left(z_1+1\right){}^2} S_1(n) \nonumber 
   \\
   &-\frac{2
    \left(2 n z_1^3-n z_1^2+2 z_1^2-5 n
   z_1+2 z_1-4 n\right)}{n z_1 \left(z_1+1\right){}^2} S_1\Big(\frac{z_1+1}{z_1};n\Big) +\frac{4 n
  }{z_1}  S_{1,1}\Big(-\frac{1}{z_1},-z_1;n\Big)\nonumber 
   \\
   &+\frac{2
    \left(2 n z_1^3+3 n z_1^2+2 z_1^2+7 n
   z_1+2 z_1+4 n\right)}{n z_1 \left(z_1+1\right){}^2} S_1\Big(-\frac{1}{z_1};n\Big)  -\frac{4 n
   }{z_1} S_{1,1}\Big(-\frac{1}{z_1},z_1+1;n\Big)\nonumber 
   \\
   &-\frac{4 n}{z_1
   \left(z_1+1\right)}  S_{1,1}\Big(\frac{z_1+1}{z_1},1;n\Big) -\frac{4 n
   S_{1,1}\left(\frac{z_1+1}{z_1},\frac{1}{z_1+1};n\right)}{z_1
   \left(z_1+1\right)}+\frac{4 n
   S_{1,1}\left(\frac{z_1+1}{z_1},\frac{z_1}{z_1+1};n\right)}{z_1
   \left(z_1+1\right)}\nonumber 
   \\
   &+\frac{2 
   \left(z_1-2\right)}{z_1
   \left(z_1+1\right)} S_{1,1}\Big(1,-\frac{1}{z_1};n\Big) \bigg]+(1+z_1)^n \bigg[ -\frac{4
    \left(n
   z_1-z_1-1\right)}{\left(z_1+1\right){}^2} S_2\Big(\frac{1}{z_1+1};n\Big)\nonumber 
   \\
   &-\frac{2
   \left(z_1+3\right)}{\left(z_1+1\right){}^2} S_{1,1}\Big(1,\frac{1}{z_1+1};n\Big)+\frac{2
    \left(3
   z_1+1\right)}{\left(z_1+1\right){}^2} S_{1,1}\Big(1,\frac{z_1}{z_1+1};n\Big)  -\frac{4
   n }{\left(z_1+1\right){}^2} S_2\Big(-\frac{1}{z_1+1};n\Big)\nonumber 
   \\
   &-\frac{2 
   \left(z_1-1\right) \left(2 z_1^2 n^2+3 z_1 n^2+2 n^2-2 z_1^2 n+3 z_1
   n-2 n-2 z_1\right)}{(n-1) n z_1 \left(z_1+1\right){}^2} S_1(n)\nonumber 
   \\
   &+\frac{2
   \left(z_1-1\right) \left(6 z_1 n^3+4 z_1^2 n^2-7 z_1 n^2+4 n^2-6 z_1^2
   n+4 z_1 n-6 n+2 z_1^2+z_1+2\right)}{(n-1) n^2 z_1
   \left(z_1+1\right){}^2}\nonumber 
   \\
   &+\frac{2 
   \left(2 n z_1^3+3 n z_1^2-2 z_1^2+7 n z_1-2 z_1+2 n\right)}{n z_1
   \left(z_1+1\right){}^2} S_1\Big(\frac{1}{z_1+1};n\Big) +\frac{4 n
   z_1}{\left(z_1+1\right){}^2}  S_2\Big(-\frac{z_1}{z_1+1};n\Big) \nonumber 
   \\
   &-\frac{2 
   \left(2 n z_1^3+7 n z_1^2-2 z_1^2+3 n z_1-2 z_1+2 n\right)}{n z_1
   \left(z_1+1\right){}^2} S_1\Big(\frac{z_1}{z_1+1};n\Big) -\frac{4 n
  }{\left(z_1+1\right){}^2}  S_{1,1}\Big(\frac{z_1}{z_1+1},1;n\Big)\nonumber 
   \\
   &-\frac{4 n}{\left(z_1+1\right){}^2}  S_{1,1}\Big(\frac{z_1}{z_1+1},-\frac{1}{z_1};n\Big)  +\frac{4 n S_{1,1}\Big(\frac{z_1}{z_1+1},\frac{z_1+1}{z_1};n\Big)
  }{\left(z_1+1\right){}^2} +\frac{4 
   \left(n-z_1-1\right)}{\left(z_1+1\right){}^2} S_2\Big(\frac{z_1}{z_1+1};n\Big) \nonumber 
   \\
   &-\frac{8 n 
   \left(z_1-1\right)}{\left(z_1+1\right){}^2} S_{-2}(n) +\frac{4 (n+1)
   \left(z_1-1\right)}{\left(z_1+1\right){}^2}  S_2(n) -\frac{6
   \left(z_1-1\right)}{\left(z_1+1\right){}^2}  S_{1,1}(n)\nonumber 
   \\
   &+\frac{4 n z_1  S_{1,1}\left(\frac{1}{z_1+1},1;n\right)}{\left(z_1+1\right){}^2}+\frac{4 n
   z_1 S_{1,1}\left(\frac{1}{z_1+1},-z_1;n\right)}{\left(z_1+1\right){}^2}-\frac{4 n
   z_1 S_{1,1}\left(\frac{1}{z_1+1},z_1+1;n\right)}{\left(z_1+1\right){}^2}\bigg] \nonumber 
   \\
   &+1^n \bigg[4 n
   S_2\left(-z_1-1;n\right)+4 n S_{1,1}\left(-z_1,1;n\right)-4 n
   S_{1,1}\Big(-z_1,-\frac{1}{z_1};n\Big) \nonumber 
   \\
   &+4 n
   S_{1,1}\Big(-z_1,\frac{z_1+1}{z_1};n\Big)+\frac{2
   \left(2 z_1-1\right)}{z_1+1}  S_{1,1}\left(1,-z_1;n\right) -\frac{8 n
   \left(2 z_1+1\right)}{z_1+1}  S_{-2}(n)\nonumber 
   \\
   &+\frac{4 (n+1) \left(2
   z_1+1\right)}{z_1+1}  S_2(n)-\frac{6 \left(2
   z_1+1\right)}{z_1+1}  S_{1,1}(n) +\frac{2 \left(2
   z_1+3\right)}{z_1+1}  S_{1,1}\left(1,z_1+1;n\right) \nonumber 
   \\
   &-\frac{4 \left(n
   z_1+1\right)}{z_1+1}  S_2\left(z_1+1;n\right)  -\frac{4  \left(z_1
   n+n-1\right)}{z_1+1} S_2\left(-z_1;n\right) \nonumber 
   \\
   &+\frac{2 \left(2 z_1+1\right) \left(z_1^2
   n^2+z_1 n^2+2 n^2-7 z_1^2 n-7 z_1 n-2 n+2 z_1^2+2 z_1\right)}{(n-1) n
   z_1 \left(z_1+1\right){}^2}  S_1(n) \nonumber 
   \\
   &+\frac{1}{(n-1) n^2 z_1 \left(z_1+1\right){}^2}\Big[2 \left(2 z_1+1\right) \big(6 z_1^2
   n^3+6 z_1 n^3-15 z_1^2 n^2-15 z_1 n^2 \nonumber
   \\
   &-4 n^2+16 z_1^2 n+16 z_1 n+6 n-3
   z_1^2-3 z_1-2\big)\Big] \nonumber 
   \\
   &-\frac{2
    \left(4 n z_1^3+5 n z_1^2-2 z_1^2+n z_1-2 z_1-2
   n\right)}{n z_1 \left(z_1+1\right){}^2} S_1\left(z_1+1;n\right) +\frac{4 n  z_1}{z_1+1} S_{1,1}\left(z_1+1,1;n\right) \nonumber 
   \\
   &-\frac{2 
   \left(4 n z_1^3+7 n z_1^2+2 z_1^2+3 n z_1+2 z_1+2 n\right)}{n z_1
   \left(z_1+1\right){}^2} S_1\left(-z_1;n\right) +\frac{4 n 
   z_1}{z_1+1} S_2\left(z_1;n\right) \nonumber 
   \\
   &-\frac{4
   n z_1 S_{1,1}\Big(z_1+1,\frac{1}{z_1+1};n\Big)}{z_1+1}+\frac{4 n z_1
   S_{1,1}\Big(z_1+1,\frac{z_1}{z_1+1};n\Big) }{z_1+1}\bigg] \,, 
   \label{eq:scalarFFresults}
\end{align}
where $$z_1 = \frac{\Delta \cdot p_2}{\Delta \cdot p_1} \,, $$ and the following 46 (generalized) harmonic sums up to weight 2 appear
\begin{align}
\bigg\{&S_ 1 (n), \, S_ 1\Big (-\frac {1} {z_ 1}; n\Big),\, 
   S_ 1\Big (-z_ 1; n\Big),\, S_ 1\Big (\frac {1} {z_ 1 + 1}; n\Big),\,
   S_ 1\Big (\frac {z_ 1} {z_ 1 + 1}; n\Big),\,
   S_ 1\Big (z_ 1 + 1; n\Big),\, \nonumber
   \\
   &S_ 1\Big (\frac {z_ 1 + 1} {z_ 1}; n\Big),\, S_ {-2} (n),\, S_ 2 (n),\, 
   S_ {1, 1} (n), \,S_ 2\Big (-z_ 1 - 1; n\Big),\, 
   S_ 2\Big (-\frac {1} {z_ 1}; n\Big),\, 
   S_ 2\Big (\frac {1} {z_ 1}; n\Big),\,\nonumber
   \\
   & S_ 2\Big (-z_ 1; n\Big),\, 
   S_ 2\Big (z_ 1; n\Big),\, S_ 2\Big (-\frac{1} {z_ 1 + 1}; n\Big),\, 
   S_ 2\Big (\frac {1} {z_ 1 + 1}; n\Big),\, 
   S_ 2\Big (-\frac {z_ 1} {z_ 1 + 1}; n\Big), \,
   \nonumber
   \\
   &
   S_ 2\Big (\frac {z_ 1} {z_ 1 + 1}; n\Big), \, 
   S_ 2\Big (z_ 1 + 1; n\Big), \,
   S_ 2\Big (-\frac {z_ 1 + 1} {z_ 1}; n\Big), \, 
   S_ 2\Big (\frac {z_ 1 + 1} {z_ 1}; n\Big), \,
   S_ {1, 1}\Big (1, -\frac {1} {z_ 1}; n\Big), \,
   \nonumber
   \\
   &
   S_ {1, 1}\Big (1, -z_ 1; n\Big), \,
   S_ {1, 1}\Big (1, \frac {1} {z_ 1 + 1}; n\Big), \,
   S_ {1, 1}\Big (1, \frac {z_ 1} {z_ 1 + 1}; n\Big), \, 
   S_ {1, 1}\Big (1, z_ 1 + 1; n\Big), \,\nonumber
   \\
   &
   S_ {1, 1}\Big (1, \frac {z_ 1 + 1} {z_ 1}; n\Big),\, 
   S_ {1, 1}\Big (-\frac {1} {z_ 1}, 1; n\Big), \,
   S_ {1, 1}\Big (-\frac {1} {z_ 1}, -z_ 1; n\Big),\, 
   S_ {1, 1}\Big (-\frac {1} {z_ 1}, z_ 1 + 1; n\Big),\,\nonumber
   \\
   & 
   S_ {1, 1}\Big (-z_ 1, 1; n\Big),\,
   S_ {1, 1}\Big (-z_ 1, -\frac {1} {z_ 1}; n\Big),\, 
   S_ {1, 1}\Big (-z_ 1, \frac {z_ 1 + 1} {z_ 1}; n\Big),\, 
   S_ {1, 1}\Big (\frac {1} {z_ 1 + 1}, 1; n\Big), \, \nonumber
   \\
   &
   S_ {1, 1}\Big (\frac {1} {z_ 1 + 1}, -z_ 1; n\Big), \,
   S_ {1, 1}\Big (\frac {1} {z_ 1 + 1}, z_ 1 + 1; n\Big), \, 
   S_ {1, 1}\Big (\frac {z_ 1} {z_ 1 + 1}, 1; n\Big), \,
  \nonumber
   \\
   & 
   S_ {1, 1}\Big (\frac {z_ 1} {z_ 1 + 1}, \,\frac {z_ 1 + 1} {z_ 1}; 
     n\Big),\, S_ {1, 1}\Big (z_ 1 + 1, 1; n\Big), \,
   S_ {1, 1}\Big (z_ 1 + 1, \frac {1} {z_ 1 + 1}; n\Big), \,\nonumber
   \\
   &
   S_ {1, 1}\Big (z_ 1 + 1, \frac {z_ 1} {z_ 1 + 1}; n\Big),\, 
   S_ {1, 1}\Big (\frac {z_ 1 + 1} {z_ 1}, 1;
     n\Big),\, 
   S_ {1, 1}\Big (\frac {z_ 1 + 1} {z_ 1}, \frac {1} {z_ 1 + 1}; 
     n\Big),\, 
     \nonumber
     \\
      &S_ {1, 1}\Big (\frac {z_ 1} {z_ 1 + 1}, -\frac {1} {z_ 1}; n\Big),\,
   S_ {1, 1}\Big (\frac {z_ 1 + 1} {z_ 1}, \frac {z_ 1} {z_ 1 + 1}; 
     n\Big) \bigg\} \,.
\end{align}
We used \texttt{HarmonicSums} to evaluate these (generalized) harmonic sums throughout. 

We stress that our counterterm Feynman rules are given in closed form with symbolic $n$ dependence. Due to the appearance of the generalized harmonic sums, by keeping the all-$n$ dependence, it is not possible to disentangle the renormalization constants (characterized by their independence of $z_1$ in this case) from the  corresponding operators (which are independent of harmonic sums). Therefore, it appears impossible to determine the individual $i=2$ operators in a closed form for symbolic $n$ in this way.

It is, however, possible to work out the operator basis $n$-by-$n$, as considered in~\cite{Falcioni:2022fdm}. Typically, more and more operators are needed as $n$ increases, and the number is expected to go to infinity when $n$ tends to infinity. For a given value of $n$, only a finite number of operators 
are contributing through their operator 
Feynman rules. 
From our all-$n$ counterterm Feynman rules, we can 
deduce the operator basis $n$-by-$n$ by setting $n$ to 
a specific number. 
 For example, the scalar products in \eqref{eq:scalarFFresults} for $n=12$ are 
\begin{align}
 &F_{-2,1}\Big|_{n=12} \, = -\left(z_1-1\right) z_1 \left(z_1+2\right) \left(2 z_1+1\right)
   \left(z_1^2+z_1+2\right) \left(2 z_1^2+z_1+1\right) \left(2 z_1^2+3
   z_1+2\right) \,, \nonumber
   \\
   &F_{-2,0}\Big|_{n=12} \, = \frac{1}{630} \left(z_1-1\right) z_1 \left(z_1+2\right) \left(2
   z_1+1\right) \bigg[34126 z_1^6+102378 z_1^5+215515 z_1^4 \nonumber
   \\
   &+260400
   z_1^3+215515 z_1^2+102378 z_1+34126\bigg]   \,, \nonumber
   \\
   & F_{-1,1}\Big|_{n=12} \, =  \frac{\left(z_1-1\right) z_1 \left(z_1+2\right) \left(2 z_1+1\right)
  }{27720}  \bigg[180899 z_1^6+542697 z_1^5+1137425 z_1^4\nonumber
  \\
  &+1370355 z_1^3+1137425
   z_1^2+542697 z_1+180899\bigg] \,, \nonumber
   \\
 &F_{-1,0}\Big|_{n=12} \,= \frac{\left(z_1-1\right) z_1 \left(z_1+2\right) \left(2 z_1+1\right)}{1164240}  \bigg[9395264 z_1^6+28185792 z_1^5 \nonumber
   \\
   &+53213569 z_1^4+59450818 z_1^3+53213569 z_1^2+28185792 z_1+9395264 \bigg]\,,
\end{align}
where the results are polynomials in $z_1$ with constant coefficients, which makes it straightforward to separate the renormalization constants (rational numbers) and the Feynman rules of the corresponding operators. To infer the operators from the corresponding Feynman rules, we need to replace a momentum with the derivative of a field, for example, $(\Delta \cdot p_1 (1+z_1))^n = (-\Delta \cdot p_3)^n \to (-\Delta \cdot \partial)^n A^a_\mu$, where $A^a_\mu$ is a gluon field.

\subsection{\texorpdfstring
{Comparison with previous fixed $n$ results}
{Comparison with previous fixed n results}}

In a recent work~\cite{Falcioni:2022fdm}, Falcioni and Herzog have 
used gauge and BRST symmetry to derive a set of constraint equations for the GV operators. By applying these constraint equations to a general 
ansatz for the GV operators at fixed values of $n$, they were able to determine the required operator bases for low Mellin moments $n=2,\,4,\,6$ and to compute the 
associated renormalization constants. 

 In the following, we compare our all-$n$ counterterm Feynman rules
 evaluated at $n=2,\,4$ and $6$ with their results. Since our approach does not allow to 
 disentangle GV operators and renormalization constants, the comparison requires 
 us to first determine the GV operator Feynman rules resulting from~\cite{Falcioni:2022fdm}, which are subsequently multiplied with corresponding 
 renormalization constants in order to determine the counterterm Feynman rules. 
  It can be easily verified from \eqref{eq:FeynRulesccgZOgVV2} and \eqref{eq:FeynRules3gZOgVV2} that the counterterm Feynman rules from $\left[Z O\right]_{g}^{\textrm{GV}}$ are non-zero only for $n\geq 6$. Therefore, for $n=2,\,4$ the  GV operator basis is contained in the operators $O_{ABC}$. We presented detailed comparisons for each $n$ separately below. All comparisons are truncated up to four legs, and we find full agreement with~\cite{Falcioni:2022fdm}.

For $n=2$, the GV operators were expressed as $\mathcal{O}_1^{(2)}$ and $\mathcal{O}_2^{(2)}$ in~\cite{Falcioni:2022fdm}, and we find that they are related to our results as follows:
\begin{align}
    O_g\Big|_{n=2} & = - \mathcal{O}_1^{(2)} \,, \nonumber
    \\
Z_{gA}  \, \big( O_{AC}\big|_{\text{FR}} \big)\Big|_{n=2,\,N_f=0} &=-  \delta Z_{12}^{(2)} \mathcal{O}_2^{(2)}|_{\text{FR}} \,, 
    \label{eq:comN=2}
\end{align}
where  we use 'FR' to indicate the Feynman rules resulting from the corresponding operators. The renormalization constant $\delta Z_{12}^{(2)}$ in the above equation is given to order $a_s^3$ with full $\xi$ dependence in~\cite{Falcioni:2022fdm}. Our result for $Z_{gA}|_{n=2,\,N_f=0}$, in \eqref{eq:zgAn2}, is in full agreement with $\delta Z_{12}^{(2)}$ presented in~\cite{Falcioni:2022fdm}. The overall minus sign in the above equation is due to different normalizations of $O_g$ and $\mathcal{O}_{1}$, 
\begin{equation}
    O_g = -i^{n-2} \mathcal{O}_1\,.
\end{equation}

For $n=4$, the GV operator basis in~\cite{Falcioni:2022fdm} consists of three elements, which relate to our results as follows:
\begin{align}
    O_g\Big|_{n=4} &=  \mathcal{O}_1^{(4)} \,, \nonumber
    \\
    Z_{gA}  \, \big( O_{AC}\big|_{\text{FR}} \big)\Big|_{n=4,\,N_f=0} &=  \delta Z_{12}^{(4)} \mathcal{O}_2^{(4)}|_{\text{FR}} + \delta Z_{13}^{(4)} \mathcal{O}_3^{(4)}\big|_{\text{FR}} \,, 
    \label{eq:comN=4}
\end{align}
where the operators $\mathcal{O}_2^{(4)},\,\mathcal{O}_3^{(4)}$ are distinguished from each other by their different color structures with $\mathcal{O}_3^{(4)}$ being proportional to $d_A^{a_1 a_2 a_3 a_4}$ as defined in \eqref{eq:d4ADef}.
In~\cite{Falcioni:2022fdm}, the constant $\delta Z_{12}^{(4)}$ was given to $a_s^3$ with $ a_s^3 (1-\xi)^2$ being dropped, and $\delta Z_{13}^{(4)}$ was given to $a_s^1$ with 
\begin{align}
  \delta Z_{13}^{(4)} = a_s \frac{C_A}{24 \epsilon} + \mathcal{O}(a_s^2)\,.
\end{align}
We find the following relations between the renormalization constants:
\begin{align}
  & Z_{gA}\Big|_{n=4,\,N_f=0} - \delta Z_{12}^{(4)} = \mathcal{O}(a_s^3 (1-\xi)^2)\,,\nonumber
  \\
  &  Z_{gA}\Big|_{n=4,\,N_f=0} + 2  \delta Z_{13}^{(4)} = \mathcal{O}(a_s^2)\,,   
\end{align}
with $Z_{gA}\big|_{n=4}$ listed in \eqref{eq:zgAn4}. Given that 
these renormalization constants 
were truncated~\cite{Falcioniprivate} at order $(1-\xi)$ in~\cite{Falcioni:2022fdm}, we thus 
find full agreement.

For $n=6$, the comparison becomes more interesting, since the counterterm operator $ \left[Z O\right]_{g}^{\textrm{GV},\,(2)}$ starts to contribute. In this case, we further decompose the comparison into three different categories, depending on the number of legs. As the first category, for counterterm Feynman rules with two legs, we find the following relation, 
\begin{align}
Z_{gA}  \, \big( O_{AC}\big|_{\text{FR}} \big)\Big|_{n=6,\,N_f=0} = - \delta Z_{12}^{(6)} \mathcal{O}_2^{(6)}|_{\text{FR}}\,.  
    \end{align}
The above relation is expected to be satisfied to all orders. It is checked explicitly to order $a_s^2$ since $\delta Z_{12}^{(6)}$ is given to $a_s^2$ in~\cite{Falcioni:2022fdm}. For counterterm Feynman rules with three legs, the following relation holds, 
\begin{multline}
Z_{gA}  \, \big( O_{AC}\big|_{\text{FR}} \big)\Big|_{n=6,\,N_f=0} +\Big[ \left[Z O\right]_{g}^{\textrm{GV},\,(2)}|_{\text{FR}}\Big]\Big|_{n=6}  \\= 
- \Big\{ \delta Z_{12}^{(6)} \mathcal{O}_2^{(6)}|_{\text{FR}} + \delta Z_{13}^{(6)} \mathcal{O}_3^{(6)}\big|_{\text{FR}} \Big\} + \mathcal{O}(a_s^3)\,. 
\end{multline}
Interestingly, both $\delta Z_{12}^{(6)}$ and $\delta Z_{13}^{(6)}$ start to contribute at order $a_s$, while only the first term in the left-hand side of the above equation contributes at order $a_s$. At order $a_s^2$, also $\left[Z O\right]_{g}^{\textrm{GV},\,(2)}$ starts to contribute. The above relation is confirmed explicitly to order $a_s^2$. For counterterm Feynman rules with four legs, the following relation holds\footnote{
We thank Falcioni and Herzog for pointing out to us, that the term $+A^{a_1} \partial^2 A^{a_2} \partial c^{a_3}$ in the third line of equation (5.31) in \cite{Falcioni:2022fdm} defining the operator $\mathcal{O}_3^{(6)}$ should be modified to $+4A^{a_1} \partial^2 A^{a_2} \partial c^{a_3}$~\protect\cite{Falcioniprivate}. After correcting this minor typesetting issue, we find full agreement between their and our results.} at order $a_s^1$,
\begin{multline}
Z_{gA}  \, \big( O_{AC}\big|_{\text{FR}} \big)\Big|_{n=6,\,N_f=0}   \\
=  - \Big\{ \delta Z_{12}^{(6)} \mathcal{O}_2^{(6)}|_{\text{FR}} + \delta Z_{13}^{(6)} \mathcal{O}_3^{(6)}\big|_{\text{FR}} + \delta Z_{14}^{(6)} \mathcal{O}_4^{(6)}|_{\text{FR}} + \delta Z_{15}^{(6)} \mathcal{O}_5^{(6)}\big|_{\text{FR}} \Big\} + \mathcal{O}(a_s^2) \,, 
\end{multline}
where the four operators on the right-hand side are assembled into a single operator $O_{AC}$ on the left-hand side. At order $a_s^2$, one more operator $\mathcal{O}_6^{(6)}$ and one more counterterm operator $\left[Z O\right]_{g}^{\textrm{GV},\,(2)}$ will appear on the right-hand side and the left-hand side of this equation, respectively.
We do not consider them in this paper, since they start to contribute to the OME renormalization first at the four-loop order. 

\section{Three-loop splitting functions from operator insertions}
\label{sec:threel}

The final goal of this paper is the application of our framework to the computation of the three-loop splitting functions.
Having worked out the Feynman rules for physical operators as well as the renormalization counterterms that originate from GV operators, the main remaining task is the computation of the two-point OMEs with different operator insertions, i.e.\ \eqref{eq:OMEsDe}. 

\begin{figure}[t]
\begin{center}
\includegraphics[scale=1]{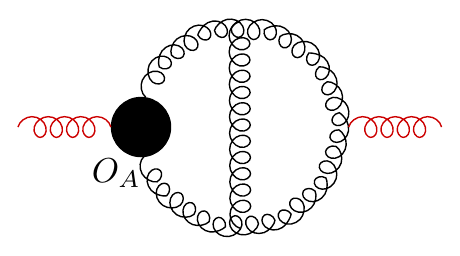} 
\hspace{1cm}
\includegraphics[scale=1]{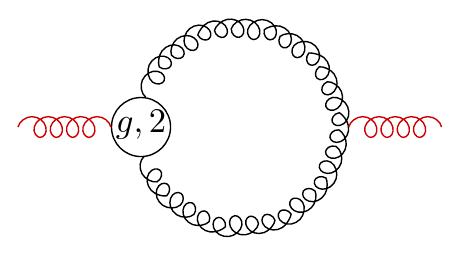} 
\end{center}
\caption{Sample diagrams with GV counterterm insertions entering the calculation of the 3-loop splitting functions.\label{fig:ctexample}}
\end{figure}

With the insertion of the operator $O_q$ or $O_g$, the two-point OMEs need to be evaluated to three loops. Regarding the insertion of $O_{ABC}$, we need to compute the corresponding two-point OMEs to two loops. 
Finally, the two-point OMEs with the 
$\left[Z O\right]_{g}^{\textrm{GV},\,(2)}$ counterterm insertion 
are needed to one loop.
In Table~\ref{Tab:summaryO}, we list all required OMEs for the  extraction of  the three-loop splitting functions, including also the multi-leg OMEs used for the determination of the counterterms and specifying the respective loop order.
Examples for two-point diagrams with counterterm insertions are depicted in Figure~\ref{fig:ctexample}.
\begin{table}[t]
\begin{center}
\begin{tabular}{|l||*{4}{c|}}
\hline
\backslashbox{Loops}{Legs}
&\makebox[6em]{2}&\makebox[6em]{3}&\makebox[6em]{4}
&\makebox[6em]{5}\\\hline 
\hline 
\makebox[6em]{0}  & & $\left[Z O\right]_{g}^{\textrm{GV},\,(2)}$  & $O_{ABC}$ & $O_q,\,O_g$ \\ 
\hline
\makebox[6em]{1} &  $\left[Z O\right]_{g}^{\textrm{GV},\,(2)}$  & $O_{ABC}$  & $O_g$ &\\\hline
\makebox[6em]{2}& $O_{ABC}$ & $O_g$ & &\\\hline
\makebox[6em]{3} & $O_q,\,O_g$  &&&\\\hline
\end{tabular}
\end{center}
 \caption{ \label{Tab:summaryO}
  Summary of all OMEs entering the calculation of 3-loop splitting functions.
OMEs with two legs have two quarks, two ghosts or two gluons in the external state.
The determination of the counterterm Feynman rules requires also OMEs with additional external gluons.
For each operator, the table denotes the maximal number of loops needed for OMEs with a given number of legs.
}
\end{table}  

For physical operators as well as for $i=1$ GV operators $O_{ABC}$, it is straightforward to turn the corresponding Feynman rules from $n$-space into $x$-parameter space using linear propagators according to \eqref{eq:sumXmethod}. However, the all-$n$ Feynman rules for the counterterm $\left[Z O\right]_{g}^{\textrm{GV},\,(2)}$, as shown in \eqref{eq:FeynRulesccgZOgVV2},~\eqref{eq:FeynRules3gZOgVV2}, leads to polylogarithms in $x$-parameter space.
 In practice, it is not clear how to perform IBP reductions of Feynman integrals involving polylogarithmic dependence on the loop kinematics.
 Therefore, the method of Section~\ref{sec:ComputationalMethods} can only be used to compute two-point OMEs with an insertion of $O_q,\, O_g$ or $O_{ABC}$, but not with 
$\left[Z O\right]_{g}^{\textrm{GV},\,(2)}$.

The contributions due to $\left[Z O\right]_{g}^{\textrm{GV},\,(2)}$
consists of three-gluon and ghost-ghost-gluon counterterms. In the following, we explain our procedure to compute the corresponding one-loop, two-gluon OMEs using the example of the ghost-ghost-gluon counterterm \eqref{eq:FeynRulesccgZOgVV2}; the procedure for the three-gluon counterterm \eqref{eq:FeynRules3gZOgVV2} follows identical steps.  
For a fixed  value of $n$, the general structure of the ghost-ghost-gluon vertex \eqref{eq:FeynRulesccgZOgVV2} reads:
\begin{align}
    i g_s f^{a_1 a_2 a_3} \Delta^{\mu_3} \,  \sum_{m=1}^{n-2} a_{ m n} (\Delta \cdot p_1)^m (\Delta \cdot p_2)^{n-1-m}\,,
    \label{eq:ghghgzo2}
\end{align}
where $a_{mn}$ can not be written as a closed rational 
form in ($m,\,n$). If $a_{mn}$ does not depend on $m$ and $n$ (this is the case for the Feynman rules of the operators $O_q,\, O_g$ as well as $O_{ABC}$), or if it has a polynomial dependence on $m$ and $n$, a single auxiliary parameter $x$ as used in Section~\ref{sec:ComputationalMethods} is sufficient to express the above equation in terms of linear propagators. For example, if $a_{mn} = m $, then
\begin{align}
    \sum_{n=3}^\infty x^n \sum_{m=1}^{n-2} m  (\Delta \cdot p_1)^m (\Delta \cdot p_2)^{n-1-m} = \frac{x^2 \Delta \cdot p_1 }{(1- x \Delta \cdot p_1)^2} \frac{x \Delta \cdot p_2}{1-x \Delta \cdot p_2}\,.
\end{align}
Without turning \eqref{eq:ghghgzo2} into a linear propagator, it is still possible to directly evaluate the corresponding OMEs for low Mellin moments $n$. However, when $n$ increases (typically values of $n$ required for an all-$n$ reconstruction range into the hundreds or thousands), 
the numerator degrees of the resulting Feynman integrals become so large that an 
IBP reduction seems inefficient.

We thus employ an alternative method to bypass this complexity. In addition to $x$, we introduce one more auxiliary parameter $t$ and replace $a_{mn}$ in \eqref{eq:ghghgzo2} with $t^m$,
\begin{align}
    h(x,t) =\sum_{n=3}^{\infty} x^n  \sum_{m=1}^{n-2}  t^m  (\Delta \cdot p_1)^m (\Delta \cdot p_2)^{n-1-m} =  \frac{x \,t \Delta \cdot p_1 }{1-x \,t \Delta \cdot p_1} \frac{x^2 \Delta \cdot p_2}{1-x \Delta \cdot p_2} \,.
        \label{eq:auxLinearxt}
\end{align}
In this way, we manage to turn \eqref{eq:ghghgzo2} into  linear propagators depending on two tracing parameters $x$ and $t$. 
By considering the matrix elements with an insertion of the above linear propagators, we have 
\begin{align}
    \braket{g(p)| h(x,t) |g(p)} = \sum_{n=3}^\infty x^{n}  \sum_{m=1}^{n-2} c_{ m n}  t^{m}\,,
\end{align}
where in this way we can perform a standard IBP reduction with the corresponding Feynman integrals without encountering high numerator degrees (at the expense of an extra symbolic parameter $t$ in the IBP relations). 
The $c_{mn}$ can be 
determined from 
differential equations in $x$ up to high values of $n$.
Using the following formula, we can then read out the results of OMEs for multiple fixed values of $n$ easily,
\begin{align}
     \braket{g(p)| \sum_{m=1}^{n-2} a_{m n} (\Delta \cdot p_1)^{m} (\Delta \cdot p_2)^{n-1-m} |g(p)} = \sum_{m=1}^{n-2} a_{m n} c_{m n}  \,.
\end{align} 
Using the all-$n$ counterterm Feynman rules, we 
can thus efficiently compute the OMEs at fixed Mellin 
moments up to high values of $n$. By introducing the auxiliary parameter $t$,  
we avoid the growth in complexity for the IBP reduction at fixed numerical $n$ with increasing $n$, and the IBP-reduced results are expressed as simple one-loop bubble integrals with a linear propagator insertion.

We compute all required two-point OMEs as listed in Table~\ref{Tab:summaryO} in a covariant gauge with full $\xi$ dependence. The all-$n$ results for OMEs with an operator insertion of $O_q,\,O_g$ or $O_{ABC}$ are computed using the method described in  Section~\ref{sec:ComputationalMethods}.
The 3-loop non-singlet splitting functions can be determined in the off-shell approach without the GV counterterms presented in this paper; the required OMEs have been computed in~\cite{Blumlein:2021enk}, and we find full agreement with them.
The OMEs with the counterterm operator $\left[Z O\right]_{g}^{\textrm{GV},\,(2)}$ insertion are computed for fixed Mellin moments up to $n=500$ with the procedure described above. These moments are then used for a reconstruction and cross-check of the all-$n$ results to order $\epsilon^0$, with the following combination contributing to the 3-loop splitting functions, 

\begin{align}
&\frac{1}{(d-2)(N_c^2-1)} \bigg[ -g_{\mu \nu} \braket{g|\left[Z O\right]_{g}^{\textrm{GV},\,(2)}|g}^{ \mu\nu, (1),\,(0)\text{}}_{\text{1PI}}  -2 \braket{c|\left[Z O\right]_{g}^{\textrm{GV},\,(2)}|c}^{ (1),\,(0)\text{}}_{\text{1PI}} \bigg]  = \nonumber 
   \\
   & \quad \frac{(1-\xi) C_A^3}{128 n(n-1)\epsilon ^2} \bigg\{-\frac{4 \left(4 n^2+13 n-13\right) S_1(n)}{(n-1) n}+\frac{4 (n+1) \left(8 n^2-7 n-3\right)}{(n-1)^2
   n^2}\nonumber 
   \\
   & \quad +(1-\xi ) \left(\frac{2 (n+1) (2 n-3)}{(n-1) n}-2 S_1(n)\right)-8 S_{-2}(n)+12 S_2(n)+12 S_{1,1}(n)\bigg\} \nonumber
   \\
   & \quad +\frac{(1-\xi) C_A^3}{128 n (n-1) \epsilon }\bigg\{  (1-\xi)^2 \bigg[ S(1,n)-\frac{(n+1) (2 n-3)}{(n-1) n} \bigg] \nonumber
   \\
   &\quad+  (1-\xi ) \bigg[\frac{2 \left(4 n^2+18 n-21\right) S(1,n)}{(n-1) n}+8 S(-2,n)-8 S(2,n) -8
   S(1,1,n)\nonumber
   \\
   &\quad+\frac{2 \left(6 n^4-28 n^3+16 n^2+17 n-3\right)}{(n-1)^2 n^2}\bigg]  + \frac{8 \left(2 n^2-n+1\right) S(-2,n)}{(n-1) n} \nonumber
   \\
   &\quad -\frac{8 \left(n^2+6
   n-6\right) S(2,n)}{(n-1) n}+\frac{4 \left(12 n^2-31 n+15\right)
   S(1,1,n)}{(n-1) n} \nonumber
   \\
   &\quad -\frac{4 \left(20 n^4-24 n^3-23 n^2+44 n-7\right)
   S(1,n)}{(n-1)^2 n^2}+4 (6 n-7) S(-3,n)\nonumber
   \\
   &\quad-8 (2 n-3) S(3,n)-8 (10 n+1)
   S(-2,1,n)+16 S(1,-2,n)+28 S(1,2,n)\nonumber
   \\
   &\quad-8 (n+1) S(2,1,n)+\frac{8 \left(4 n^5+2
   n^4-13 n^3+12 n^2-10 n+3\right)}{(n-1)^3 n^2}  \bigg\}  + {\cal O} (\epsilon^0)  \,. 
   \label{eq:combinationOgVV}
\end{align}
The quantity in the above equation is zero in Feynman gauge with $\xi=1$. It implies that the correct three-loop splitting functions can be extracted without considering the contribution from the counterterm operator $\left[Z O\right]_{g}^{\textrm{GV},\,(2)}$ if we work in Feynman gauge, which is exactly what we found in~\cite{Gehrmann:2022euk}. 
We point out that we compute all two-parton OMEs to one order higher in $\epsilon$ than what is needed for the extraction of the three-loop splitting functions, since those contributions will enter the renormalization at the four-loop level.
That is, we compute the OMEs at three-loop, two-loop and one-loop order to $\epsilon^0,\, \epsilon^1$ and $\epsilon^2$, respectively. We collect results of all two-parton OMEs in ancillary files and provide instructions for their usage in Appendix~\ref{sec:Ancillary}.

Having the results for all required OMEs at hand, it is straightforward to extract the physical renormalization constants according to the renormalization procedure in \eqref{eq:RenorGVSimQ} and \eqref{eq:RenorGVSimG}. We find that the $\xi$ dependence indeed cancels for the physical renormalization constants, stressing that the inclusion of \eqref{eq:combinationOgVV} is crucial for this cancellation. According to \eqref{eq:ZfactorIntermsofGamma}, we extract the physical singlet anomalous dimensions $\gamma_{ij}(n)$ to three loops. Our results are in full agreement with the results in~\cite{Vogt:2004mw}. Through a similar procedure, we also extract non-singlet anomalous dimension $\gamma_{\text{ns}}(n)$ to three loops. Separating even and odd moments, $\gamma_{\text{ns}}$ can be decomposed as follows:
\begin{align}
    \gamma_{\text{ns}} = \frac{1+(-1)^n}{2} \gamma_{\text{ns}}^+ + \frac{1-(-1)^n}{2} \left(  \gamma_{\text{ns}}^- + \gamma^{\text{s}}_{\text{ns}} \right), 
\end{align}
where the definitions of $\gamma_{\text{ns}}^{\pm}$ and $\gamma_{\text{ns}}^{\text{s}}$ can be found for example in~\cite{Moch:2004pa}. The anomalous dimension $\gamma_{\text{ns}}^{\text{s}}$ starts to contribute at the three-loop level and is proportional to the color structure $d^{abc} d_{abc}$ at this loop order. Our results for these quantities agree with those in~\cite{Moch:2004pa}. Through the equations
\begin{align}
\gamma_{ij}(n)  &= - \int_0^1 \,dz \, z^{n-1} \,P_{ij}(z) ,\,\\ 
\gamma_{\text{ns}}^{\pm,\,\text{s}} &= - \int_0^1 \,dz \, z^{n-1} \,P_{\text{ns}}^{\pm,\text{s}}(z), \,  
\end{align} 
the physical anomalous dimensions are related to the splitting functions in momentum fraction $z$ space. Using \texttt{HarmonicSums}, the splitting functions $P_{ij}(z)$ ($P_{\text{ns}}^{\pm,\,\text{s}}(z)$) are obtained from $\gamma_{ij}$ ($\gamma_{\text{ns}}^{\pm,\,\text{s}}$) by an inverse  Mellin transformation. Also for the splitting functions, we find complete agreement with the results in~\cite{Moch:2004pa,Vogt:2004mw}.

\section{Summary and conclusions}
\label{sec:conc}
The operator-product expansion (OPE) allows to systematically separate short-distance and long-distance contributions to hadron-induced processes in QCD. The anomalous dimensions of the resulting quark and gluon operators determine the scale evolution of parton distributions. 
More specifically, these anomalous dimensions are directly related to 
the Altaralli-Parisi splitting functions by a Mellin transformation. 

The quantization of QCD in a covariant gauge enlarges the particle spectrum by   
ghosts and induces an a priori infinite number of gauge-variant (GV) operators, which contribute to the renormalization  of the physical quark singlet and gluon operators (\ref{eq:RenorGV}).  
Only a finite number 
of these GV operators contribute to the renormalization of the quark and gluon operators for a fixed Mellin moment and at a given loop order. The fully general structure of the GV operators is only poorly understood at present, thus preventing 
the calculation of anomalous dimensions beyond the two-loop order in the OPE up to now.

In this setup, the anomalous dimensions of the quark and gluon operators are determined from the divergences of the simplest operator matrix elements (OMEs) with two external, off-shell partons.
Compared to alternative methods to determine anomalous dimensions or splitting functions, the calculation of off-shell OMEs is generally considered to involve lower computational complexity at the level of amplitudes and Feynman integrals. It is therefore highly desirable to extend the applicability of the OPE to higher loop orders by developing a consistent renormalization framework for the quark and gluon operators. 

In this paper, we developed a new approach to systematically determine the counterterm Feynman rules that result from the GV operators. These counterterm Feynman rules are sufficient to determine the renormalization of operator matrix elements (OMEs) for a fixed number of external partons at a given loop order, even without full knowledge of the GV operators themselves. As an example, the first line of Table~\ref{Tab:summaryO} summarizes which Feynman rules 
for operators or counterterms are required for the 
computation of the three-loop, two-parton OMEs and their renormalization. 
With increasing loop order, OMEs with an increasing number of additional gluons need to be considered, corresponding to vertices with increasing multiplicity entering the bare matrix elements.

Our approach is based on the extraction of the counterterm Feynman rules from the divergent parts of multi-leg OMEs in general off-shell kinematics. 
The renormalization of these OMEs for two partons and an arbitrary number of gluons is illustrated at two-loop order in
(\ref{eq:counterterm2cmg}), (\ref{eq:counterterm2qmg}), (\ref{eq:counterterm2gmg}). The central observation is that the computation of the divergent parts of the OME for a given external state at a given loop order can be used to infer the counterterm Feynman rules for this external state at this loop order.

A special role is played by the GV operators that 
contribute to the 
renormalization of the quark and gluon operators already at one loop. At this order, only three operators contribute, and they are highly constrained in their 
structure. We collectively denote them by 
$O_{ABC}$, and their contribution to the one-loop renormalization of the physical operators is controlled by a single renormalization constant (\ref{eq:zgA1}). 
It is thus possible to extract the 
operator Feynman rules for $O_{ABC}$ for a fixed 
multiplicity of external partons. These turn into 
the respective counterterm Feynman rules only upon 
multiplication with the operator renormalization constant. 
The $O_{ABC}$ operator Feynman rules had  been 
computed previously~\cite{Hamberg:1991qt,Matiounine:1998ky,Blumlein:2022ndg} for external states with up to three partons. We extend their determination now to four partons in (\ref{eq:OA2q2g}), (\ref{eq:OA2c2g}), (\ref{eq:OA4g}). 
The $O_{ABC}$ operators are particularly special, since they already exhaust all allowed Lorentz structures
in GV operators 
at the lowest multiplicities (two-gluon, two-ghost and two-quark-plus-gluon), as we demonstrate in 
Appendix~\ref{sec:FeynmanRulesGV2les}. 
Consequently, the counterterm Feynman rules 
for the lowest multiplicities must be proportional 
to the $O_{ABC}$ operator Feynman rules, multiplied 
with a higher-loop correction to the operator 
renormalization constant (\ref{eq:zgA2Res}), which can be determined 
from a low-multiplicity OME calculation. 

At two loops, an infinite number of GV operators besides $O_{ABC}$ 
contribute to the operator renormalization. 
These other GV operators yield, however, only counterterm Feynman 
rules starting at higher multiplicities than $O_{ABC}$. 
By computing the respective three-parton OMEs in general kinematics, 
we explicitly determined the two-loop counterterm Feynman rules 
relevant to the renormalization of the gluon operator 
$\left[Z O\right]_{g}^{\textrm{GV},\,(2)}$
for the ghost-ghost-gluon
(\ref{eq:FeynRulesccgZOgVV2}) and three-gluon 
(\ref{eq:FeynRules3gZOgVV2}) operator vertices. 
From the structural properties of these counterterm Feynman rules 
(their dependence on $n$ and on the external kinematics), we can infer that they can not 
originate from a single operator, but only from a linear combination 
of an unknown number of operators. 

The computations of multi-leg OMEs are performed for fixed 
numerical values of the external kinematics. 
A finite number of these numerical samples is then sufficient to reconstruct the full kinematical dependence of the resulting counterterm Feynman rules, exploiting their symmetries and their dimensional scaling. The full $n$-dependence of the OMEs is usually retained through the 
introduction of resummed propagators (\ref{eq:sumXmethod}). The two-loop counterterm Feynman rules can not be cast into this resummed form. Instead, OMEs with these counterterm insertions are evaluated repeatedly for multiple integer values of $n$ (based on the all-$n$ Feynman rules as well as a generalized resummed form \eqref{eq:auxLinearxt}, such that the computational effort does not increase with $n$), allowing subsequently their all-$n$ reconstruction. 

We applied our newly computed
GV operator and counterterm Feynman rules to rederive the three-loop anomalous dimensions of the unpolarized quark singlet and gluon operators in a general covariant gauge using the OPE method. 
These three-loop anomalous dimensions (or the corresponding splitting functions) 
were previously computed with several other approaches~\cite{Vogt:2004mw,Mistlberger:2018etf,Duhr:2020seh,Luo:2019szz,Ebert:2020yqt,Ebert:2020unb,Luo:2020epw,Baranowski:2022vcn}, and we find full agreement with the literature. Our result establishes the independence of the anomalous dimensions on the gauge parameter, which was expected and supported by calculations at fixed $n$, but not proven up to now. 

A method for the systematic construction of GV operators and counterterms for the renormalization of the gluon operator based on BRST symmetry has recently been outlined in Reference~\cite{Falcioni:2022fdm}. The resulting constraint equations were, however, formulated only for fixed low values of $n$ up to now, and they show a substantial growth in complexity with increasing $n$. Our method enables the determination of the counterterm Feynman rules for symbolic $n$, thereby allowing the computation of counterterm OMEs for multiple values of $n$ without an increase in complexity towards larger $n$.  Our results reproduce the two-loop counterterm Feynman rules for low values of $n$ that were obtained in~\cite{Falcioni:2022fdm}.

Owing to the relative computational simplicity of two-parton OMEs at high loop orders, the OPE method holds the potential for computing the four-loop corrections to anomalous dimensions or splitting functions, which are of paramount importance to precision collider physics. While first results on the quark non-singlet~\cite{Moch:2017uml} and most recently on low-$n$ moments of the quark singlet and gluon anomalous 
dimensions~\cite{Moch:2021qrk,Falcioni:2022fdm} were obtained, their all-$n$ calculation still remains an outstanding challenge. The developments made in this paper lay out a strategy to determine the renormalization counterterms that are required in this context, thereby paving the way for a future derivation of the four-loop anomalous dimensions in the OPE method.

\begin{acknowledgments}
We thank Vasily Sotnikov for useful discussions and Kay Sch\"onwald for helpful comments on the manuscript. We would like to thank the European Research Council (ERC)
for funding of this work under the European Union's Horizon 2020 research and innovation programme grant agreement 101019620 (ERC Advanced Grant TOPUP) and the National Science Foundation (NSF) for support under grant number 2013859.
\end{acknowledgments}

\appendix

\section{Feynman rules for physical operators}
\label{sec:FeynOqOg}
As outlined in Subsection~\ref{subsec:twist2O}, the Feynman rules for physical operators can be obtained from \eqref{eq:nonsingletOP} and \eqref{eq:singletOP} by a functional variation. Equivalently, one can read out the Feynman rules directly from the definitions of the operators by replacing a derivative of a field with $-i$ times the associated momentum. As an example, we derive the Feynman rule for one of the terms in $O_g$, 
\begin{align}
   &-\frac{i^{n-2}}{2} \mathcal{S} \left[ \left( \Delta \cdot \partial  A_{\mu}^{a_1} \right) ( \Delta \cdot \partial)^{n-1} A^{\mu,\,a_2}  \delta^{a_1 a_2} \right]  \nonumber  
   \\
   & \qquad \to  - \frac{i^{n-2}}{2} (-i)^n  \delta^{a_1 a_2} (\Delta \cdot p_2)^{n-1} \Delta \cdot p_1 g^{\mu_1 \mu_2} + \{ p_1 \leftrightarrow p_2,\, \mu_1 \leftrightarrow \mu_2,\,a_1 \leftrightarrow a_2 \}    \nonumber
   \\
   & \qquad \quad =  -\delta^{a_1 a_2} (\Delta \cdot p_1)^{n} g^{\mu_1 \mu_2}\,,
\end{align}
where in the last line we use momentum conservation $p_2 =-p_1$ to simplify the result. 
In the following, the Feynman rules for the physical operators $O_q$ and $O_g$ and up to 5 legs are listed for completeness, with the convention of all momenta flowing into the vertices.

\subsection{\texorpdfstring
{Feynman rules for the operator $O_q$}
{Feynman rules for the operator O_q}}

\begin{align}
&\includegraphics[scale=1.0]{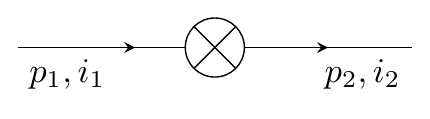} \nonumber
\\
 & \quad \to  \frac{1}{2} \slashed{\Delta}  \delta_{i_1 i_2}\left(\Delta \cdot p_1\right){}^{n-1} \,, 
 \label{eq:FeynOq2q}
 \\
 & \includegraphics[scale=1.0]{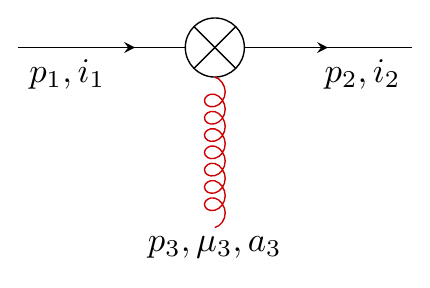} \nonumber
 \\
  \label{eq:FeynOq2q1g}
 & \quad \to \frac{1}{2} g_s \Delta ^{\mu _3}  T_{i_2 i_1}^{a_3}  \slashed{\Delta }\sum_{j_1=0}^{n-2}\left(\left(-\Delta
   \cdot p_2\right){}^{j_1} \left(\Delta \cdot p_1\right){}^{-j_1+n-2}\right)\,, 
   \\
&\includegraphics[scale=1.0]{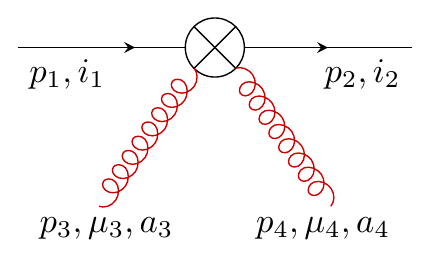} \nonumber
\\
 & \quad \to  \frac{1}{2} g_s^2 \Delta ^{\mu _3} \Delta ^{\mu _4}   \slashed{\Delta} \bigg\{  \sum_{j_1=0}^{n-3} \sum_{j_2=0}^{j_1} 
  \left(-\Delta \cdot p_2\right){}^{j_1-j_2} 
   \left(\Delta \cdot p_1\right){}^{-j_1+n-3} \nonumber
   \\
   & \qquad \times \bigg[ \left(T^{a_3}T^{a_4}\right)_{i_2 i_1} \left(\Delta \cdot \left(p_1+p_4\right)\right){}^{j_2}   +\left(T^{a_4} T^{a_3}\right)_{i_2 i_1} \left(\Delta \cdot \left(p_1+p_3\right)\right){}^{j_2} \bigg] \bigg\}\,, 
   \\
   & \includegraphics[scale=1.0]{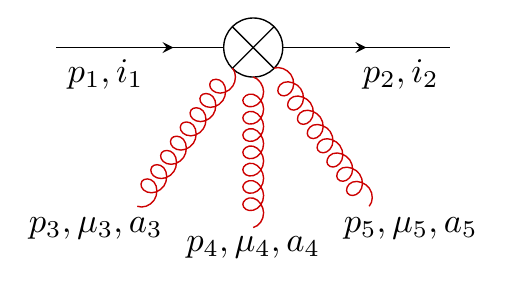} \nonumber
   \\
& \quad \to  \frac{1}{2} g_s^3 \Delta ^{\mu _3} \Delta ^{\mu _4} \Delta ^{\mu _5} \slashed{\Delta}
   \bigg\{ \sum_{j_1=0}^{n-4} \sum_{j_2=0}^{j_1} \sum_{j_3=0}^{j_2}  \left(\Delta \cdot
   p_1\right){}^{-j_1+n-4} \left(-\Delta \cdot p_2\right){}^{j_1-j_2} \nonumber
   \\
   & \qquad \times  \bigg[ \left(T^{a_3} T^{a_4} T^{a_5}\right)_{i_2 i_1} \left(\Delta \cdot
   \left(-p_2-p_3\right)\right){}^{j_2-j_3} \left(\Delta \cdot \left(p_1+p_5\right)\right){}^{j_3} \nonumber
   \\
   & \qquad  +   \left(T^{a_3} T^{a_5} T^{a_4}\right)_{i_2 i_1} \left(\Delta \cdot
   \left(-p_2-p_3\right)\right){}^{j_2-j_3} \left(\Delta \cdot \left(p_1+p_4\right)\right){}^{j_3} \nonumber
   \\
   & \qquad  +   \left(T^{a_4} T^{a_3} T^{a_5}\right)_{i_2 i_1} \left(\Delta \cdot
   \left(-p_2-p_4\right)\right){}^{j_2-j_3} \left(\Delta \cdot \left(p_1+p_5\right)\right){}^{j_3}\nonumber
   \\
   & \qquad +   \left(T^{a_4} T^{a_5} T^{a_3}\right)_{i_2 i_1} \left(\Delta \cdot
   \left(-p_2-p_4\right)\right){}^{j_2-j_3} \left(\Delta \cdot \left(p_1+p_3\right)\right){}^{j_3}\nonumber
   \\
   &  \qquad +   \left(T^{a_5} T^{a_3} T^{a_4}\right)_{i_2 i_1} \left(\Delta \cdot
   \left(-p_2-p_5\right)\right){}^{j_2-j_3} \left(\Delta \cdot \left(p_1+p_4\right)\right){}^{j_3}\nonumber
   \\
   & \qquad  +   \left(T^{a_5} T^{a_4} T^{a_3}\right)_{i_2 i_1} \left(\Delta \cdot
   \left(-p_2-p_5\right)\right){}^{j_2-j_3} \left(\Delta \cdot \left(p_1+p_3\right)\right){}^{j_3}  \bigg] \bigg\}\,.
   \label{eq:FeyRulesOq}
\end{align}

\subsection{\texorpdfstring
{Feynman rules for the operator $O_g$}
{Feynman rules for the operator O_g}}

\begin{align}
& \includegraphics[scale=1.0]{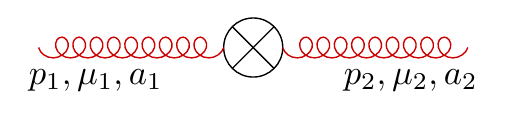} \nonumber
\\
\label{eq:FeynOg2g}
& \quad \to - \delta ^{a_1 a_2} \left(\Delta \cdot p_1\right){}^{n-2} \left[ (\Delta \cdot p_1)^2 g^{\mu_1 \mu_2} - \Delta \cdot p_1 \left( p_1^{\mu_1} \Delta^{\mu_2} + \Delta^{\mu_1} p_1^{\mu_2} \right) + \Delta^{\mu_1} \Delta^{\mu_2} p_1 \cdot p_1 \right] \,, 
   \\ 
& \includegraphics[scale=1.0]{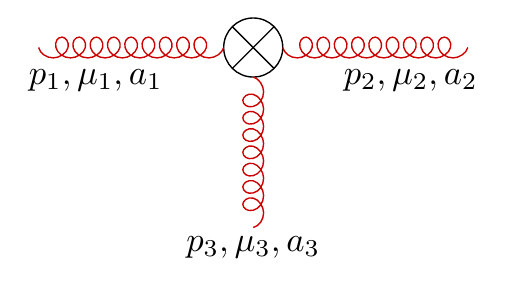} \nonumber
\\
& \quad \to  \frac{-i}{4} g_s f^{a_1 a_2 a_3} \bigg\{\left(\Delta \cdot p_1\right){}^{n-2} \Big(-\Delta ^{\mu _2} g^{\mu_1 \mu _3} \Delta \cdot p_1 \nonumber
\\
& \qquad +\Delta ^{\mu _3} g^{\mu _1 \mu _2} \Delta \cdot p_1-\Delta ^{\mu _1} \Delta ^{\mu _3}p_1^{\mu _2}+\Delta ^{\mu _1} \Delta ^{\mu _2} p_1^{\mu _3}\Big)\nonumber
   \\
    &\qquad+\left(-\Delta \cdot p_3\right){}^{n-2} \left(-\Delta ^{\mu _1} g^{\mu _2 \mu _3} \Delta \cdot p_3+\Delta ^{\mu _2} g^{\mu
   _1 \mu _3} \Delta \cdot p_3-\Delta ^{\mu _2} \Delta ^{\mu _3} p_3^{\mu _1}+\Delta ^{\mu _1} \Delta ^{\mu _3} p_3^{\mu
   _2}\right)    \nonumber 
   \\
   &\qquad -2 \Delta ^{\mu _2} \left(-g^{\mu _1 \mu _3} \Delta \cdot p_1 \Delta \cdot
   p_3+\Delta ^{\mu _3} p_3^{\mu _1} \Delta \cdot p_1+\Delta ^{\mu _1} p_1^{\mu _3} \Delta \cdot p_3-\Delta ^{\mu _1}
   \Delta ^{\mu _3} p_1\cdot p_3\right) \nonumber
   \\
   & \qquad \times \sum_{j_1=0}^{n-3} \left(\left(\Delta \cdot \left(p_1+p_2\right)\right){}^{j_1}
   \left(\Delta \cdot p_1\right){}^{-j_1+n-3}\right)   \bigg\} +\textit{permutations}\,, 
\\
& \includegraphics[scale=1.0]{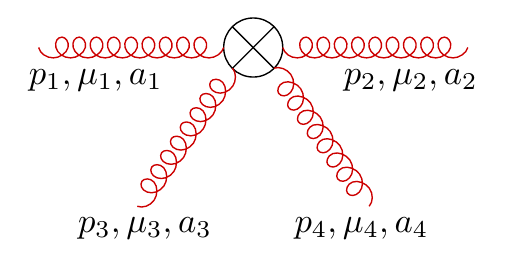} \nonumber
\\
& \quad \to \frac{1}{8} g_s^2 f^{a a_1 a_2} f^{a a_3 a_4}  \bigg\{2 \Delta ^{\mu _3} \Big(\Delta ^{\mu _1} g^{\mu _2 \mu
   _4} \Delta \cdot p_4-\Delta ^{\mu _2} g^{\mu _1 \mu _4} \Delta \cdot p_4+\Delta^{\mu
   _2} \Delta ^{\mu _4} p_4^{\mu _1}\nonumber 
   \\
   &\qquad -\Delta ^{\mu _1} \Delta ^{\mu _4} p_4^{\mu
   _2}\Big) \sum_{j_1=0}^{n-3}\left(\left(\Delta \cdot \left(p_1+p_2+p_3\right)\right){}^{j_1}
   \left(\Delta \cdot \left(p_1+p_2\right)\right){}^{-j_1+n-3}\right) \nonumber 
   \\
   & \qquad +4 \Delta ^{\mu _2}
   \Delta ^{\mu _3} \left(-g^{\mu _1 \mu _4} \Delta \cdot p_1 \Delta \cdot p_4+\Delta^{\mu _4} p_4^{\mu _1} \Delta \cdot p_1+\Delta ^{\mu _1} p_1^{\mu _4} \Delta \cdot
   p_4-\Delta ^{\mu _1} \Delta ^{\mu _4} p_1\cdot p_4\right) \nonumber 
   \\
   & \qquad \times \sum_{j_1=0}^{n-4} \sum_{j_2=0}^{j_1}\left(\left(\Delta
   \cdot \left(p_1+p_2\right)\right){}^{j_1-j_2} \left(\Delta \cdot
   \left(p_1+p_2+p_3\right)\right){}^{j_2} \left(\Delta \cdot
   p_1\right){}^{-j_1+n-4}\right)  \nonumber 
   \\
   & \qquad +\Big(-\Delta ^{\mu _2} \Delta ^{\mu _4} g^{\mu _1 \mu
   _3}+\Delta ^{\mu _1} \Delta ^{\mu _4} g^{\mu _2 \mu _3}\nonumber
   \\
   & \qquad +\Delta ^{\mu _2} \Delta^{\mu
   _3} g^{\mu _1 \mu _4}-\Delta ^{\mu _1} \Delta ^{\mu _3} g^{\mu _2 \mu _4}\Big)
   \left(\Delta \cdot \left(p_1+p_2\right)\right){}^{n-2}  \nonumber 
   \\
   & \qquad  +2 \Delta^{\mu
   _2} \left(\Delta ^{\mu _3} g^{\mu _1 \mu _4} \Delta \cdot p_1-\Delta ^{\mu _4} g^{\mu
   _1 \mu _3} \Delta \cdot p_1+\Delta ^{\mu _1} \Delta ^{\mu _4} p_1^{\mu _3}-\Delta^{\mu _1} \Delta ^{\mu _3} p_1^{\mu _4}\right) \nonumber 
   \\
   & \qquad\times  \sum_{j_1=0}^{n-3} \left(\left(\Delta \cdot
   \left(p_1+p_2\right)\right){}^{j_1} \left(\Delta \cdot
   p_1\right){}^{-j_1+n-3}\right)  \bigg\} +\textit{permutations}\,, 
\\ 
& \includegraphics[scale=1.0]{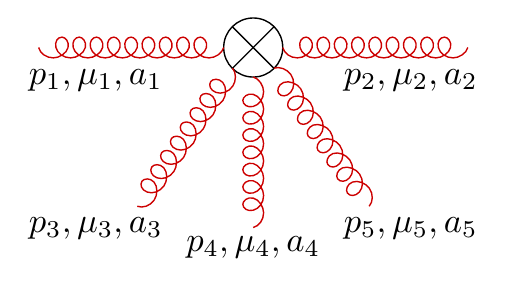} \nonumber 
\\
& \quad \to  \frac{-i}{8} g_s^3 f^{a a_1 a_2} f^{a b a_3} f^{b a_4 a_5}
   \bigg\{-2 \Delta ^{\mu _2}
   \Delta ^{\mu _3} \Big(\Delta ^{\mu _4} g^{\mu _1 \mu _5} \Delta \cdot p_1-\Delta^{\mu _5} g^{\mu _1 \mu _4} \Delta \cdot p_1+\Delta ^{\mu _1} \Delta ^{\mu _5}
   p_1^{\mu _4} \nonumber 
   \\
   & \qquad -\Delta ^{\mu _1} \Delta ^{\mu _4} p_1^{\mu _5}\Big) \sum_{j_1= 0}^{n-4} \sum_{j_2=0}^{j_1}\left(\left(\Delta \cdot \left(p_1+p_2\right)\right){}^{j_1-j_2}
   \left(\Delta \cdot \left(p_1+p_2+p_3\right)\right){}^{j_2} \left(\Delta \cdot
   p_1\right){}^{-j_1+n-4}\right) \nonumber 
   \\
   & \qquad -2 \Delta ^{\mu _3} \Delta ^{\mu _4} \left(\Delta^{\mu
   _1} g^{\mu _2 \mu _5} \Delta \cdot p_5-\Delta ^{\mu _2} g^{\mu _1 \mu _5} \Delta \cdot
   p_5+\Delta ^{\mu _2} \Delta ^{\mu _5} p_5^{\mu _1}-\Delta ^{\mu _1} \Delta ^{\mu _5}
   p_5^{\mu _2}\right) \nonumber 
   \\
   &  \qquad \times  \sum_{j_1= 0}^{n-4} \sum_{j_2=0}^{j_1} \left(\left(\Delta \cdot
   \left(p_1+p_2+p_3\right)\right){}^{j_1-j_2} \left(-\Delta \cdot
   p_5\right){}^{j_2} \left(\Delta \cdot
   \left(p_1+p_2\right)\right){}^{-j_1+n-4}\right) \nonumber 
   \\
   &  \qquad  -4 \Delta ^{\mu _2} \Delta ^{\mu _3}
   \Delta ^{\mu _4} \left(-g^{\mu _1 \mu _5} \Delta \cdot p_1 \Delta \cdot p_5+\Delta^{\mu _5} p_5^{\mu _1} \Delta \cdot p_1+\Delta ^{\mu _1} p_1^{\mu _5} \Delta \cdot
   p_5-\Delta ^{\mu _1} \Delta ^{\mu _5} p_1\cdot p_5\right)  \nonumber 
   \\
   &  \qquad \times \sum_{j_1=0}^{n-5} \sum_{j_2=0}^{j_1} \sum_{j_3=0}^{j_2} \Big[\left(\Delta
   \cdot \left(p_1+p_2\right)\right){}^{j_1-j_2} \nonumber 
   \\
   &\qquad \times \left(\Delta \cdot
   \left(p_1+p_2+p_3\right)\right){}^{j_2-j_3} \left(-\Delta \cdot
   p_5\right){}^{j_3} \left(\Delta \cdot
   p_1\right){}^{-j_1+n-5}\Big] \nonumber 
\\
&  \qquad +\Delta ^{\mu _3} \left(\Delta ^{\mu _2} \Delta ^{\mu _5} g^{\mu _1 \mu
   _4}-\Delta ^{\mu _1} \Delta ^{\mu _5} g^{\mu _2 \mu _4}-\Delta ^{\mu _2} \Delta^{\mu
   _4} g^{\mu _1 \mu _5}+\Delta ^{\mu _1} \Delta ^{\mu _4} g^{\mu _2 \mu _5}\right)  \nonumber 
   \\
   &  \qquad \times \sum_{j_1=0}^{n-3}\left(\left(\Delta \cdot \left(p_1+p_2+p_3\right)\right){}^{j_1}
   \left(\Delta \cdot \left(p_1+p_2\right)\right){}^{-j_1+n-3}\right)
   \bigg\}  + \textit{permutations}\,,
      \label{eq:FeyRulesOg}
\end{align}
where plus {\it{permutations}} indicates the summation over all the 
external gluon indices (simultaneous permutation of $\mu_i,a_i,p_i$).

\section{Feynman rules for all GV operators at lowest multiplicity}
\label{sec:FeynmanRulesGV2les}
As discussed at the end of Subsection~\ref{subsec:FormulaZOgVV}, {the two-point vertex Feynman rules for all GV operators except $O_A$ and $O_C$ are zero}. We prove this statement in the following. For the vertex Feynman rules with two quark or two ghost legs, the only possible forms are:
\begin{align}
    &\braket{q|O|q}^{(0),(0)} = h_1 \slashed{\Delta} (\Delta \cdot p)^{n-1} \,, 
    \label{eq:2legsqqO}
    \\
    & \braket{c|O|c}^{(0),(0)} = h_2    (\Delta \cdot p)^{n} \,,
    \label{eq:2legsccO}
\end{align}
where $O$ is an arbitrary twist-two operator, and $h_1$ and $h_2$ are constants. We recognize that the above two equations are exactly the Feynman rules for $O_q$ and $O_C$ respectively, as in \eqref{eq:FeynOq2q} and \eqref{eq:FeynOC2c}. Therefore, 
the vertex Feynman rules with two quarks for all GV operators are zero, and the vertex Feynman rules with two ghosts for all GV operators except $O_C$ are zero. 

For the case of the vertex Feynman rules with two gluons, there are four possible tensor structures $T^{\mu \nu}_i$ with $i=1\ldots 4$ as shown in \eqref{eq:tensorStructure}. From Subsection~\ref{subsec:Lorentzstructure}, $T_2^{\mu \nu}$ can not be the Feynman rule of a twist-two operator. Moreover, the Feynman rules with two gluons are expected to satisfy the condition of being transverse, which is not the case for $T^{\mu \nu}_4$, i.e., 
\begin{align}
   p_\mu p_\nu T^{\mu \nu}_4  \neq 0 \,. 
\end{align}
The remaining tensor structures $T_1^{\mu \nu}$ and $T_3^{\mu \nu}$ are just the Feynman rules of $O_g$ and $O_A$ separately, as shown in \eqref{eq:FeynOg2g} and \eqref{eq:FeynOA2g}. Therefore, all GV operators involving only two gluon fields except $O_A$ are zero. 

Based on the symmetry from the exchange of two quarks for a singlet twist-two operator, and from the Feynman rules in \eqref{eq:FeynOq2q1g}, \eqref{eq:FeynOB2q1g} and \eqref{eq:FeynOgVV2q1g}, we conjecture but not prove the statement: {the $q\bar{q}g$ vertex Feynman rules for all GV operators except $O_B$ are zero}.

\section{\texorpdfstring
{Renormalization constants $Z_{qA}$ and $Z_{gA}$ to three loops}
{Renormalization constants Z_qA and Z_gA to three loops}}

\label{sec:zqAzgA}
We give the all-$n$ results for the renormalization constants $Z_{qA}$ and $Z_{gA}$ to three loops, which can be extracted from the OMEs with two-ghost external states. $Z_{qA}$ starts at $a_s^2$ and reads as follows,
\begin{align}
    Z_{qA} = &  \frac{a_s^2 C_A N_f}{ (n-1) n^2 (n+1) (n+2)} \bigg\{ \frac{\left(n^2+n+2\right) }{\epsilon
   ^2}-\frac{\left(n^2+5 n+2\right) \left(n^3+n^2+n+2\right)}{n (n+1) (n+2) \epsilon } \bigg\} \nonumber
   \\
   & + \frac{a_s^3 N_f}{ (n-1) n^2 (n+1) (n+2)} \bigg\{ \frac{1}{\epsilon^3} \bigg[ \frac{8}{9} \left(n^2+n+2\right) C_A N_f+C_A C_F
   \bigg(\frac{4}{3} \left(n^2+n+2\right)
   S_1(n)  \nonumber
   \\
   & -\frac{\left(n^2+n+2\right) \left(3 n^2+3
   n+2\right)}{3 n (n+1)}\bigg)+C_A^2 \bigg(\frac{11}{6}
   \left(n^2+n+2\right) S_1(n) \nonumber
   \\
   & -\frac{\left(n^2+n+2\right)
   \left(100 n^4+203 n^3-52 n^2-173 n+30\right)}{18 (n-1) n
   (n+1) (n+2)}\bigg) \bigg]  \nonumber
   \\
   & 
   + \frac{1}{\epsilon^2} \bigg[ C_A C_F \bigg(\frac{8}{3} \left(n^2+n+2\right)S_{1,1}(n)-\frac{8}{3} \left(n^2+n+2\right) S_2(n) \nonumber
   \\
   & -\frac{4
   \left(n^5+8 n^4+20 n^3+35 n^2+36 n+12\right) S_1(n)}{3 n
   (n+1) (n+2)} \nonumber
   \\
   & +\frac{13 n^7+71 n^6+176 n^5+309 n^4+333
   n^3+182 n^2+76 n+24}{3 n^2 (n+1)^2 (n+2)}\bigg) \nonumber
   \\
   & +C_A^2
   \bigg(-\frac{11}{6} \left(n^2+n+2\right)
   S_{1,1}(n)-\frac{10}{3} \left(n^2+n+2\right) S_{-2}(n)-2
   \left(n^2+n+2\right) S_2(n) \nonumber
   \\
   & +(1-\xi ) \Big(\frac{1}{12}
   \left(n^2+n+2\right) S_1(n)-\frac{\left(n^2+n+2\right)
   \left(2 n^3+n^2-1\right)}{12 (n-1) n
   (n+1)}\Big) \nonumber
   \\
   & -\frac{\left(53 n^6+219 n^5+33 n^4+77 n^3+310
   n^2-284 n-120\right) S_1(n)}{18 (n-1) n (n+1)
   (n+2)} \nonumber
   \\
   & +\frac{1}{54
   (n-1)^2 n^2 (n+1)^2 (n+2)^2}\Big( 577 n^{10}+3302 n^9+5318 n^8+1049 n^7\nonumber
   \\
   & -4127
   n^6-3430 n^5-3270 n^4+291 n^3+4742 n^2-780 n-1080\Big)\bigg)\nonumber
   \\
   &+C_A N_f
   \left(\frac{4}{9} \left(n^2+n+2\right) S_1(n)-\frac{4
   \left(11 n^5+50 n^4+80 n^3+91 n^2+68 n+12\right)}{27 n
   (n+1) (n+2)}\right) \bigg]  \nonumber
   \\
   & 
   + \frac{1}{\epsilon} \bigg[ C_A^2 \bigg( \frac{1}{162 (n-1)^3 n^3 (n+1)^3 (n+2)^3} \Big( 353 n^{14}-943 n^{13}-5075 n^{12}+1993
   n^{11}\nonumber
   \\
   & -6788 n^{10}-54586 n^9-38153 n^8+57445 n^7+61625
   n^6-49753 n^5-42370 n^4\nonumber
   \\
   & +54772 n^3+8808 n^2-25344
   n-8640\Big)+\frac{2 \left(9
   n^3+19 n^2+24 n+12\right) S_{-3}(n)}{3 (n+1)}\nonumber
   \\
   & +\frac{2
   \left(6 n^6+47 n^5+105 n^4+129 n^3+137 n^2+96 n+20\right)
   S_{-2}(n)}{3 n (n+1)^2 (n+2)}\nonumber
   \\
   & -\frac{1}{54 (n-1) n^2 (n+1)^2
   (n+2)^2} \Big(28 n^9-144
   n^8-172 n^7+1210 n^6\nonumber
   \\
   & -379 n^5-6950 n^4-6539 n^3-290
   n^2+1068 n+72\Big) S_1(n)\nonumber
   \\
   & +\frac{\left(5 n^5+37 n^4+18 n^3-14 n^2+40
   n+8\right) S_2(n)}{3 n (n+1) (n+2)}+\frac{\left(11 n^3+20
   n^2+35 n+30\right) S_3(n)}{3 (n+1)}\nonumber
   \\
   & -\frac{8}{3} n (n+1)
   S_{-2,1}(n)+\frac{4 \left(n^3+n^2+2 n+4\right)
   S_{1,-2}(n)}{3 (n+1)}\nonumber
   \\
   & +\frac{\left(53 n^6+219 n^5+321
   n^4+221 n^3-122 n^2-284 n-120\right) S_{1,1}(n)}{18 (n-1)
   n (n+1) (n+2)}\nonumber
   \\
   & +(1-\xi ) \Big(-\frac{3 n^9-7 n^7-34 n^6-50
   n^5+3 n^4+26 n^3+31 n^2-8 n-12}{12 (n-1)^2 n^2 (n+1)^2
   (n+2)}\nonumber
   \\
   & +\frac{\left(n^5+5 n^3+6 n^2-12 n-8\right)
   S_1(n)}{12 n (n+1) (n+2)}+\frac{1}{12}
   \left(-n^2-n-2\right) S_{1,1}(n)\Big)\nonumber
   \\
   & +\frac{\left(11
   n^3+21 n^2+32 n+24\right) S_{1,2}(n)}{3 (n+1)}+\frac{n
   \left(n^2+3 n+4\right) S_{2,1}(n)}{3 (n+1)}\nonumber
   \\
   & -\frac{13}{6}
   \left(n^2+n+2\right) S_{1,1,1}(n)-6 \left(n^2+n+2\right)
   \zeta_3\bigg) \nonumber
   \\
   & +N_f C_A \bigg(-\frac{4 \left(13 n^8+13
   n^7+45 n^6+223 n^5-51 n^4-833 n^3-718 n^2-60
   n+72\right)}{81 n^2 (n+1)^2 (n+2)^2}\nonumber
   \\
   &+\frac{8 (n-2)
   \left(n^4+4 n^2+13 n+6\right) S_1(n)}{27 n (n+1)
   (n+2)}-\frac{4}{9} \left(n^2+n+2\right) S_{1,1}(n)\bigg)
  \nonumber
   \\
   &+C_F C_A \bigg(-\frac{2}{{3 n^3 (n+1)^3 (n+2)^2}}\Big( 11 n^{10}+101 n^9+384
   n^8+875 n^7+1523 n^6  \nonumber
   \\
   &+2068 n^5+1886 n^4+1072 n^3+448
   n^2+160 n+32 \Big)  \nonumber
   \\
   &-\frac{2
   \left(n^8-47 n^6-202 n^5-466 n^4-726 n^3-708 n^2-360
   n-80\right) S_1(n)}{3 n^2 (n+1)^2 (n+2)^2}  \nonumber
   \\
   &+\frac{8
   \left(n^5+6 n^4+11 n^3+20 n^2+24 n+8\right) S_2(n)}{3 n
   (n+1) (n+2)}-\frac{4}{3} \left(n^2+n+2\right)
   S_3(n)  \nonumber
   \\
   &-\frac{4 \left(2 n^5+13 n^4+22 n^3+31 n^2+36
   n+12\right) S_{1,1}(n)}{3 n (n+1) (n+2)}-\frac{8}{3}
   \left(n^2+n+2\right) S_{1,2}(n)  \nonumber
   \\
   &+\frac{4}{3}
   \left(n^2+n+2\right) S_{1,1,1}(n)+8 \left(n^2+n+2\right)
   \zeta_3\bigg) \bigg]
   \bigg\} + {\cal O} (a_s^4) \,.
\end{align}
 $Z_{gA}$ starts at $a_s^1$ and we gave the first order $Z_{gA}^{(1)}$ and the second order $Z_{gA}^{(2)}$ in \eqref{eq:zgA1} and \eqref{eq:zgA2Res}, respectively. The third order contribution to $Z_{gA}$ is obtained as
 \begin{align}
     Z_{gA}^{(3)} = &\frac{1}{\epsilon^3} \frac{1}{n(n-1)} \bigg[ C_A^3 \bigg(-\frac{391}{48} S_{1,1}(n)+\frac{\left(2636 n^4+5407 n^3-698 n^2-4051
   n+1242\right) S_1(n)}{144 (n-1) n (n+1) (n+2)} \nonumber
   \\
   &-\frac{1}{144 (n-1)^2 n^2 (n+1)^2
   (n+2)^2}\Big(2736 n^8+11192 n^7+9977
   n^6\nonumber
   \\
   &-11572 n^5-14586 n^4+8032 n^3+6901 n^2-3020 n+1140\Big) \nonumber
   \\
   &+(1-\xi ) \left(\frac{S_1(n)}{96}-\frac{(n+1) (2 n-3)}{96 (n-1)
   n}\right)+\frac{1}{24} S_{-2}(n)+\frac{191 S_2(n)}{48}\bigg) \nonumber
   \\
   &+ N_f C_A^2
   \bigg(\frac{2 \left(26 n^4+53 n^3-6 n^2-39 n+14\right)}{9 (n-1) n (n+1)
   (n+2)}-\frac{26 S_1(n)}{9}\bigg)\nonumber
   \\
   &-\frac{2 \left(n^2+n+2\right)^2  C_A C_F N_f }{3 (n-1)
   n^2 (n+1)^2 (n+2)}  -\frac{4}{9} C_A N_f^2 \bigg] \nonumber
   \\
   &+\frac{1}{\epsilon^2} \frac{1}{n(n-1)} \bigg[C_A^3 \bigg( (1-\xi
   )^2\Big(\frac{9 n^3+6 n^2+n-4}{192 (n-1) n (n+1)}-\frac{S_1(n)}{32}\Big)\nonumber
   \\
   &+(1-\xi ) \Big(-\frac{356 n^7+863 n^6+45 n^5-912 n^4-130 n^3+619 n^2+401 n-90}{288
   (n-1)^2 n^2 (n+1)^2 (n+2)}\nonumber
   \\
   &-\frac{1}{12} S_{-2}(n)+\frac{\left(343 n^4+764 n^3+59
   n^2-374 n-72\right) S_1(n)}{288 (n-1) n (n+1) (n+2)}+\frac{5
   S_2(n)}{16}-\frac{55}{96} S_{1,1}(n)\Big) \nonumber
   \\
   & +\frac{1}{432 (n-1)^3 n^3 (n+1)^3
   (n+2)^3}\Big(21598 n^{12}+130274
   n^{11}+217650 n^{10}-97011 n^9 \nonumber
   \\
   &-545042 n^8-168389 n^7+540328 n^6+320207 n^5-220384
   n^4-157997 n^3+51298 n^2 \nonumber
   \\
   &-9804 n-20520\Big)+\frac{1}{4} (-n-20) S_{-3}(n)-\frac{\left(63 n^4+133 n^3+102
   n^2+128\right) S_{-2}(n)}{12 (n-1) n (n+1) (n+2)} \nonumber
   \\
   &-\frac{1}{216 (n-1)^2 n^2 (n+1)^2 (n+2)^2}\Big( 9787 n^8+41377
   n^7+30596 n^6-55169 n^5\nonumber
   \\
   &-61462 n^4+36244 n^3+57835 n^2-852 n-6516 \Big)
   S_1(n) \nonumber
   \\
   &-\frac{\left(385 n^4+872 n^3+158 n^2-527
   n+138\right) S_2(n)}{18 (n-1) n (n+1) (n+2)}+\frac{1}{48} (8 n-477)
   S_3(n) \nonumber
   \\
   &+\frac{1}{6} (5 n+14) S_{-2,1}(n)+\frac{47}{6} S_{1,-2}(n) \nonumber
   \\
   &+\frac{\left(2818
   n^4+7001 n^3+932 n^2-4901 n-1170\right) S_{1,1}(n)}{144 (n-1) n (n+1)
   (n+2)} \nonumber
   \\
   &+\frac{187}{12} S_{1,2}(n)+\frac{1}{24} (2 n+377) S_{2,1}(n)-\frac{137}{16}
   S_{1,1,1}(n)\bigg) +\left(\frac{32}{27}-\frac{4 S_1(n)}{9}\right) N_f^2
   C_A \nonumber
   \\
   &+  \bigg(-\frac{1}{108 (n-1)^2 n^2 (n+1)^2 (n+2)^2}\Big(1791 n^8+7217 n^7+4685 n^6-11071 n^5\nonumber
   \\
   &-11004 n^4+6286
   n^3+6040 n^2-1880 n-768\Big)\nonumber
   \\
   &+\frac{2}{3}
   S_{-2}(n)+(1-\xi ) \Big(\frac{2 n^3+n^2-1}{18 (n-1) n
   (n+1)}-\frac{S_1(n)}{18}\Big)\nonumber
   \\
   &+\frac{\left(1205 n^4+2575 n^3-668 n^2-2344
   n-120\right) S_1(n)}{108 (n-1) n (n+1) (n+2)}+\frac{61 S_2(n)}{18}-\frac{71}{18}
   S_{1,1}(n)\bigg)C_A^2 N_f  \nonumber
   \\
   &-\frac{2 \left(n^2+n+2\right) \left(2 n^7+10 n^6+13 n^5-12
   n^4-30 n^3-15 n^2-20 n-12\right) C_A C_F N_f}{3 (n-1) n^3 (n+1)^3 (n+2)^2}    \bigg]  \nonumber
   \\
   &+\frac{1}{\epsilon} \frac{1}{n(n-1)} \bigg[    C_A^3 \bigg(\Big\{-\frac{30 n^6-n^5-27 n^4-31 n^3+9 n^2+20 n-8}{192 (n-1)^2 n^2
   (n+1)^2} \nonumber
   \\
   &+\frac{\left(15 n^3+6 n^2-7 n-6\right) S_1(n)}{192 (n-1) n
   (n+1)}-\frac{S_2(n)}{192}-\frac{5}{192} S_{1,1}(n)\Big\} (1-\xi
   )^2  \nonumber
   \\
   & +(1-\xi ) \Big\{\frac{1}{864 (n-2) (n-1)^3
   n^3 (n+1)^3 (n+2)^2} \big( 1589 n^{12}+4922 n^{11}-6625 n^{10}\nonumber
   \\
   &-26755 n^9-7874 n^8+40335
   n^7+31081 n^6-35477 n^5-16025 n^4+15763 n^3\nonumber
   \\
   &+14 n^2-8508 n-1512\big)-\frac{\left(39 n^2+39 n+16\right) S_{-3}(n)}{96 n
   (n+1)}\nonumber
   \\
   &-\frac{\left(35 n^6-11 n^5-85 n^4-61 n^3-14 n^2+8 n+32\right) S_{-2}(n)}{48
   (n-2) (n-1) n^2 (n+1)^2}\nonumber
   \\
   &-\frac{1}{864 (n-1)^2 n^3 (n+1)^2
   (n+2)^2}\Big(1180 n^9+6049 n^8+8192 n^7+1843 n^6-730
   n^5\nonumber
   \\
   &-2948 n^4-5456 n^3+4866 n^2+4392 n-2808 \Big)S_1(n)\nonumber
   \\
   &-\frac{\left(281 n^4+345 n^3-164 n^2+306 n-234\right) S_2(n)}{288 (n-1)
   n^2 (n+1)}-\frac{\left(17 n^2+17 n-8\right) S_3(n)}{48 n (n+1)}\nonumber
   \\
   &+\frac{1}{12}
   S_{-2,1}(n)+\frac{\left(21 n^2+21 n+16\right) S_{1,-2}(n)}{48 n
   (n+1)}\nonumber
   \\
   &+\frac{\left(60 n^4+201 n^3+135 n^2-82 n-152\right) S_{1,1}(n)}{96 (n-1) n
   (n+1) (n+2)}\nonumber
   \\
   &+\frac{\left(35 n^3+71 n^2+23 n+39\right) S_{1,2}(n)}{48 n
   (n+1)^2}+\frac{\left(67 n^3+132 n^2+91 n-78\right) S_{2,1}(n)}{96 n
   (n+1)^2}\nonumber
   \\
   &-\frac{35}{96} S_{1,1,1}(n)+\frac{\left(n^3+4 n^2-31 n+70\right) \zeta_3}{16 n (n+1)^2}\Big\} \nonumber
   \\
   &-\frac{1}{1296 (n-2) (n-1)^4 n^4 (n+1)^4
   (n+2)^4}\Big(57083 n^{17}+341102 n^{16}+177962
   n^{15}\nonumber
   \\
   &-2219135 n^{14}-3448638 n^{13}+3989078 n^{12}+11054452 n^{11}-1139746
   n^{10}\nonumber
   \\
   &-17966031 n^9-7357576 n^8+14022462 n^7+10545037 n^6\nonumber
   \\
   &-4729434 n^5-5907736
   n^4+1571152 n^3+1706640 n^2-477216 n-359424\Big)\nonumber
   \\
   &+\frac{\left(12 n^6+869 n^5+3303 n^4+6881
   n^3+5709 n^2+1154 n-504\right) S_{-3}(n)}{144 (n-1) n (n+1)^2
   (n+2)} \nonumber
   \\
   & +\frac{1}{36 (n-2) (n-1)^2 n^2 (n+1)^3
   (n+2)^2}\Big(629 n^{10}+1839 n^9-2772 n^8-9854 n^7\nonumber
   \\
   &-1227 n^6+12939 n^5+8026
   n^4-7548 n^3-7536 n^2+2336 n+1440 \Big)S_{-2}(n) \nonumber
   \\
   &+\frac{1}{648 (n-1)^3 n^3 (n+1)^3 (n+2)^3}\Big(17788 n^{12}+127947 n^{11}+281430 n^{10}+30563 n^9\nonumber
   \\
   &-617061
   n^8-517548 n^7+589151 n^6+1063392 n^5+160500 n^4-539150 n^3\nonumber
   \\
   &-192804 n^2+82008
   n+44496 \Big) S_1(n) +\frac{1}{16} (6 n+197) S_{-4}(n)\nonumber
   \\ 
   &+\frac{1}{432 (n-1)^2 n^2 (n+1)^2 (n+2)^2}\Big(19838
   n^8+80657 n^7+42469 n^6-130108 n^5 \nonumber
   \\ 
   &-109184 n^4+98303 n^3+133133 n^2-1440
   n-11196 \Big) S_2(n)\nonumber
   \\ 
   &+\frac{\left(197 n^5+1036
   n^4+2214 n^3+1526 n^2+177 n+154\right) S_3(n)}{48 (n-1) n (n+1)^2
   (n+2)}+\frac{1}{24} (257-6 n) S_4(n)\nonumber
   \\ 
   &+\frac{1}{6} (-8 n-41)
   S_{-3,1}(n)+\frac{\left(3 n^2-10 n-17\right) S_{-2,-2}(n)}{6 (n+1)}\nonumber
   \\ 
   &+\frac{\left(3
   n^5-52 n^4-185 n^3-296 n^2-34 n+276\right) S_{-2,1}(n)}{18 (n-1) n (n+1)
   (n+2)}+\frac{1}{6} (n-22) S_{-2,2}(n)\nonumber
   \\ 
   &-\frac{\left(48 n^2+937 n+913\right)
   S_{1,-3}(n)}{48 (n+1)}\nonumber
   \\ 
   &-\frac{\left(767 n^5+2949 n^4+2771 n^3-933 n^2-3274
   n-2136\right) S_{1,-2}(n)}{72 (n-1) n (n+1)^2 (n+2)}\nonumber
   \\ 
   &-\frac{1}{432 (n-1)^2 n^2 (n+1)^2 (n+2)^2}\Big(8524 n^8+38614
   n^7+37691 n^6-28388 n^5\nonumber
   \\ 
   &-48148 n^4+12310 n^3+47149 n^2+12780 n+468\Big) S_{1,1}(n)\nonumber
   \\ 
   & -\frac{\left(1085 n^5+4209 n^4+3803
   n^3-1713 n^2-3340 n-1236\right) S_{1,2}(n)}{72 (n-1) n (n+1)^2
   (n+2)}\nonumber
   \\ 
   &+\frac{\left(12 n^2-325 n-321\right) S_{1,3}(n)}{24 (n+1)}+\frac{\left(3
   n^2-123 n-130\right) S_{2,-2}(n)}{12 (n+1)}\nonumber
   \\ 
   &+\frac{\left(12 n^6-2213 n^5-8175
   n^4-7625 n^3+1995 n^2+6238 n+1992\right) S_{2,1}(n)}{144 (n-1) n (n+1)^2
   (n+2)}\nonumber
   \\ 
   &+\frac{\left(n^2-119 n-122\right) S_{2,2}(n)}{6 (n+1)}-\frac{\left(5 n^2+347
   n+346\right) S_{3,1}(n)}{24 (n+1)}+\frac{1}{3} (5-n) S_{-2,1,1}(n)\nonumber
   \\ 
   &+\frac{2}{3} (3
   n+10) S_{1,-2,1}(n)+\frac{(135 n+143) S_{1,1,-2}(n)}{24 (n+1)}\nonumber
   \\ 
   &+\frac{\left(989
   n^4+2725 n^3+373 n^2-2161 n-1278\right) S_{1,1,1}(n)}{144 (n-1) n (n+1)
   (n+2)}+\frac{(173 n+165) S_{1,1,2}(n)}{24 (n+1)}\nonumber
   \\ 
   &+\frac{(389 n+381)
   S_{1,2,1}(n)}{48 (n+1)}-\frac{\left(4 n^2-389 n-417\right) S_{2,1,1}(n)}{48
   (n+1)}-\frac{7}{3} S_{1,1,1,1}(n)\nonumber
   \\ 
   & +\Big(\frac{19 n^5+12 n^4-27 n^3+18 n^2-30
   n-4}{4 (n-1) n (n+1)^2 (n+2)}-\frac{(n+3) S_1(n)}{n+1}\Big) \zeta_3\bigg)\nonumber
   \\ 
   &+\left(\frac{32 S_1(n)}{27}-\frac{4}{9} S_{1,1}(n)-\frac{16}{27}\right) N_f^2 C_A\nonumber
   \\ 
   &
   + C_A^2 N_f \bigg(\frac{1}{648 (n-1)^3 n^3 (n+1)^3 (n+2)^3}\Big(8065 n^{12}+48004 n^{11}+65499 n^{10}\nonumber
   \\ 
   &-81506 n^9-224797
   n^8-10924 n^7+311121 n^6+187594 n^5-180704 n^4-169280 n^3\nonumber
   \\ 
   &+19376 n^2+38496
   n+4608\Big)-\frac{8}{9} S_{-3}(n)\nonumber
   \\
   &-\frac{\left(29
   n^4+58 n^3-21 n^2-50 n-64\right) S_{-2}(n)}{9 (n-1) n (n+1)
   (n+2)}\nonumber
   \\ 
   &-\frac{1}{324 (n-1)^2 n^2 (n+1)^2 (n+2)^2}\Big(3701 n^8+16697 n^7+13876 n^6-23791 n^5 \nonumber
   \\ 
   &-34295 n^4+3380 n^3+21920
   n^2+6432 n-144 \Big) S_1(n)\nonumber
   \\ 
   &-\frac{\left(236
   n^3+27 n^2-278 n-9\right) S_2(n)}{27 (n-1) n (n+1)}-\frac{S_3(n)}{2}+\frac{8}{9}
   S_{-2,1}(n)+\frac{16}{9} S_{1,-2}(n)\nonumber
   \\ 
   &+(1-\xi ) \Big\{-\frac{14 n^6+16 n^5+n^4-23
   n^3-6 n^2+7 n+3}{54 (n-1)^2 n^2 (n+1)^2}\nonumber
   \\
   &+\frac{\left(19 n^3+6 n^2-7 n-6\right)
   S_1(n)}{54 (n-1) n (n+1)}\nonumber
   \\ 
   &+\frac{S_2(n)}{9}-\frac{1}{6}
   S_{1,1}(n)\Big\}+\frac{\left(547 n^4+1445 n^3+116 n^2-1172 n-576\right)
   S_{1,1}(n)}{108 (n-1) n (n+1) (n+2)}\nonumber
   \\ 
   &+\frac{\left(28 n^2+59 n+25\right)
   S_{1,2}(n)}{9 (n+1)^2}+\frac{\left(29 n^2+55 n+32\right) S_{2,1}(n)}{9
   (n+1)^2}-\frac{25}{18} S_{1,1,1}(n)\nonumber
   \\ 
   &+\frac{2 \left(2 n^2+5 n+1\right) \zeta_3}{(n+1)^2}\bigg)  +C_F C_A N_f \bigg(\frac{1}{9
   (n-1) n^4 (n+1)^4 (n+2)^3}\Big(67 n^{12}+603 n^{11}\nonumber
   \\ 
   &+2168 n^{10}+3846
   n^9+3021 n^8-387 n^7-3292 n^6-4758 n^5-4748 n^4-3384 n^3\nonumber
   \\ 
   &-2544 n^2-1536 n-384\Big)-\frac{32 S_{-2}(n)}{3 (n-1) n (n+1) (n+2)}-\frac{4
   S_1(n)}{3}-\frac{16 \zeta_3}{3}\bigg)   \bigg] \,.
 \end{align}
In view of applications at fixed $n$, we also summarize
 $Z_{qA}$ and $Z_{gA}$ for $n=2,\ldots,10$,
\begin{align}
    \frac{Z_{qA}\Big|_{n=2}}{C_A} =&a_s^2 N_f 
   \left(\frac{1}{6 \epsilon ^2}-\frac{2 }{9 \epsilon}\right) + a_s^3 N_f \bigg\{\frac{1}{\epsilon^3} \left(\frac{4  N_f}{27}+\frac{4 C_F}{27}-\frac{7
   C_A}{12}\right)\nonumber
   \\ 
   &+\frac{1}{\epsilon^2} \bigg[-\frac{5  N_f}{27}+\frac{7 
   C_F}{27}+\left(\frac{61}{72}-\frac{5 (1-\xi )}{216}\right)
   C_A\bigg]\nonumber
   \\ 
   &+\frac{1}{\epsilon} \bigg[-\frac{67  N_f}{486}+\left(\frac{4 \zeta_3}{3}-\frac{895}{486}\right)  C_F\nonumber
   \\ 
   &+C_A \left(\frac{7 (1-\xi )}{432}-\zeta_3+\frac{329}{432}\right)\bigg]\bigg\} \,,
   \\ 
 \frac{Z_{qA}\Big|_{n=4}}{C_A} =&a_s^2 N_f\left(\frac{11 }{720 \epsilon
   ^2}-\frac{817 }{43200 \epsilon }\right) +a_s^3 N_f  \bigg\{\frac{1}{\epsilon
   ^3}\bigg[ \frac{11  N_f}{810}+\frac{1727 
   C_F}{64800}-\frac{3773 C_A}{129600}\bigg] \nonumber
   \\ 
   &+\frac{1}{\epsilon
   ^2}\bigg[-\frac{601  N_f}{48600}+\frac{209633 
   C_F}{3888000}+\left(-\frac{11 (1-\xi
   )}{28800}-\frac{14989}{2592000}\right) C_A \bigg]  \nonumber
   \\ 
   & 
   + \frac{1}{\epsilon
   } \bigg[ -\frac{58907 
   N_f}{2916000}+\left(\frac{11 \zeta
   _3}{90}-\frac{7092241}{29160000}\right) C_F\nonumber
   \\ 
   & +C_A
   \left(-\frac{6179 (1-\xi )}{1728000}-\frac{11 \zeta
   _3}{120}+\frac{4233367}{29160000}\right)\bigg]  \bigg\} \,, 
   \\
   \frac{Z_{qA}\Big|_{n=6}}{C_A} = &a_s^2 N_f \bigg\{\frac{11 }{2520 \epsilon^2 }-\frac{221 }{42336 \epsilon }\bigg\}+a_s^3 N_f \bigg\{\frac{1}{\epsilon^3} \bigg[\frac{11
   N_f}{2835}+\frac{7799  C_F}{793800}-\frac{2123
   C_A}{423360}\bigg]\nonumber
   \\ 
   &+\frac{1}{\epsilon^2} \bigg[-\frac{3263
    N_f}{1190700}+\frac{1585219 
   C_F}{66679200}+\left(\frac{209 (1-\xi
   )}{2540160}-\frac{2683601}{177811200}\right)
   C_A\bigg]\nonumber
   \\ 
   &+\frac{1}{\epsilon} \bigg[-\frac{3611983 
   N_f}{500094000}+\left(\frac{11 \zeta
   _3}{315}-\frac{11494299613}{140026320000}\right) 
   C_F\nonumber
   \\ 
   &+C_A  \left(-\frac{1533541 (1-\xi
   )}{1066867200}-\frac{11 \zeta
   _3}{420}+\frac{699651481}{14002632000}\right)\bigg]\bigg\} \,, 
   \\
   \frac{Z_{qA}\Big|_{n=8}}{C_A} = & a_s^2 N_f
   \left(\frac{37  }{20160 \epsilon^2 }-\frac{15529 }{7257600 \epsilon}\right) + a_s^3 N_f \bigg\{\nonumber
   \\
   &\frac{1}{\epsilon^3} \bigg[\frac{37  N_f}{22680}+\frac{365671
   C_F}{76204800}-\frac{35039 C_A }{30481920}\bigg] \nonumber
   \\ 
   &+\frac{1}{\epsilon^2}
   \bigg[-\frac{10393  N_f}{11430720}+\frac{119144021 
   C_F}{9144576000}+\left(\frac{3737 (1-\xi
   )}{43545600}-\frac{1352619283}{128024064000}\right) C_A\bigg]\nonumber
   \\ 
   &+\frac{1}{\epsilon}
   \bigg[-\frac{506429639  N_f}{144027072000}+\left(\frac{37 \zeta
   _3}{2520}-\frac{574815689173}{15122842560000}\right)  C_F \nonumber
   \\ 
   &+C_A
   \left(-\frac{26006693 (1-\xi )}{36578304000}-\frac{37 \zeta
   _3}{3360}+\frac{2769495158803}{120982740480000}\right)\bigg]\bigg\}  \,, 
   \\
  \frac{Z_{qA}\Big|_{n=10}}{C_A} =&a_s^2  N_f
   \left(\frac{7  }{7425 \epsilon^2 }-\frac{2641 }{2450250 \epsilon }\right)+ a_s^3  N_f \bigg\{ \nonumber
   \\ 
   &\frac{1}{\epsilon^3} \bigg[\frac{56  N_f}{66825}+\frac{2411 
   C_F}{882090}-\frac{37199 C_A }{176418000}\bigg] \nonumber
   \\ 
   & +\frac{1}{\epsilon^2}
   \bigg[-\frac{24421  N_f}{66156750}+\frac{820067201 
   C_F}{101881395000}+\left(\frac{2491 (1-\xi
   )}{39204000}-\frac{35096312513}{4890306960000}\right) C_A\bigg]\nonumber
   \\ 
   &+\frac{1}{\epsilon}
   \bigg[-\frac{3690429931  N_f}{1833865110000}+\left(\frac{56 \zeta
   _3}{7425}-\frac{352770630827569}{16944913616400000}\right)  C_F \nonumber
   \\ 
   & +C_A
   \left(-\frac{9807967 (1-\xi )}{24149664000}-\frac{14 \zeta_3}{2475}+\frac{23337173703212867}{1897830325036800000}\right)\bigg] \bigg\}\,,
\end{align}

\begin{align}
    \frac{Z_{gA}\Big|_{n=2}}{C_A} =&-\frac{a_s}{2 \epsilon} +a_s^2
   \bigg\{\frac{1}{\epsilon^2} \left(\frac{19 C_A}{24}-\frac{N_f}{3}\right)+\frac{1}{\epsilon}
   \bigg[\left(\frac{5 (1-\xi )}{48}-\frac{35}{48}\right) C_A+\frac{7
   N_f}{18}\bigg]\bigg\} \nonumber
   \\ 
   & +  a_s^3 \bigg\{\frac{1}{\epsilon^3} \bigg[ \frac{71 C_A N_f}{54}-\frac{779
   C_A^2}{432}-\frac{4 C_F N_f}{27}-\frac{2 N_f^2}{9}\bigg] \nonumber
   \\ 
   &+\frac{1}{\epsilon^2}
   \bigg[\left(\frac{5 (1-\xi )}{108}-\frac{73}{36}\right) C_A
   N_f \nonumber
   \\ 
   &+\left(\frac{5}{288} (1-\xi )^2-\frac{35 (1-\xi )}{216}+\frac{2807}{864}\right)
   C_A^2-\frac{16 C_F N_f}{27}+\frac{7 N_f^2}{27}\bigg]\nonumber
   \\ 
   & +\frac{1}{\epsilon} \bigg[C_A N_f
   \bigg(-\frac{7}{216} (1-\xi )+\frac{19 \zeta _3}{9}-\frac{194}{243}\bigg)+C_A^2
   \bigg(-\frac{65 (1-\xi )^2}{1728}-\frac{16759}{7776}  \nonumber
   \\ 
   &-\frac{11 \zeta
   _3}{72} +(1-\xi ) \Big(\frac{5 \zeta
   _3}{72}+\frac{377}{1728}\Big)\bigg)+\left(\frac{3059}{972}-\frac{8 \zeta
   _3}{3}\right) C_F N_f+\frac{11 N_f^2}{54}\bigg] \bigg\} \,, 
   \label{eq:zgAn2}
    \\
    \frac{Z_{gA}\Big|_{n=4}}{C_A} =&-\frac{ a_s}{12 \epsilon } +a_s^2
   \bigg\{\frac{1}{\epsilon^2} \bigg[-\frac{97
   C_A}{1440}-\frac{N_f}{18}\bigg] +\frac{1}{\epsilon} \bigg[\left(\frac{1-\xi
   }{320}-\frac{8641}{86400}\right) C_A+\frac{7
   N_f}{216}\bigg] \bigg\} \nonumber
   \\ 
   &+  a_s^3 \bigg\{  \frac{1}{\epsilon^3} \bigg[\frac{C_A N_f}{324}+\frac{9437
   C_A^2}{86400}-\frac{121 C_F N_f}{32400}-\frac{N_f^2}{27}\bigg]\nonumber
   \\ 
   &+\frac{1}{\epsilon^2}
   \bigg[\left(\frac{1-\xi }{720}-\frac{277}{129600}\right) C_A N_f \nonumber
   \\ 
   & +\left(-\frac{13
   (1-\xi )^2}{23040}+\frac{853 (1-\xi )}{86400}-\frac{1520341}{15552000}\right)
   C_A^2-\frac{224719 C_F N_f}{1944000}+\frac{7 N_f^2}{324}\bigg]\nonumber
   \\ 
   &  +\frac{1}{\epsilon}
   \bigg[C_A N_f \left(\frac{457 (1-\xi )}{43200}+\frac{53 \zeta
   _3}{150}-\frac{5748673}{23328000}\right)+C_A^2 \bigg(-\frac{2357 (1-\xi
   )^2}{460800} \nonumber
   \\ 
   & +(1-\xi ) \Big(\frac{37 \zeta
   _3}{9600}+\frac{37199}{648000}\Big)-\frac{\zeta_3}{2400}-\frac{166178237}{466560000}\bigg) \nonumber
   \\ 
   & +\left(\frac{11814181}{29160000}-\frac
   {4 \zeta _3}{9}\right) C_F N_f+\frac{193 N_f^2}{3888}\bigg] \bigg\}\,, 
    \label{eq:zgAn4}
    \\
     \frac{Z_{gA}\Big|_{n=6}}{C_A} =& -\frac{ a_s}{30 \epsilon} +a_s^2 \bigg\{\frac{1}{\epsilon^2} \left(-\frac{653
   C_A}{10080}-\frac{N_f}{45}\right)\nonumber
   \\ 
   &+\frac{1}{\epsilon} \bigg[\left(-\frac{19 (1-\xi
   )}{20160}-\frac{185093}{4233600}\right) C_A+\frac{13
   N_f}{2700}\bigg] \bigg\} \nonumber
   \\ 
   &   + a_s^3 \bigg\{\frac{1}{\epsilon^3} \bigg[-\frac{22 C_A N_f}{567}+\left(\frac{7 (1-\xi
   )}{57600}+\frac{600331}{12700800}\right) C_A^2-\frac{121 C_F N_f}{198450}-\frac{2
   N_f^2}{135}\bigg] \nonumber
   \\ 
   & +\frac{1}{\epsilon^2} \bigg[\left(\frac{11761}{381024}-\frac{19
   (1-\xi )}{45360}\right) C_A N_f  -\frac{253187 C_F
   N_f}{5556600}+\frac{13 N_f^2}{4050}\nonumber
   \\ 
   & +\left(-\frac{37 (1-\xi )^2}{48384}+\frac{585157
   (1-\xi )}{169344000}-\frac{275019229}{1778112000}\right) C_A^2\bigg] \nonumber
   \\ 
   & +\frac{1}{\epsilon} \bigg[C_A N_f
   \left(\frac{88439 (1-\xi )}{19051200}+\frac{103 \zeta
   _3}{735}-\frac{203978813}{2000376000}\right)+C_A^2 \bigg(-\frac{195683 (1-\xi
   )^2}{101606400}\nonumber
   \\ 
   & +(1-\xi ) \Big(\frac{61 \zeta
   _3}{35280}+\frac{4629063223}{213373440000}\Big)+\frac{1219 \zeta
   _3}{176400}-\frac{385478104231}{2240421120000}\bigg)\nonumber
   \\ 
   &  +\left(\frac{4970689127}{3500
   6580000}-\frac{8 \zeta _3}{45}\right) C_F N_f+\frac{5231
   N_f^2}{243000}\bigg] \bigg\}\,,    \label{eq:zgAn6}
    \\
    \frac{Z_{gA}\Big|_{n=8}}{C_A} =& -\frac{ a_s}{56 \epsilon}+a_s^2 \bigg\{\frac{1}{\epsilon^2} \left(-\frac{2749
   C_A}{56448}-\frac{N_f}{84}\right)  \nonumber
   \\ 
   &  +\frac{1}{\epsilon} \bigg[ \left(-\frac{101 (1-\xi
   )}{80640}-\frac{18855769}{711244800}\right) C_A-\frac{43
   N_f}{70560}\bigg]  \bigg\}  + a_s^3 \bigg\{ \frac{1}{\epsilon^3} \bigg[\nonumber
   \\ 
   &-\frac{11281 C_A N_f}{317520} +\left(\frac{11
   (1-\xi )}{94080}+\frac{26350343}{2133734400}\right) C_A^2-\frac{1369 C_F
   N_f}{7620480}-\frac{N_f^2}{126}\bigg]\nonumber
   \\ 
   &  +\frac{1}{\epsilon^2}
   \bigg[\left(\frac{68655941}{3200601600}-\frac{101 (1-\xi )}{181440}\right) C_A
   N_f+ C_A^2 \bigg(-\frac{1151 (1-\xi )^2}{1935360} \nonumber
   \\ 
   &+\frac{112333 (1-\xi
   )}{213373440}-\frac{25862266399}{199148544000}\bigg)-\frac{22126259 C_F
   N_f}{914457600}-\frac{43 N_f^2}{105840}\bigg]\nonumber
   \\ 
   & +\frac{1}{\epsilon} \bigg[C_A N_f
   \left(\frac{1022971 (1-\xi )}{457228800}+\frac{169 \zeta
   _3}{2268}-\frac{48152267873}{896168448000}\right) \nonumber
   \\ 
   & +C_A^2 \bigg(-\frac{4988429
   (1-\xi )^2}{4877107200}+(1-\xi ) \Big(\frac{295 \zeta
   _3}{290304}+\frac{333783455249}{32262064128000}\Big) \nonumber
   \\ 
   &+\frac{14029 \zeta
   _3}{2540160}-\frac{730008394755263}{6775033466880000}\bigg) \nonumber
   \\ 
   & +\left(\frac{835100678
   591}{12098274048000}-\frac{2 \zeta _3}{21}\right) C_F N_f+\frac{1028201
   N_f^2}{88905600}\bigg] \bigg\} \,, 
    \\
     \frac{Z_{gA}\Big|_{n=10}}{C_A} =&-\frac{1}{90} \frac{1}{\epsilon} a_s  +a_s^2 \bigg\{\frac{1}{\epsilon^2} \left(-\frac{ N_f}{135}-\frac{369361 C_A}{9979200}\right) \nonumber
   \\ 
   &+\frac{1}{\epsilon} \bigg[\left(-\frac{2491 (1-\xi
   )}{2217600}-\frac{1030030931}{55324684800}\right) C_A-\frac{661 
   N_f}{340200}\bigg]\bigg\} \nonumber
   \\ 
   &+ a_s^3 \bigg\{\frac{1}{\epsilon^3} \bigg[-\frac{784  C_F N_f}{11026125}-\frac{327427
   C_A N_f}{11226600}-\frac{2}{405}  N_f^2\nonumber
   \\ 
   &+\left(\frac{143 (1-\xi
   )}{1451520}-\frac{2899927799}{829870272000}\right) C_A^2\bigg]+\frac{1}{\epsilon^2}
   \bigg[-\frac{27254752  C_F
   N_f}{1819310625}\nonumber
   \\ 
   &+\left(\frac{2071551877}{155600676000}-\frac{2491 (1-\xi
   )}{4989600}\right) C_A N_f-\frac{661  N_f^2}{510300}\nonumber
   \\ 
   &+\left(-\frac{12121
   (1-\xi )^2}{26611200}-\frac{100407731 (1-\xi
   )}{147532492800}-\frac{15087045196453}{142000024320000}\right)
   C_A^2 \bigg]\nonumber
   \\ 
   &+\frac{1}{\epsilon}
   \bigg[ \left(\frac{167914071891373}{4236228404100000}-\frac{8 \zeta _3}{135}\right)
    C_F N_f\nonumber
   \\ 
   & +C_A  N_f \left(\frac{158792849 (1-\xi )}{138311712000}+\frac{251
   \zeta _3}{5445}-\frac{560285942701061}{17253002954880000}\right) \nonumber
   \\ 
   &+C_A^2 \bigg( -\frac{79286117 (1-\xi )^2}{122943744000}+(1-\xi )
   \Big(\frac{29 \zeta
   _3}{43560}+\frac{67751391198259}{12268802101248000}\Big)\nonumber
   \\ 
   &+\frac{1333 \zeta
   _3}{338800}-\frac{24201974825379990307}{318835494606182400000}\bigg) +\frac{8972717 
   N_f^2}{1285956000}\bigg]\bigg\} \,.
\end{align}

\section{Standard QCD renormalization constants}
\label{sec:QCDren}
In our computations, we need the QCD beta function to two-loop order~\cite{Caswell:1974gg,Jones:1974mm},
\begin{align}
  &\beta_0 = \frac{11 C_A}{3} - \frac{2 N_f}{3} \,,  
  \\
   & \beta_1 = -\frac{10 C_A N_f}{3}+\frac{34 C_A^2}{3}-2 C_F N_f \,,
\end{align}
where $\beta_i$ is defined with the following convention, 
\begin{align}
    \frac{d a_s}{ d \ln \mu} = \beta(a_s) = -2 a_s \sum_{i=0}^\infty a_s^{i+1} \beta_i. 
\end{align}

The computations of off-shell OMEs requires also wave function renormalizations. The quark and gluon field renormalization constants in the $\overline{\text{MS}}$ scheme are needed up to three-loop order~\cite{Tarasov:1980au,Larin:1993tp}, and the ghost field renormalization constant is required up to two-loop order. We list all these ingredients in our convention~(\ref{eq:zexp}):
\begin{align}
 Z_q^{(0)}  = &1 \,, 
     \\
     Z_q^{(1)} = &- C_F \frac{\xi}{\epsilon}  \,, 
    \\
      {Z_q^{(2)}} = &  \frac{C_F}{\epsilon} \bigg[\left(-\frac{\xi ^2}{8}-\xi -\frac{25}{8}\right)
   C_A+\frac{3 C_F}{4}+\frac{N_f}{2}\bigg]+\frac{C_F}{\epsilon^2} \bigg[\left(\frac{\xi
   ^2}{4}+\frac{3 \xi }{4}\right) C_A+\frac{\xi ^2 C_F}{2}\bigg] \,, 
   \\
    {Z_q^{(3)}} = & \frac{C_F}{\epsilon^3} \bigg[C_A \left(\Big(-\frac{\xi ^3}{4}-\frac{3 \xi
   ^2}{4}\Big) C_F+\frac{\xi  N_f}{6}\right)+\left(-\frac{\xi ^3}{12}-\frac{3 \xi
   ^2}{8}-\frac{31 \xi }{24}\right) C_A^2-\frac{1}{6} \xi ^3
   C_F^2\bigg] \nonumber 
   \\
   &+\frac{C_F}{\epsilon^2} \bigg[C_A \left(\left(\frac{\xi ^3}{8}+\xi
   ^2+\frac{25 \xi }{8}-\frac{11}{6}\right) C_F+\left(-\frac{\xi
   }{2}-\frac{47}{18}\right) N_f\right) \nonumber 
   \\
   &+\left(\frac{\xi ^3}{8}+\frac{3 \xi
   ^2}{4}+\frac{73 \xi }{24}+\frac{275}{36}\right) C_A^2+\left(\frac{1}{3}-\frac{\xi
   }{2}\right) C_F N_f-\frac{3 \xi  C_F^2}{4}+\frac{2 N_f^2}{9}\bigg]  \nonumber 
   \\
   & +\frac{C_F}{\epsilon}
   \bigg[C_A \left(\Big(\frac{143}{12}-4 \zeta _3\Big) C_F+\Big(\frac{17 \xi
   }{24}+\frac{287}{54}\Big) N_f\right)-\frac{C_F N_f}{2}-\frac{C_F^2}{2} -\frac{5
   N_f^2}{27}\nonumber 
   \\
   & +C_A^2 \bigg\{-\frac{5 \xi ^3}{48}+\xi ^2
   \left(-\frac{\zeta _3}{8}-\frac{13}{32}\right)+\xi  \left(-\frac{\zeta
   _3}{4}-\frac{263}{96}\right)+\frac{23 \zeta
   _3}{8}-\frac{9155}{432}\bigg\}\bigg] \,,
\\[2ex]
 Z_g^{(0)}  = &1 \,, 
     \\
    Z_g^{(1)}  = &\frac{1}{\epsilon}
   \bigg[ \left(\frac{13}{6}-\frac{\xi }{2}\right) C_A-\frac{2 N_f}{3}\bigg] \,, 
   \\
     Z_g^{(2)}  =  &\frac{1}{\epsilon} \bigg[-\frac{5 C_A N_f}{4}+\left(-\frac{\xi
   ^2}{8}-\frac{11 \xi }{16}+\frac{59}{16}\right) C_A^2-C_F
   N_f\bigg]\nonumber 
   \\
   & +\frac{1}{\epsilon^2} \bigg[\left(\frac{\xi }{3}+\frac{1}{2}\right) C_A
   N_f+\left(\frac{\xi ^2}{4}-\frac{17 \xi }{24}-\frac{13}{8}\right)
   C_A^2\bigg] \,, 
   \\
    Z_g^{(3)}  =  & \frac{1}{\epsilon^2} \bigg[C_F \bigg\{\left(\frac{\xi
   }{2}+\frac{31}{18}\right) C_A N_f-\frac{2 N_f^2}{9}\bigg\}+\left(\frac{\xi
   ^2}{12}+\frac{19 \xi }{24}+\frac{481}{108}\right) C_A^2 N_f-\frac{25}{54} C_A
   N_f^2\nonumber 
   \\
   &+\left(\frac{7 \xi ^3}{48}+\frac{13 \xi ^2}{24}-\frac{143 \xi
   }{96}-\frac{7957}{864}\right) C_A^3\bigg]+\frac{1}{\epsilon} \bigg[C_F \bigg\{\left(-4
   \zeta _3-\frac{5}{108}\right) C_A N_f+\frac{11 N_f^2}{27}\bigg\}\nonumber 
   \\
   & +C_A^2 N_f
   \left(\frac{\xi }{3}+3 \zeta _3-\frac{911}{108}\right)+\frac{19}{27} C_A
   N_f^2+C_A^3 \bigg\{-\frac{7 \xi ^3}{96}+\xi ^2 \left(-\frac{\zeta
   _3}{16}-\frac{11}{32}\right)\nonumber 
   \\
   &+\xi  \left(-\frac{\zeta
   _3}{4}-\frac{167}{96}\right)-\frac{3 \zeta
   _3}{16}+\frac{9965}{864}\bigg\}+\frac{1}{3} C_F^2 N_f\bigg]+\frac{1}{\epsilon^3}
   \bigg[\left(-\frac{\xi ^2}{6}-\frac{5 \xi }{12}-\frac{11}{9}\right) C_A^2
   N_f\nonumber 
   \\
   &+\frac{1}{9} C_A N_f^2+\left(-\frac{\xi ^3}{8}+\frac{\xi ^2}{6}+\frac{47 \xi
   }{48}+\frac{403}{144}\right) C_A^3\bigg] \,, 
\\[2ex]
     Z_c^{(0)}  = &1 \,, 
     \\
      Z_c^{(1)}  = &\frac{1}{\epsilon}
   \left(\frac{3}{4}-\frac{\xi }{4}\right) C_A\,, 
   \\
   Z_c^{(2)}  = &\frac{1}{\epsilon^2} \bigg[\frac{C_A
   N_f}{4}+\left(\frac{3 \xi ^2}{32}-\frac{35}{32}\right)
   C_A^2\bigg]+\frac{1}{\epsilon} \bigg[ \left(\frac{\xi
   }{32}+\frac{95}{96}\right) C_A^2-\frac{5 C_A
   N_f}{24}\bigg]\,. 
\end{align}

\section{Instructions for ancillary files}
\label{sec:Ancillary}
The \texttt{Mathematica} files '\textbf{NSingletOMEs.m}' and '\textbf{SingletOMEs.m}' contain all results for two-parton OMEs in the non-singlet and singlet case, respectively. All OMEs are normalized according to \eqref{eq:sumoverSpinColorF}, and expanded to order $a_s^j \epsilon^{3-j}$ ($j \leq 3$), where the highest order is relevant only at the four-loop level. A \texttt{Mathematica} notebook '\textbf{ExtractSpFromOMEs.nb}' is used to combine all OMEs and to derive the physical anomalous dimensions. In the same notebook, we also compare our results with the literature results, which we assembled in '\textbf{RefNSsp.m}' and '\textbf{RefSingletsp.m}', finding perfect agreement for both, the non-singlet and the singlet physical anomalous dimensions. The file '\textbf{FRzogGV2.m}' contains the result for the last line of equation~\eqref{eq:FeynRules3gZOgVV2}. We also provide the results for the renormalization constants $Z_{qA}$ and $Z_{gA}$ shown in Appendix~\ref{sec:zqAzgA} in the files '\textbf{zqA.m}' and '\textbf{zgA.m}', respectively.  

The generic notation for two-parton OME with an insertion of a general twist-two operator is defined in \eqref{eq:OMEsDe}.
More explicitly, for OMEs with an operator insertion of $O_q$ or $O_g$, we follow this notation closely, for example,
\begin{align}
    A_{qg} &= \braket{g(p)|O_q|g(p)},\,\nonumber \\
    A_{gc} &= \braket{c(p)|O_g|c(p)}\,,
\end{align}
while for OMEs with the insertion of GV counterterms, we use slightly different notations, for example,
\begin{align}
    A_{ABC,\,g} &= \braket{g(p)|O_{ABC}|g(p)},\,\nonumber\\
    A_{\left[Z O\right]_{g}^{\textrm{GV},\,(2)},\, c} &= \braket{c(p)|\left[Z O\right]_{g}^{\textrm{GV},\,(2)}|c(p)}\,.
\end{align}
Similarly, in \eqref{eq:sumoverSpinColorF} we use $\mathcal{F}_{ABC,\, q},\,
\mathcal{F}_{\left[Z O\right]_{g}^{\textrm{GV},\,(2)},\, q}$, and so on to represent form factors with the insertion of a GV counterterm. The above OMEs apply to the singlet case only.
In the non-singlet case, we need a single OME, 
\begin{align}
    \mathcal{F}_{ns} = \frac{1}{2 N_c} \frac{\text{Tr} \left(  \slashed{p} \,  \braket{q(p)|O_{q,k}|q(p)} \right)}{(\Delta \cdot p )^n}\,.
\end{align}
In the following tables, we list all objects that appear in this paper and their notation in the ancillary files. 
\begin{center}
\begin{minipage}{7cm}
\begin{tabular}{ |c | c | } 
\hline
    Object & In ancillary files \\
    \hline
      $\mathcal{F}_{ns}$ & \textbf{Fns[q, q]} \\
   $\mathcal{F}_{qq}$ & \textbf{Fs[q, q]} \\ 
   $\mathcal{F}_{qg}$ & \textbf{Fs[q, g]}  \\ 
  $\mathcal{F}_{qc}$ & \textbf{Fs[q, c]} \\
     $\mathcal{F}_{gq}$ & \textbf{Fs[g, q]} \\ 
   $\mathcal{F}_{gg}$ & \textbf{Fs[g, g]}  \\ 
  $\mathcal{F}_{gc}$ & \textbf{Fs[g, c]} \\
     $\mathcal{F}_{ABC,\,q}$ & \textbf{Fs[ABC, q]} \\ 
   $\mathcal{F}_{ABC,\,g}$ & \textbf{Fs[ABC, g]}  \\ 
  $\mathcal{F}_{ABC,\,c}$ & \textbf{Fs[ABC, c]} \\
     $\mathcal{F}_{\left[Z O\right]_{g}^{\textrm{GV},\,(2)},\,q}$ & \textbf{Fs[zogGV2, q]} \\ 
   $\mathcal{F}_{\left[Z O\right]_{g}^{\textrm{GV},\,(2)},\,g}$ & \textbf{Fs[zogGV2, g]}  \\ 
  $\mathcal{F}_{\left[Z O\right]_{g}^{\textrm{GV},\,(2)},\,c}$ & \textbf{Fs[zogGV2, c]} \\
  \hline
\end{tabular}
\end{minipage}
\begin{minipage}{7cm}
\begin{tabular}{ |c | c | } 
\hline
    Object & In ancillary files \\
    \hline
    $z_1$ & \textbf{z1} \\
  $a_s$ & \textbf{as} \\
  $a_s^b$ & \textbf{asb} \\
    $\xi$ & \textbf{xi} \\
  $\xi^b$ & \textbf{xib} \\
    $\epsilon$ & \textbf{eps} \\
  $C_A$ & \textbf{ca} \\
    $C_F$ & \textbf{cf} \\
  $N_f$ & \textbf{nf} \\
      $16 d^{abc}d_{abc}$ & \textbf{d33c} \\
        $T_F$ & \textbf{tf} \\
  Harm.\ sum $S_{...}(n)$ & \textbf{S[...,\,n]} \\
  G. H.\ sum $S_{...}(..;n)$ & \textbf{S[...,\,\{..\},\,n]}\\
   Riemann $\zeta_n$ & \textbf{zeta[n]} \\
    \hline
\end{tabular}
\end{minipage}
\end{center}  

\bibliographystyle{JHEP}
\bibliography{Renormalization}

\end{document}